\documentclass[11pt,a4paper]{article}
\usepackage{epsfig}
\usepackage[utf8]{inputenc}
\usepackage{graphics}
\usepackage{graphicx}
\usepackage{pstricks,pst-coil,pst-fill,pst-plot}
\usepackage{cite,indentfirst}
\usepackage[fleqn]{amsmath}    
\usepackage{amssymb}    
\usepackage{amsfonts}   
\usepackage{verbatim}   
\usepackage{mathrsfs}   
\usepackage{dsfont}
\usepackage{euscript}
\usepackage{yfonts}
\usepackage{enumerate}     
\usepackage{amsthm}         
\usepackage{txfonts}
\usepackage{marvosym}
\usepackage{stmaryrd}
\usepackage{vmargin}        
\usepackage{wasysym}		
\usepackage{mathtools}      

\setmarginsrb{1.8cm}{2cm}{1.8cm}{2cm}{1cm}{1cm}{1cm}{1.6cm}
 \makeatletter
 \@addtoreset{equation}{section}
 \makeatother

\mathtoolsset{showonlyrefs}   

\providecommand{\bysame}{\leavevmode\hbox to3em{\hrulefill}\thinspace}
\providecommand{\MR}{\relax\ifhmode\unskip\space\fi MR }

\providecommand{\href}[2]{#2}

\let\tend=\rightarrow

\long\def\symbolfootnote[#1]#2{\begingroup%
\def\thefootnote{\fnsymbol{footnote}}\footnote[#1]{#2}\endgroup}

\newtheorem*{theorem*}{Theorem}

\newcommand\beq{\begin{equation}}
\newcommand\enq{\end{equation}}
\newcommand\bem{\begin{multline}}
\newcommand\enm{\end{multline}}

\def\beqa{\begin{eqnarray}}
\def\eeqa{\end{eqnarray}}
\def\ba{\begin{array}}
\def\ea{\end{array}}
\def\det{\operatorname{det}}

\newcommand{\f}[2]{{\ensuremath{%
    \mathchoice%
    {\dfrac{#1}{#2}}
    {\dfrac{#1}{#2}}
    {\frac{#1}{#2}}
    {\frac{#1}{#2}}
}}}
\newcommand{\tf}[2]{\ensuremath{#1/#2}}

\def\a{\alpha}

\def\be{\beta}
\def\ga{\gamma}
\def\Ga{\Gamma}

\def\de{\delta}

\def\De{\Delta}
\def\eps{\epsilon}
\def\veps{\varepsilon}
\def\la{\lambda}
\def\La{\Lambda}

\def\sg{\sigma}

\def\Ups{\Upsilon}
\def\ups{\upsilon}
\def\th{\theta}

\def\vth{\vartheta}

\def\Om{\Omega}
\def\om{\omega}
\def\vp{\varphi}

\newcommand{\mc}[1]{\ensuremath{\mathcal{#1}}}
\newcommand{\mf}[1]{\ensuremath{\mathfrak{#1}}}
\newcommand{\msc}[1]{\ensuremath{\mathscr{#1}}}

\newcommand{\bs}[1]{\ensuremath{\boldsymbol{#1}}}

\DeclareFontFamily{OT1}{pzc}{}
\DeclareFontShape{OT1}{pzc}{m}{it}{<-> s * [1.10] pzcmi7t}{}
\DeclareMathAlphabet{\mathpzc}{OT1}{pzc}{m}{it}

\def \i{ \mathrm i}

\newcommand{\wt}[1]{\ensuremath{\widetilde{#1}}}
\newcommand{\wh}[1]{\ensuremath{\widehat{#1}}}

\newcommand{\Int}[2]{\ensuremath{\int\limits_{#1}^{#2}}}

\newcommand{\Fint}[2]{\ensuremath{\fint\limits_{#1}^{#2}}}

\newcommand{\sul}[2]{\ensuremath{\sum\limits_{#1}^{#2}}}
\newcommand{\pl}[2]{\ensuremath{\prod\limits_{#1}^{#2}}}

\newcommand{\R}{\ensuremath{\mathbb{R}}}
\newcommand{\Cx}{\ensuremath{\mathbb{C}}}

\newcommand{\Dp}[1]{\ensuremath{\partial_{#1}}}

\newcommand{\limit}[2]{\ensuremath{\underset{#1 \tend #2}{\longrightarrow} }}

\newcommand{\ex}[1]{\ensuremath{\e{e}^{#1}}}

\newcommand{\op}[1]{ \boldsymbol{ \texttt{#1} } }

\newcommand{\norm}[1]{\ensuremath{  || #1 || }}

\newcommand{\dd}{\mathrm{d}}
\newcommand{\e}[1]{\ensuremath{\mathrm{#1}}}

\newcommand{\intff}[2]{\ensuremath{ [  #1 \,; #2 ] }}

\begin{document}

\begin{center}
\begin{LARGE}
{\bf Correlation functions of the Lieb--Liniger Bose gas in
equilibrium states: a microscopic approach to space-time~asymptotics}
\end{LARGE}

\vspace{1cm}

\vspace{4mm}
{\large Frank Göhmann \footnote{e-mail: goehmann@uni-wuppertal.de}} \\[1ex]
Fakult\"at f\"ur Mathematik und Naturwissenschaften, Bergische Universit\"at Wuppertal, 42097 Wuppertal, Germany \\[2.5ex]

{\large Karol K. Kozlowski \footnote{e-mail: karol.kozlowski@ens-lyon.fr}} \\[1ex]
Univ Lyon, ENS de Lyon, Univ Claude Bernard Lyon 1, CNRS, Laboratoire de Physique, F-69342 Lyon, France \\[2.5ex]

{\large Mikhail D. Minin \footnote{e-mail: minin@uni-wuppertal.de}} \\[1ex]
Fakult\"at f\"ur Mathematik und Naturwissenschaften, Bergische Universit\"at Wuppertal, 42097 Wuppertal, Germany \\[2.5ex]

\par

\vspace{40pt}

\centerline{\bf Abstract} \vspace{1cm}
\parbox{12cm}{\small
We develop an approximation-free approach to the analysis of form-factor
series for the correlation functions of quantum integrable
models in the thermodynamic limit and in equilibrium states described
by a class of stationary density matrices. This allows us to obtain
the long-time, large-distance asymptotics of the correlation functions
in thermal equilibrium and in stationary states after a quench,
including such states that can be described by generalized Gibbs
ensembles. Our approach will be exemplified with the Lieb--Liniger
Bose gas or non-linear Schrödinger model. It applies to the correlation
functions of all local operators. For the two-point functions of
particle-density or particle-current density operators our findings confirm
the phenomenological predictions based on generalised hydrodynamics.
We extract as well the long-time, large-distance asymptotics of a field- conjugated field
correlation function. At present its behaviour is outside the reach of any
heuristic method. It is governed by a novel non-linear integral equation that
depends on space and time.

}

\vspace{20pt}

\end{center}

\vspace{40pt}

\tableofcontents

\section{Introduction}
The calculation of dynamical correlation functions of many-body quantum
systems in thermal equlibrium or in states described by more general
stationary density matrices is a notorious problem. Not too much is
known about it, even within the generally well-understood class of
integrable quantum systems in one spatial dimension. As far as the
dynamical correlation functions at finite temperature are concerned,
exact results are so far mostly available for the sub-class of models
that can be mapped to free Fermions
\cite{ItsIzerginKorepinVarzuginTimeSpaceAsymptImpBoseGaz,%
ItsIzerginKorepinSlavnovTempCorrFctSpinsXY,%
JieLargetABofDynCorrFctsinXX,%
KozGohmannSuzukiXXTransverseDynamicInfiniteT,%
KozGohmannSuzukiXXTransverseSpaceLikeAsymptotics,%
KozGohmannMininWeisseImpenetrableBoseGasAsymp}.
The only results we are aware of outside this sub-class is the
calculation of the long-time, large-distance asymptotics of
the transverse two-point correlation function of the XXZ chain in the
space-like regime in \cite{KozGohmannSpaceLikeAEFor2PtsFctsThermalXXZ} as well as the
calculation of the correlation length for the field-conjugated field correlator in the non-linear
Schrödinger model \cite{SlavnovComputationDualFieldVaccumExpLongTimeDistTempeRedDensNLSE}.

Even on the phenomenological level it took a long time before a
pratically satisfactory approach could be devised. This approach
comes under the name ``generalized hydrodynamics''. It was developed
about ten years ago in
\cite{BertiniColluraDeNardisFagottiGHDIntroonXXZEG,Castro-AlvaredoDoyonYoshimuraFirstIntroGDH}
and is reviewed, \textit{e.g.}, in
\cite{CortesCubertoSpohnYoshimuraFormFactorsAndGHDforIS,%
DoyonMedenjakNardisPanfilReviewonGHDAndCorrFcts,SpohnGGEBook}.
Unlike the phenomenological approaches designed for the description
of ground state and low-temperature correlation functions this
approach does not rely on any effective quantum field theory, but
is based on the thermodynamic Bethe Ansatz (TBA) and on the
existence of local conserved charges in the integrable models.
Among other things it provides simple and rather general formula
for the dynamical two-point functions of the densities and
current densities of the conserved charges
\cite{DoyonSpohnGDHPredictionsForNLSMDrudeWeight}. These are
reviewed below and are a reference point for our work.

We shall present an exact approach that allows us to obtain those
formulae from a microscopic reasoning. It will allow us to go to
higher orders, and it is not restricted to ``hydrodynamic variables''.
The latter will be illustrated with the example of the two-particle
reduced density matrix of the non-linear Schr\"odinger model for
which we shall obtain an asymptotic formula characterized by a
(dynamical) correlation length, its logarithmic correction, and
by an amplitude which are all represented in a form suitable for
a precise numerical evaluation. In fact, as common in integrable
models, they are determined by the solutions of linear and
nonlinear integral equations.

\subsection{The non-linear Schrödinger model and generalised hydrodynamics}

%

For definiteness and since this is the setting which is of interest to this work,
we shall describe the implications of the generalised hydrodynamic approach for the
non-linear Schrödinger model.  The associated Hamiltonian acts on the Fock space
$\mf{h}\, = \, \bigoplus_{N \geq 0} \mc{V}_N$ with $\mc{V}_N= L^2( \intff{0}{L}^N)$
and reads
\beq
\op{H} \,  = \, \Int{0}{L}\! \dd y \, \Big\{ \Dp{y} \Phi^{\dagger}(y)  \cdot  \Dp{y} \Phi(y) + c \, \Phi^{\dagger}(y)\Phi^{\dagger}(y) \Phi(y) \Phi(y)  \Big\} \; .
\label{definition Hamiltonien Bose}
\enq
In \eqref{definition Hamiltonien Bose}, $\Phi$ and $\Phi^{\dagger}$
are canonical Bose fields subject to periodic boundary conditions on $\intff{0}{L}$.

The Hamiltonian commutes with the total particle-number operator
\beq
\op{N}\, = \, \Int{0}{L} \! \dd y \: \Phi^{\dagger}(y)  \Phi(y)   \; .
\enq
$\op{H}$ is thus block diagonal with respect to the Fock space decomposition  $\bigoplus_{N \geq 0} \mc{V}_N$.
In fact, it also commutes with the total momentum operator
\beq
\op{P}\; = \; - \f{ \i }{ 2 } \Int{0}{L} \! \dd y \, \Big\{\Phi^{\dagger}(y)  \cdot  \Dp{y} \Phi(y) \, - \,   \Dp{y} \Phi^{\dagger}(y)  \cdot  \Phi(y)  \Big\}  \;,
\label{definition Total momentum}
\enq
which amounts to the existence of three local, mutually commuting charges
in the model:
\beq
\op{H}_0 = \op{N}\;, \quad \op{H}_1 \, = \, \op{P}, \qquad \e{and} \qquad \op{H}_2 \, = \,  \op{H}   \;.
\enq

As follows from taking the limit of lattice spacing to zero in the lattice
discretisation of the model as given in
\cite{DavidKorepinHigherCOnservationLawsNLSM,IzerginKorepinLatticeVersionsofQFTModelsABANLSEandSineGordon},
the model admits higher-order local conserved charges $\op{H}_g$, $g \geq 3$, which can be formally thought of as given by the below expression,
\beq
\op{H}_g \, = \, \Int{0}{L} \! \dd y \: \op{q}_g(y)  \qquad \e{with} \qquad  \op{q}_g(y) \, = \, \mc{H}_g\Big( \Phi(y), \Phi^{\dagger}(y),  \dots , \Dp{y}^{g-1}\Phi(y),  \Dp{y}^{g-1}\Phi^{\dagger}(y)\Big) \;,
\enq
for some explicit, non-commutative polynomial functions of the field and its
adjoint. The crucial property of those expressions is that the integrand
is \textit{local}, namely, it only involves the values of the fields along with
a finite number of its derivatives at the same point $y$.
While, by definition, the conserved charges $\op{H}_g $ are unaffected
by the Heisenberg time-evolution generated by $\op{H}$, their densities do evolve in time
$\op{q}_g(y) \hookrightarrow \op{q}_g(y,t)$. These time-dependent densities permit
to define their currents $\op{j}_g(y,t)$ through the continuity equation
\beq
\Dp{t}\op{q}_g(y,t) \, + \, \Dp{y}\op{j}_g(y,t) \, = \, 0 \;.
\enq

We shall consider the non-linear Schrödinger model in an equilibrium
state assumed after the relaxation following a quench.
This equilibrium is described by a density matrix $\bs{\varrho}$.
In any experimental realization of the model the unavoidable
coupling to an environment will finally drive the system toward
{\it thermal} equilibrium described by the Boltzmann statistical operator
$ \tf{ \ex{-\tf{\op{H}}{T}} }{ \mc{Z}_{B} }$. At intermediate time
scales its statistical operator may be of generalised Gibbs type,
$\tf{ \exp\Big\{-   \sul{k=1}{M} \tfrac{a_k }{T} \op{H}_k  \Big\} }{ \mc{Z}_{GGE} }$
for some $M>0$. The corresponding partition functions are defined as
\beq
\mc{Z}_{B} \, = \, \e{tr}_{\mf{h}}\Big[  \ex{-\tf{\op{H}}{T}} \Big]  \qquad \e{and} \qquad
\mc{Z}_{GGE} \, = \, \e{tr}_{\mf{h}}\Big[ \exp\Big\{-    \sul{k=1}{M} \tfrac{a_k }{T}  \op{H}_k  \Big\} \Big] \, .
\enq
More generally, one may even deal with density matrices outside of this setting,
see, \textit{e.g.}, \cite{BrockmannCauxDeNardisWoutersQuenchActionAndNonGGEeqDistrNLSM},
in particular, also with density matrices  that do not admit higher moments
(cf.\ the recent discussion in \cite{KozGohmannMininWeisseImpenetrableBoseGasAsymp}).
The correlation functions of the model are then given by
\beq
\big< \op{O} \big>_{ \bs{\varrho}  } \, = \, \lim_{L\tend + \infty} \Big\{ \e{tr}_{\mf{h}}\big[  \op{O}     \bs{\varrho}  \big]   \Big\} \;.
\label{definition moyenne contre etat densite rho}
\enq

In the above cases, generalised hydrodynamics allows one to predict the
leading space-time asymptotic behaviour of all connected correlation functions
involving the densities of conserved charges and the corresponding current 
densities, for instance, the connected two-point functions
\beq
\big<\op{q}_{k}(x,t) \op{q}_{\ell}(0,0) \big>_{ \bs{\varrho} ; \e{c} }  \;, \quad  \big<\op{q}_{k}(x,t) \op{j}_{\ell}(0,0) \big>_{ \bs{\varrho}  ; \e{c}   }  \quad \e{or} \quad
\big<\op{j}_{k}(x,t) \op{j}_{\ell}(0,0) \big>_{ \bs{\varrho} ; \e{c}   }  \;.
\enq
Here it is understood that, given two local operators $\op{O}_1(x,t)$,
$\op{O}_2(x,t)$, their connected two-point function reads
\beq
\big<\op{O}_{1}(x,t) \op{O}_{2}(0,0) \big>_{ \bs{\varrho} ; \e{c} }
\, = \,  \big<\op{O}_{1}(x,t) \op{O}_{2}(0,0) \big>_{ \bs{\varrho} }
\, - \,  \big<\op{O}_{1}(0,0) \big>_{ \bs{\varrho}   }  \big< \op{O}_{2}(0,0) \big>_{ \bs{\varrho}   } \; .
\enq
Generalised hydrodynamics provides predictions \cite{DoyonSpohnGDHPredictionsForNLSMDrudeWeight}
for the large-scale space-time asymptotics along rays, \textit{i.e.}, for 
$(x,t)\tend \infty$ with $\tf{x}{t}$ fixed. These take the form
\beqa
\big<\op{q}_{k}(x,t) \op{q}_{\ell}(0,0) \big>_{ \bs{\varrho} ; \e{c}  }   & = &
 \Int{\R}{} \hspace{-1mm} \dd s \: \rho(s) \big( 1-\vth(s)\big) \op{h}^{(\e{dr})}_{ \op{q}_{k} }(s)  \op{h}^{(\e{dr})}_{ \op{q}_{\ell} }(s) \de\big( x - \ups_{\e{eff}}(s) t)
\, + \, \e{o}\Big( \f{1}{x}\Big) \;,  \label{ecriture GHD pour densite densite}\\
\big<\op{q}_{k}(x,t) \op{j}_{\ell}(0,0) \big>_{ \bs{\varrho}  ; \e{c} }   & = &
\Int{\R}{} \hspace{-1mm}  \dd s \: \rho(s) \big( 1-\vth(s)\big) \ups_{\e{eff}}(s) \op{h}^{(\e{dr})}_{ \op{q}_{k} }(s)  \op{h}^{(\e{dr})}_{ \op{q}_{\ell} }(s) \de\big( x - \ups_{\e{eff}}(s) t)
\, + \, \e{o}\Big( \f{1}{x}\Big) \;,  \label{ecriture GHD pour densite courant} \\
\big<\op{j}_{k}(x,t) \op{j}_{\ell}(0,0) \big>_{ \bs{\varrho}  ; \e{c} }   & = &
\Int{\R}{} \hspace{-1mm}  \dd s \: \rho(s) \big( 1-\vth(s)\big) \ups_{\e{eff}}^2(s) \op{h}^{(\e{dr})}_{ \op{q}_{k} }(s)  \op{h}^{(\e{dr})}_{ \op{q}_{\ell} }(s) \de\big( x - \ups_{\e{eff}}(s) t)
\, + \, \e{o}\Big( \f{1}{x}\Big)   \;.
\label{ecriture GHD pour courant courant}
\eeqa

There are several building blocks occurring in these formulae. Since all
of them will be defined precisely in the core of the paper, we state their
definition only in words. $\rho$ refers to the density of roots in the
Bethe momentum space, which build up the representative state -- in
the sense first described by Yang and Yang \cite{YangYangNLSEThermodynamics} --
that catches all the features of the density matrix $\bs{\varrho}$.
$\vartheta$ is the associated filling fraction, \textit{aka}.\ occupation number.
Within the thermodynamic Bethe Ansatz description
\cite{DorlasLewisPuleRigorousProofYangYangThermoEqnNLSE,YangYangNLSEThermodynamics},
one introduces the thermodynamic Lieb operator $\op{K}_{\vth}$, which
allows to define dressed quantities that determine the thermodynamic
properties of the model subject to equilibrium conditions governed by
the density matrix $\bs{\varrho}$. The Bethe Ansatz description comes
naturally with a set of bare quantities, \textit{e.g.}, the functions
$ h_{ \op{q}_{k} }(s)$ associated with the densities of the local charges
$\op{q}_{k}$. Their dressed counterparts are then defined as solutions
to the linear integral equation $\big(\e{id}-\op{K}_{\vth}\big)
[\op{h}_{\op{q}_k}^{\e{dr}}]= h_{ \op{q}_{k} }$. Finally, $\ups_{\e{eff}}$
is the propagation velocity of the dressed excitations.

The higher-order corrections gathered in the remainder $\e{o}\big( \tf{1}{x} \big)$
go beyond the scope of generalised hydrodynamics. In order to predict their
structure and overall form, one needs to recourse to more advanced arguments
by incorporating diffusive corrections to the hydrodynamic regime. This was
developed in the works
\cite{BernardDoyonDeNardisDiffusionInGDHFullPaper,%
BernardDoyonDeNardisDiffusionInGDHAnnouncement}.
Calculations supporting the predictions for the explicit expression of the
diffusive corrections were carried out for the non-linear Schrödinger model
in \cite{Panfil2phDSFGGENLSM}. We shall not discuss this sub-leading regime
further, because it would lead us beyond the scope of the present work.

The predictions provided by generalised hydrodynamic that are
outlined above are quite explicit and neat. The reasoning at their root
is expected to apply to a broad class of integrable systems, be it classical
or quantum. It is more than desirable to provide an explicit verification
or at least some support for the predictions
\eqref{ecriture GHD pour densite densite}--\eqref{ecriture GHD pour courant courant}
through some first-principle approach based on a microscopic model.
Basically, two attempts were pursued in the literature. One approach
developed the concept of thermodynamic representative form factors
\cite{DeNardisPanfilDensityFFForGGEStates}, which led to the reproduction
\cite{DeNardisPanfilDensityFFNLSMinGGEDetailerBalance} of
\eqref{ecriture GHD pour densite densite} translated to the level of the
small momentum limit of the density structure factor. The other approach,
developed in \cite{GranetEsslerGGEFFSumIncMinus1}, relied on a perturbative
expansion of the correlation functions in $\tf{1}{c}$ around $c=+\infty$.

\subsection{The thermodynamic Bethe Ansatz}
The thermodynamic form-factor approach, see, \textit{e.g.}, the reviews
\cite{BernardDoyonDeNardisDiffusionInGDHAnnouncement,CortesCubertoSpohnYoshimuraReviewFFandGHD},
is based on the thermodynamic Bethe Ansatz (TBA) paradigm. The latter
takes its roots in the visionary reasoning of Landau and Lifshitz in the
case of the ideal Fermi gas. It was developed for integrable models with
the example of the non-linear Schrödinger model in \cite{YangYangNLSEThermodynamics}
and allows one to compute the thermodynamic observables of the model, the
\textit{per}-site free energy in particular:
\beq
\op{f}_{B} \; = \; \lim_{L\tend + \infty} \Big\{ \f{1}{L}  \ln [ \mc{Z}_{B}  ]  \Big\}
\qquad \text{and, more generally,} \qquad
\op{f}_{GGE} \; = \; \lim_{L\tend + \infty} \Big\{ \f{1}{L}  \ln [ \mc{Z}_{GGE}  ]  \Big\} \;.
\enq

Within the thermodynamic Bethe Ansatz approach, one can
express\symbolfootnote[1]{The formula for $\op{f}_{B}$ follows upon specialisation.}
$\op{f}_{GGE}$ in terms of the solution to a non-linear integral equation of Yang--Yang type,
\beq
\veps(\la) \, = \, V(\la) \, - \, T \Int{ \R }{}\hspace{-1mm} \dd \mu \: K(\la-\mu) \ln \big[ 1\, + \, \ex{- \f{ \veps(\mu) }{T} }\big] \;
\quad \e{with} \quad K(\la) \, = \, \f{ \th^{\prime}(\la) }{ 2\pi } \;,
\enq
in which the potential $V$, growing sufficiently fast at infinity, is to be 
understood as the limit of the sequence of polynomials $\sul{s=1}{M} a_s \la^s$
on compact subsets of $\R$. The above encompasses polynomial potentials and includes,
in particular, $V(\la)=\la^2-h$, which corresponds to the grand-canonical thermal
distribution with chemical potential~$h$.

The TBA approach was subsequently generalised to address the thermodynamics
of the XXZ chain in the easy-axis ($\De \geq 1$) \cite{GaudinTBAXXZMassiveInfiniteSetNLIE} and
easy-plane ($1 > \De > - 1$) \cite{TakahashiTBAforXXZFiniteTinfiniteNbrNLIE} anisotropy regimes.
While the thermodynamic Bethe Ansatz for the non-linear Schrödinger model
has established on rigorous grounds within large deviation theory in
\cite{DorlasLewisPuleRigorousProofYangYangThermoEqnNLSE},
the status of its applicability with respect to the XXZ chain, not
to even mention the more complex integrable models, is much less clear.

Indeed, the structure of the Bethe Ansatz equations for the XXZ chain,
just as for most integrable models, is way more complicated than that
of those for the non-linear Schrödinger model. While there are only real-valued
solutions possible in the latter case, the former admits genuine
complex-valued solutions as well. In some regimes, \textit{e.g.}, for
the finite-energy excitations above the \textit{ground state} of the
XXZ chain, one is able to predict the patterns taken by the Bethe roots
in the complex plane. These form so-called strings (see
\cite{KozProofOfAsymptoticsofFormFactorsXXZBoundStates} for a thorough
description thereof). A detailed analysis shows that not all string
lengths are allowed \cite{KorepinAnalysisofBoundStateConditionMassiveThirring,%
KozProofOfStringSolutionsBetheeqnsXXZ,TakahashiSuzukiFiniteTXXZandStrings}.
Note that the wrong counting of strings led to a mistake in the expression
for the XXZ chain's free energy; it was corrected subsequently
\cite{TakahashiSuzukiFiniteTXXZandStrings}.

The set of TBA equations for the XXZ chain was shown to be equivalent
\cite{KlumperShiroishiTakahashiEquivalenceTBAandQTM} to a single non-linear
integral equation that determines the free energy through the
\textit{ab initio} and \textit{rigorous} approach to the thermodynamics
of the XXZ chain based on the quantum transfer matrix
\cite{KlumperNLIEfromQTMDescrThermoXYZOneUnknownFcton,%
KozGohmannGoomaneeSuzukiRigorousApproachQTMForFreeEnergy,%
KozFaulmanGohmannLowTNLIERigourousAnalysisForQTMMasslessRegime}.
In some sense, this is supporting that the TBA approach provides a reasonable way to
handle thermodynamic calculations of extensive observables. However, we
would like to point out that string counting is not the only issue with
the TBA approach to the XXZ chain. The main criticism addresses
the non-exact nature of the string solutions. The string Bethe
root patterns solely emerge in certain regimes as the volume tends to
infinity: complex solutions with bounded real parts, sectors away from
zero magnetisation ground state, etc. It is, for instance, well known
that strings are strongly deformed when their centres are located towards
the $\ln L$ edges of the support of the real-valued Bethe roots
\cite{CauxHagemansDeformedStringsInXXX,%
EsslerKorepinSchoutensBAE2ParticleSectorXXXAndSomeNonStandardSols}.
A second feature of string solutions is that their number and allowed
lengths are utterly depending on the anisotropy $\De$:
when the latter is algebraic -- in the so-called root of unity case -- only
strings up to a certain maximal length are allowed, while, when $\De$
is transcendental the string length may go to infinity
\cite{TakahashiThermodynamics1DSolvModels}. Still, in the root of unity case,
Bethe equations only produce one representative of a cyclic family
of Eigenvectors: all others would have to be generated by coadjoint action,
but the explicit construction of their generators is still open
\cite{KorffHilbertSpaceStructureXXZAtRootOfUnity}. The number of complex
solutions to the Bethe Ansatz deviating from the string hypothesis,
just as the summed dimensions of the coadjoint orbits are expected
to be sub-exponential in the system size $L$, and thus generate
sub-dominant contributions to those of the \textit{bona fide} strings.
In the light of this, it seems plausible that, \textit{ad extremum}, the
TBA approach provides an non-rigorous but sound approach to dealing with
the thermodynamics of extensive observables. Still, TBA's validity in properly
characterising non-extensive observables is limited (see, \textit{e.g.}, the
discussion in \cite{BortzFrahmGohmannSurfaceFreeEnergy}).

\subsection{The thermodynamic form-factor approach}

The idea behind the effective thermodynamic form factors is as follows.
Similarly to the case of the \textit{per}-site free energy, see \textit{e.g.},
\cite{BogoliubiovIzerginKorepinBookCorrFctAndABA}, one argues that
the expectation value of any local operator with respect to the density
matrix $\bs{\varrho}$ as in \eqref{definition moyenne contre etat densite rho}
reduces, in the thermodynamic limit, to an effective average
\beq
\big< \op{O} \big>_{ \bs{\varrho}} \, = \,  \lim_{N,L \tend + \infty} \big< \op{O} \big>_{L, \bs{\varrho} }^{(\e{eff})}
\qquad \e{with}  \qquad
 \big< \op{O} \big>_{L, \bs{\varrho} }^{(\e{eff})} \; = \;     \f{  \big(  \Psi_{\bs{\varrho}} , \op{O} \,    \Psi_{\bs{\varrho}}  \big) }
{  \big(   \Psi_{\bs{\varrho}} ,   \Psi_{\bs{\varrho}} \big)   }   \;.
\enq
In this expression, $\Psi_{\bs{\varrho}}$ is a density matrix
$\bs{\varrho}$-dependent, highly excited Eigenstate of the
Hamiltonian $\op{H}$. The associated one-dimensional projector
onto $\e{Vect}\big\{ \Psi_{\bs{\varrho}} \big\}$ can be seen
as an effective approximation to the density matrix in the
large-$L$ limit.  There are, in principle, infinitely many
choices of states $\Psi_{\bs{\varrho}}$ leading to the same
thermodynamic limit of the expectation values. In the thermodynamic
form-factor approach, one fixes \textit{one} representative
$\Psi_{\bs{\varrho}}$. Now given, say, a finite-volume two-point
function built up from particle-number preserving local operators
$\op{O}_1$, $\op{O}_2$, one may expand it into the form-factor,
\textit{aka}.\ Käll\'{e}n--Lehmann, series
\beq
 \big< \op{O}_1(x,t) \op{O}_2(0,0) \big>_{L, \bs{\varrho} }^{(\e{eff})}  \, = \, \sul{ n }{}   \ex{ \i x \wh{\mc{P}}_{\e{ex}}^{(n)}-\i t \wh{\mc{E}}_{\e{ex}}^{(n)} }
\mc{F}_{n} \qquad \e{with} \qquad
 \mc{F}_{n}  \, = \, \f{  \big(  \Psi_{\bs{\varrho}} , \op{O}_1(0,0) \,    \Psi_{\bs{\varrho}}^{(n)}  \big)\big(  \Psi_{\bs{\varrho}}^{(n)}  , \op{O}_2(0,0) \,    \Psi_{\bs{\varrho}}   \big) }
{  \big(   \Psi_{\bs{\varrho}} ,   \Psi_{\bs{\varrho}} \big)   \big(   \Psi_{\bs{\varrho}}^{(n)}  ,   \Psi_{\bs{\varrho}}^{(n)}  \big)  } \;.
\enq
Above, $ \big\{ \Psi_{\bs{\varrho}}^{(n)} \big\}_{n \geq 0}$ is a
basis of Eigenstates of $\op{H}$. For convenience, the labelling is
such that $\Psi_{\bs{\varrho}}^{(0)} =\Psi_{\bs{\varrho}}$ and, as $n$
grows, the $ \Psi_{\bs{\varrho}}^{(n)} $ have larger and larger
excitation energies relative to $\Psi_{\bs{\varrho}}$.
$\wh{\mc{P}}_{\e{ex}}^{(n)}$, resp.\ $\wh{\mc{E}}_{\e{ex}}^{(n)}$,
is the excitation momentum, resp.\ energy, of the state
$\Psi_{\bs{\varrho}}^{(n)}$ relative to $\Psi_{\bs{\varrho}}$.

The Eigenvalues of a many-body Hamiltonian in finite volume $L$
form a dense distribution as $L\tend + \infty$ and gather in an
``energy band''. As the non-linear Schrödinger model is a continuum
model, not a lattice model, there is only a single semi-infinite band,
continuously connecting the ground state to all other excited states.
Within Lieb's classification of excitations
\cite{LiebExcitationStructureBoseGas}, the excited states above
$\Psi_{\bs{\varrho}}$ are obtained by creating $m$ holes at rapidities
$\bs{\vth}_{m}=\big( \vth_1,\dots, \vth_m)$ in the distribution of
Bethe roots realising $\Psi_{\bs{\varrho}}$ and adding $m$ particles
at rapidities $\bs{\nu}_{m}=\big( \nu_1,\dots, \nu_m)$ to the
distribution of Bethe roots realising  $\Psi_{\bs{\varrho}}$.
In the thermodynamic limit the relative excitation energies and
momenta will only depend on these rapidities:
\beq
\lim_{L\tend + \infty} \Big\{ \wh{\mc{P}}_{\e{ex}}^{(n)}  \Big\} \; = \; \mc{P}_m\big(\bs{\vth}_{m}, \bs{\nu}_{m} \big) \qquad \e{and} \qquad
\lim_{L\tend + \infty} \Big\{ \wh{\mc{E}}_{\e{ex}}^{(n)}  \Big\} \; = \; \mc{E}_{m}\big( \bs{\vth}_{m}, \bs{\nu}_{m} \big) \;.
\label{ecriture limite lisse energie et impulsion relative}
\enq
However, the data $\big(m; \bs{\vth}_{m}, \bs{\nu}_{m} )$
is only macroscopic: there are many ``microscopic'' Eigenstates which,
as $L\tend + \infty$, will be characterised by the \textit{same} set of
particle-hole rapidities $\big(m; \bs{\vth}_{m}, \bs{\nu}_{m} )$. Still,
for ``global'' observables, such as the relative excitation energy and momentum,
the description of their leading $L \tend + \infty$ behaviour will be
possible solely in terms of the macroscopic data $\big(m; \bs{\vth}_{m}, \bs{\nu}_{m} )$.
There are, however, numerous other observables which do ``feel'' the
microscopic structure of the Hilbert space and have no chance of
admitting a purely macroscopic description of their leading $L\tend + \infty$
behaviour. This is, in particular, the case for the form factors.
Due to their specific form,
\cite{KorepinSlavnovFormFactorsNLSEasDeterminants,
KozFFConjFieldNLSELatticeSpacingGoes0,SlavnovFormFactorsNLSE},
form factors do not admit, in general, smooth leading asymptotics
solely parameterised by $\big(m; \bs{\vth}_{m}, \bs{\nu}_{m} )$.
This holds already \cite{KozKitMailSlaTerThermoLimPartHoleFormFactorsForXXZ}
for the simplest possible density matrices $\bs{\varrho}_{G.S.}$
corresponding to the zero-temperature thermal Gibbs ensemble,
for which the density matrix is the projector onto the ground state.
In this case, form factors involving particle-hole excitations
collapsing at the Fermi points exhibit anomalous behaviour in the volume,
that is different from the naively expected $L^{-2m}$ decay, and
have leading asymptotics described by the microscopic particle-hole
integers which characterise the given excited state
$\Psi_{\bs{\varrho}_{G.S.}}^{(\e{n})}$. The large-$L$ behaviour of
the form factor $\big(  \Psi_{\bs{\varrho}} , \op{O}_a(0,0) \,
\Psi_{\bs{\varrho}}^{(n)}  \big)$ in the case of a generic
$\bs{\varrho}$-density matrix can be of yet more complex type
and might even depend on the specific representative
$\Psi_{\bs{\varrho}} $ chosen to approximate $\bs{\varrho}$.
In fact, so far, no one was ever able to extract, in any mathematically
meaningful way, the large-$L$ behaviour of form factors in case of a
generic $\bs{\varrho}$-density matrix. The idea behind thermodynamic
form factors consists in sweeping the microscopic details -- and all
the problems inherent to their treatment -- of form factors under
the carpet and consider effective form factors only.

To achieve this, one partitions the form-factor expansion as
\beq
 \big< \op{O}_1(x,t) \op{O}_2(0,0) \big>_{L, \bs{\varrho} }^{(\e{eff})}  \, = \, \sul{m \geq 0}{} \sul{\bs{\nu}_{m}, \bs{\vth}_{m}   }{} \sul{ n : \bs{\nu}_{m} , \bs{\vth}_{m} }{}
 \ex{ \i x \wh{\mc{P}}_{\e{ex}}^{(n)}-\i t \wh{\mc{E}}_{\e{ex}}^{(n)} } \mc{F}_{n}  \;.
\enq
The first sum is over the particle-hole excitation number $m$. The second
sum pertains to summing up the classes of excited states giving rise to
different particle-hole momenta. Finally, the remaining sum runs over
those Eigenstates which, in the thermodynamic limit, give rise to the
particle-hole momenta $\bs{\nu}_{m} , \bs{\vth}_{m}$. Owing to the
smoothness of the limit \eqref{ecriture limite lisse energie et impulsion relative},
for any Eigenstate $n$ leading to the given set of particle-hole momenta
$\bs{\nu}_{m} , \bs{\vth}_{m}$, one has
\beq
 \wh{\mc{P}}_{\e{ex}}^{(n)} \simeq \mc{P}_m\big(\bs{\vth}_{m}, \bs{\nu}_{m} \big)  \qquad \e{and} \qquad  \wh{\mc{E}}_{\e{ex}}^{(n)} \simeq \mc{E}_m\big(\bs{\vth}_{m}, \bs{\nu}_{m} \big)
\enq
which allows one to write
\beq
 \sul{ n : \bs{\nu}_{m} , \bs{\vth}_{m} }{}
 \ex{ \i x \wh{\mc{P}}_{\e{ex}}^{(n)}-\i t \wh{\mc{E}}_{\e{ex}}^{(n)} } \mc{F}_{n}  \; \simeq \;
 \ex{ \i x \mc{P}_m(\bs{\vth}_{m}, \bs{\nu}_{m} )  -\i t \mc{E}_m(\bs{\vth}_{m}, \bs{\nu}_{m})   }
 \sul{ n : \bs{\nu}_{m} , \bs{\vth}_{m} }{}  \mc{F}_{n}  \;.
\enq

The next crucial approximation consists in assuming that the remaining
microscopic sum over form factors could, in fact, be represented by a
single, typical form factor corresponding to the configuration
$\bs{\nu}_{m} , \bs{\vth}_{m}$:
\beq
 \sul{ n : \bs{\nu}_{m} , \bs{\vth}_{m} }{}  \mc{F}_{n}  \; \simeq \;  \f{ \ex{ \de S  } }{ L^{2m} } \mc{F}_{\op{O}_1}(\bs{\nu}_{m} , \bs{\vth}_{m})
 \mc{F}_{\op{O}_2}^*(\bs{\nu}_{m} , \bs{\vth}_{m}) \;.
\enq
Here $\de S$ corresponds to the number of such microscopic configurations,
while $L^{-2m}$ is a geometric factor measuring the typical phase-space volume
of microscopic configurations leading to $\bs{\nu}_{m} , \bs{\vth}_{m}$.
Finally, $\mc{F}_{\op{O}_k}(\bs{\nu}_{m} , \bs{\vth}_{m})$ is the
thermodynamic form factor. It is regular, except for coinciding particle-hole
momenta $\nu_a=\vth_b$, for which it has a simple kinematic pole.
The pole is only apparent as it does not appear on the level of the
finite-$L$ form factor, so that any apparent singularity present in
the expression for $\mc{F}_{\op{O}_k}(\bs{\nu}_{m} , \bs{\vth}_{m})$
when $\vth_a \not=\nu_s$ should, in fact, disappear from the limiting
expression as $\vth_a =\nu_s$. Such a summation-based appropriate
regularisation of form factors is very hard to achieve and, so far, was
only performed for the massless regime of the zero-temperature XXZ
chain \cite{KozMasslessFFSeriesXXZ}. In particular, no such procedure 
is proposed within the thermodynamic form-factor approach. Still, in
order to be able to cope with the non-integrable singularities of
the thermodynamic form factors, one then argues that, in the thermodynamic
limit, the outer summations go to regularised integrals, leading to
\beq
\big< \op{O}_1(x,t) \op{O}_2(0,0)  \big>_{ \bs{\varrho}} \, = \, \sul{m \geq 0}{} \Int{\R^m}{} \dd^m \nu \Fint{\R^m}{} \dd^m \vth \:
 \ex{ \i x \mc{P}_m(\bs{\vth}_{m}, \bs{\nu}_{m} )  -\i t \mc{E}_m(\bs{\vth}_{m}, \bs{\nu}_{m})   }
 \mc{F}_{\op{O}_1}(\bs{\nu}_{m} , \bs{\vth}_{m})
 \mc{F}_{\op{O}_2}^*(\bs{\nu}_{m} , \bs{\vth}_{m}) \;.
\enq
Above $\Fint{}{}$ refers to a regularisation of the double poles that we do not discuss further.

Building on the expression for the thermodynamic form factors of the
density operator in the non-linear Schrödinger model obtained in
\cite{DeNardisPanfilDensityFFForGGEStates}, the authors of
\cite{DeNardisPanfilDensityFFForGGEStates} argued that, due to the
highly oscillating nature of the exponentials, only the first non-trivial
term of the series should contribute to the leading order of the
$(x,t)$ asymptotics of their thermodynamic form-factor expansion.
It turns out that the two-dimensional integral appearing at $m=1$
does not need the regularisation procedure and does reproduce the
hydrodynamic prediction for $\big< \op{n}(x,t) \op{n}(0,0)  \big>_{ \bs{\varrho}}$.
The analysis of the $m=2$ integral reproduced the diffusive
predictions \cite{Panfil2phDSFGGENLSM}. However, the singularities
become worse and worse as $m$ grows. For non-particle-preserving
operators, such as field operators or their conjugates, the approach
based on thermodynamic form factors did not produce any meaningful
representation, not to speak of a prediction for the
$(x,t)\tend \infty$ asymptotics of the corresponding two-point
function.

\subsection{The previous microscopic approaches}

In the literature different routes were taken which attempted to
circumvent the various problems arising within the effective
thermodynamic form-factor approach. A fruitful idea was to perform
the microscopic summations directly in the form-factor expansion in
a limiting regime. The work \cite{GranetEsslerGGEFFSumIncMinus1} built
on the relation of the nonlinear Schr\"odinger model to non-interacting
fermions at $c=+\infty$. The authors computed, perturbatively in $\tf{1}{c}$
and \textit{uniformly} in the space-time variables, the two-point
functions of particle densities. This provided a full characterisation
of these two-point functions up to and including the order $\tf{1}{c^2}$.
The microscopic summation over the various excitations
was carried out with full care, albeit the contributions were only
estimated implicitly, \textit{i.e.}, not in closed form. However, it
turned out that the unknown contributions were eventually cancelling out
each other on the level of the \textit{per se} full correlation function.
As a result and as expected, the final expression was only depending
on the overall thermodynamic properties of the averaging
representative state $\Psi_{\bs{\varrho}}$. Those final expressions were
further analysed by the steepest-descent method. This confirmed,
up to $\e{O}(c^{-3})$ corrections, the $\asymp t^{-1}$ predictions
from generalised hydrodynamics and provided a closed expression for
the $\asymp t^{-2}$ contribution.


Last but not least, the calculation of the dynamical two-point functions
of the field-conjugated field operators was already addressed in the early
days of the quantum inverse scattering method. Starting with a so-called
dual-field representation for the on-shell/off-shell scalar products of
Bethe vectors \cite{KorepinFirstIntroDualFields} and exploiting the form
of the action of the field operator on a Bethe state obtained in
\cite{IzerginKorepinReshetikhinCommRelationFieldOpAndMonoMatrix}, the work
\cite{KojimaKorepinSlavnovNLSEDeterminatFormFactorAndDualFieldTempeAndTime}
provided a dual-field based representation for the mentioned correlator.
Here, we recall that dual fields are certain commutative operator-valued
functions. These operators act on an infinite dimensional Hilbert space. Up to trivial factors,
the representation obtained in
\cite{KojimaKorepinSlavnovNLSEDeterminatFormFactorAndDualFieldTempeAndTime}
was given in terms of a dual-field vacuum average of a Fredholm minor of
a dual-field valued integrable integral operators. While rather intricate,
it is important to stress that the dual-field based representation
does not rely on any assumption or simplification other than the existence
of the thermodynamic limit. Thus, it does not suffer from any ambiguities,
infinities or \textit{ad hoc} regularisations. Fredholm determinants of
integrable integral operators are known to be characterisable in terms
of a solution to an associated operator-valued Riemann--Hilbert problem
\cite{KorepinSlavnovRHPForNLSMOpertorValuedSettingInto}. The Riemann--Hilbert
problem associated with the field-conjugated field setting was analysed,
in the long-time and large-distance regime, by Its and Slavnov
\cite{ItsSlavnovNLSTimeAndSpaceCorrDualFields}. One should note, however,
that the analysis was performed by treating the dual fields as classical,
sufficiently regular functions. In order to be applicable to the
\textit{per se} correlation function setting, one should raise the
asymptotic formulae issuing from the Riemann--Hilbert analysis to
the operator-valued setting and then perform the dual-field vacuum
average. This procedure is known to be extremely tricky, in particular
due to the non-commutativity of the asymptotics with the dual-field averages.
Still, overcoming these to a certain extend, the work
\cite{SlavnovComputationDualFieldVaccumExpLongTimeDistTempeRedDensNLSE}
calculated the correlation length that governs the exponential decay
rate of the field-conjugated field correlator in thermal equilibrium.
The latter was described in terms of a solution to a coupled set of
three non-linear integral equations.

\subsection{Main results and outline of the multidimensional free-flow approach}
In the present work we developed an \textit{ab initio} approach to the
analysis of form-factor expansions of dynamical correlation functions
in \textit{interacting} integrable models in equilibrium states characterized
by a generic stationary density matrix $\bs{\varrho}$, represented by a filling
fraction $\vth$ that is an analytic on $\R$ and rapidly decaying at $\pm \infty$.
The method allows us to deal efficiently and without any \textit{ad hoc}
approximation with the microscopic structure of the model. It further 
allows us to recast the form-factor expansions of dynamical two-point
functions in such a way that we can directly \textit{read-off}
the long-time and large-distance asymptotics with $\tf{x}{t}$ kept
fixed. The method is \textit{significantly} different from the
methods previously applied in the literature to deal with an equilibrium
setting in that
\begin{itemize}
\item
it is genuinely non-perturbative;
\item
it avoids the various internal problems inherent to the
use of dual fields (\textit{e.g.}, the non-commutativity of limits and
the operator-valued setting for the associated Riemann--Hilbert problem);
\item
it does not rely on effective thermodynamic form factors
and therefore avoids all \textit{ad hoc} assumptions connected with those;
it has no regularization problem, since -- in first place and from
the outset -- it is not plagued by any emergent singularities in the thermodynamic limit;
\item
it produces representations for which the thermodynamic limit 
can be directly taken;
\item
it reduces the problem of the large-$(x,t)$ asymptotic analysis to standard
methods.
\end{itemize}
We would like to stress the effectiveness of the method, in that
it not only enables to extract, on the basis of first-principle
calculations, the $(x,t) \tend \infty$ asymptotics of correlators
involving the local particle densities and/or currents thereof,
but also of particle-number non-conserving local operators
such as the reduced density matrix $\big< \Phi(x,t) \Phi^{\dagger}(0,0)
\big>_{\bs{\varrho}}$. Such correlation functions could not be addressed
by heuristic methods so far. Thus, our approach produces results going
well-beyond the present state-of-the-art.

\subsubsection{The hydrodynamic behaviour of particle densities and currents}

In order to state our results, we need to introduce several notations.
To start with, introduce the bare phase and the Lieb kernel
\beq
\th(\la) \, = \, \i  \ln \bigg( \f{ \i c + \la  }{ \i c - \la  } \bigg)  \qquad \e{and} \qquad
K(\la) \, = \, \f{ \th^{\prime}(\la) }{ 2\pi } \;.
\label{definition bare phase}
\enq
The representative state $\Psi_{\bs{\varrho}}$ for the density matrix
$\bs{\varrho}$ is described by a filling fraction $\vartheta \in \intff{0}{1}$
which encodes the macroscopic overall ``filling'' of the Bethe vacancies
by the Bethe roots. Altogether, this allows to introduce the ``thermal''
Lieb operator
\beq
\op{K}_{\vth}[f](\la) \, =  \Int{\R}{} \!\! \dd \mu \: \vth(\mu) K(\la-\mu) f(\mu) \;.
\label{definition operateur integral Kvth}
\enq

The operator $\e{id}-\op{K}_{\vth}$ is invertible, which makes it possible
to define the dressed charge $Z$ along with the derivative of the dressed
energy $\veps$ as solutions to the linear integral equations
\beq
\big(\e{id}-\op{K}_{\vth}\big)\big[ Z \big](\la) \,= \, 1   \quad \e{and} \quad \big(\e{id}-\op{K}_{\vth}\big)\big[ \veps^{\prime} \big](\la) \,= \, 2 \la \;.
\enq
Due to the specific nature of the bare momentum in the non-linear
Schrödinger model, the derivative of the dressed momentum coincides
with the dressed charge $p^{\prime}(\la) = Z(\la)$. The effective
velocity is defined as $\ups_{\e{eff}}(\la)=\tf{ \veps^{\prime}(\la) }{ p^{\prime}(\la) }$.
This velocity is expected to be a strictly increasing diffeomorphism
from $\R$ onto $\R$. This ensures the existence of a unique saddle point
$\la_0\in \R$ such that $\ups_{\e{eff}}(\la_0) \, = \, \tf{x}{t}$.
Finally, one denotes by $\xi^{\prime}(\la)=p^{\prime}(\la)/(2\pi)$
the overall density of particle-hole solutions in presence of a
filling fraction $\vth$. The latter defines the density of particles
as $\rho(\la) \, = \, \xi^{\prime}(\la) \vth(\la)$. Let $\op{n}(x,t)$
be the particle density and $\op{j}_{\op{n}}(x,t)$ the particle-%
current density. Our approach yields the following results:
\beqa
\big<\op{n}(x,t) \op{n}(0,0) \big>_{ \bs{\varrho}   }   & = &  D^2 \, + \,  \f{1}{t}
 \f{ \rho(\la_0)   }{   \ups_{\e{eff}}^{\prime}(\la_0)  } \big( 1-\vth(\la_0)\big) Z^2(\la_0)
\, + \, \e{o}\Big( \f{1}{t}\Big) \;,    \\
\big<\op{n}(x,t) \op{j}_{ \op{n} }(0,0) \big>_{ \bs{\varrho}   }   & = & D J \, + \,
 \f{1}{t} \f{ \rho(\la_0)   }{   \ups_{\e{eff}}^{\prime}(\la_0)  } \big( 1-\vth(\la_0)\big) Z^2(\la_0)  \ups_{\e{eff}}(\la_0)
\, + \, \e{o}\Big( \f{1}{t}\Big) \;,    \\
\big< \op{j}_{ \op{n} } (x,t) \op{j}_{ \op{n} }(0,0) \big>_{ \bs{\varrho}   }   & = & J^2 \, + \,
 \f{1}{t} \f{ \rho(\la_0)   }{   \ups_{\e{eff}}^{\prime}(\la_0)  } \big( 1-\vth(\la_0)\big) Z^2(\la_0)  \ups_{\e{eff}}^2(\la_0)   \, + \, \e{o}\Big( \f{1}{t}\Big) \;.
\eeqa
This exactly reproduces the hydrodynamic predictions. For the time being,
our analysis is meant as a proof of concept. In particular, we did not
push the asymptotic calculations far enough to cover the diffusive
corrections expected to govern the next order $\asymp x^{-2}$.
We would like to stress, however, that our approach would allow us
to extend the asymptotic expansion, through order-by-order calculations,
to any desired precision in $x^{-1}$.

\subsubsection{Long-time and large-distance asymptotics of the reduced density matrix}

We now describe our second main result which goes beyond the paradigm of
generalised hydrodynamics, namely the $(x,t)\tend \infty$ asymptotics of
the reduced density matrix. We need to discuss several ingredients of the
asymptotic formula. First of all, as it is customary in all
finite-temperature calculations,\symbolfootnote[2]{The setting
of the two-point functions of conserved quantities, where only linear
integral equations, albeit driven by a non-equilibrium Lieb kernel, appear
is exceptional.} one needs to introduce the solution $\mf{f}_1^{\star}$
to a non-linear integral equation,
\beq
  \mf{f}^{\star}_1(\om) \, = \,  - \, \f{1}{2}  \, - \, \f{1}{2\pi} \th(\om-\la_0)
\, - \, \Int{\R}{}  \hspace{-1mm} \f{ \dd \la }{2\i\pi}  K(\om-\la) \, \e{sgn}(\la-\la_0)  \,
\ln \Big[ 1 + \vth(\la) \Big( \ex{-2\i\pi \e{sgn}(\la-\la_0) \mf{f}^{\star}_1(\la) } -1 \Big) \Big]  \;.
\enq
The reduced density matrix admits the asymptotic behaviour
\beq
\big< \Phi(x,t) \Phi^{\dagger}(0,0) \big>_{\bs{\varrho}} \, = \,  \ex{    \i x\, \Xi(\la_0)   }  \f{ \msc{A}(\la_0)  }{  x^{  \tf{ \mc{T}^2 (\la_0)  }{2}   } }
 \cdot \Big( 1+ \e{O}\Big( \f{\ln^2 x }{ x } \Big)   \Big) \;.
\enq
All constants in this formula are known explicitly. The one that
governs the exponential decay reads
\beq
\Xi(\la_0) \, = \, u_0(\la_0)   \, + \,  \Int{\R}{} \hspace{-1mm} \dd \mu  \,\e{sgn}(\la-\la_0)  \,
\ln \Big[ 1 + \vth(\la) \big( \ex{-2\i\pi \e{sgn}(\la-\la_0) \mf{f}^{\star}_1(\la) } -1 \big) \Big]  u^{\prime}_0(\mu)
\enq
with $u_0(\la) \, = \, \la - t \la^2/x$.
The exponent describing the power-law behaviour in $x$ takes the form
\beq
 \mc{T} (\la_0) \; = \;  - \f{1}{2 \pi \i}  \bigg\{
    \ln \Big[ 1 + \vth(\la_0) \big( \ex{2\i\pi \mf{f}^{\star}_1(\la_0) } -1 \big) \Big]
    + \ln \Big[ 1 + \vth(\la_0) \big( \ex{- 2\i\pi \mf{f}^{\star}_1(\la_0) } -1 \big) \Big]
    \bigg\}  \, - \, 1\;.
\enq
The expression for the amplitude, while explicit, is much more involved.
We do not show it here, but refer the interested reader to equation
\eqref{ecriture Amplitude reduced density matrix} in the core of the paper.

It is important for us to stress that the previously mentioned
dual-field approach should be thought of as a precursor to
the multidimensional free-flow approach developed in this work.
The dual-field approach was an incredible conceptual jump. However,
it was plagued by various technical obstacles due to the use
of non-efficient representations for the correlation functions.
These could be overcome at the expense of amazing efforts,
crowning the achievement of the approach with the above-mentioned
calculation of the correlation length for the reduced density
matrix in thermal equilibrium. Due to the intricacy of the approach,
the answer was paramaterised by solutions to three coupled non-linear
integral equations. Because of this very different description, we
were, so far, unfortunately unable to compare our result directly
with that of \cite{SlavnovComputationDualFieldVaccumExpLongTimeDistTempeRedDensNLSE}.
In its turn, the multidimensional free-flow approach simplifies
all technical problems related to the dual-field based techniques
and leads to a structurally simpler result.

\subsubsection{The multidimensional free-flow approach}

As it was discussed at length above, the central problem
with the thermodynamic limit of form-factor expansions
for models described by a density matrix $\bs{\varrho}$ pertains to
dealing in an appropriate way, avoiding any \textit{ad hoc}
argument, with the summations over the microscopic eigenstates
of the model. This problem is fully resolved by the multidimensional
free-flow approach. The multidimensional free-flow approach was
devised in the works
\cite{KozReducedDensityMatrixAsymptNLSE,KozTerrasNatteSeriesNLSECurrentCurrent},
where the ground state dynamical two-point functions of the
non-linear Schrödinger model were studied.

The multidimensional free-flow approach is the concrete realisation
of the adiabatic continuity at the root of Landau's Fermi liquid theory.
In a nutshell, within a given phase of matter,\symbolfootnote[3]{Any two
such phases are separated by a phase transition, where the dependence on
the coupling constants becomes non-smooth. The continuity of a deformation
only holds, if one is able to connect the given state of the interacting
system with one of an non-interacting theory, \textit{both} in the same
phase.} it should be possible to connect all observables of a certain 
\textit{interacting} model of interest to observables of a
non-interacting, \textit{aka}.\ free model, in the same phase
through a smooth flow in the coupling constant. While sounding very
reasonable and being supported by a plethora of examples from
\textit{few}-body classical and quantum mechanics, the adiabatic
deformation in the coupling constant turns out to be hard
to realise in practice in the context of \textit{many}-body physics.

The first observation underlying the multidimensional free-flow
method is the following. In a free theory, such as realised at the free-%
fermion point $c=+\infty$ of the non-linear Schrödinger model,
the form-factor expansion of all physical correlation functions,
\textit{i.e.}, of the two \cite{KorepinSlavnovTimeDepCorrImpBoseGas}
or multi-point \cite{SlavnovPDE4MultiPtsFreeNLSM} functions, can
be exactly summed up into a compact, well-defined representation.
This representation holds as much for the ground state expectation
values as for the cases of thermal or, more generally, generalised
Gibbs-ensemble averages. The correlators are given in terms of Fredholm
determinants or minors of operators of the form $\e{id} + \wh{\op{V}}_{x,t}$,
where $\wh{\op{V}}_{x,t}$ is an integral operator of integrable type
\cite{ItsIzerginKorepinSlavnovDifferentialeqnsforCorrelationfunctions,%
SakhnovichFirstStepsOfElementsOfIntegrableIntOps}.
The operator depends parametrically on the space-time variables $(x,t)$
present in the original correlator and act on $\ell^2(\mathbb{I}_N)$,
where $\mathbb{I}_N$ is a subset of $\mathbb{Z}$. Thus, for two-point
functions of ``particle-number preserving operators'', say, in free
fermion equivalent models, one has a relation of the type
\beq
\big<\op{O}(x,t) \op{O}^{\dagger}(0,0) \big>_{L, \bs{\varrho}}^{(\e{eff})} \; = \; \sul{n \geq 0 }{} \ex{ \i x \wh{\mc{P}}^{(n)}_{\e{ex}} - \i  t \wh{\mc{E}}^{(n)}_{\e{ex}}}
 \bigg|  \f{ \big(\Psi_{\bs{\varrho}},\op{O}(0,0)  \Psi_{\bs{\varrho}}^{(n)} \big) }{ \norm{ \Psi_{\bs{\varrho}} } \cdot \norm{ \Psi_{\bs{\varrho}}^{(n)} }  }    \bigg|^2 \; = \;
 \underset{\ell^2(\mathbb{I}_L) }{\det}\big[ \e{id} + \wh{\op{V}}_{x,t} \big] \;.
\label{ecriture dvpment en series FF Fct 2 pts}
\enq

Taking the thermodynamic limit on the level of the intermediate form-factor
expansion is extremely hard owing to the highly intricate and singular
form of the form-factor summand. However, the form of the operator
$ \wh{\op{V}}_{x,t}$ arising \textit{after} the summation of the \textit{full}
form-factor series expansion, is such that taking the thermodynamic
limit is trivial and simply amounts to replacing the operator $\wh{\op{V}}_{x,t}$
with the limiting operator $\op{V}_{x,t}$,
\beq
 \underset{\ell^2(\mathbb{I}_N) }{\det} \big[ \e{id} + \wh{\op{V}}_{x,t} \big] \, \tend \,  \underset{ L^2(\mc{C}) }{\det}\big[ \e{id} + \op{V}_{x,t} \big] \, ,
\label{ecriture cvgence limite thermo F det}
\enq
and likewise for the Fredholm minor-based representations.

Above, the limiting operator $\op{V}_{x,t}$ acts on functions supported
on some curve $\mc{C}$ in the complex plane, most often the real line
or a slight deformation thereof. We stress that the representation in
terms of a Fredholm determinant is perfectly well-defined. It is given
in terms of a series of multiple integrals. For instance, in the
discussed scenario one obtains 
\beq
\big<\op{O}(x,t) \op{O}^{\dagger}(0,0) \big>_{\bs{\varrho}} \; = \;  \sul{n \geq 0}{} \f{1}{n!} \Int{\mc{C}^n }{} \dd^n \la \det_n\big[ \op{V}_{x,t}(\la_a,\la_b) \big] \;.
\enq

We would like to stress that all multiple integrals appearing above
are well-defined in the usual sense: no distributional regularisation
of any sort is necessary, in striking contrast with the thermodynamic
form-factor approach. The representation of the correlator in terms
of a Fredholm series may or may not be optimal for a given purpose.
For instance, it is very useful for computing the small $(x,t)$
behaviour, since each summand contributes to higher orders with 
respect to the previous one. However, the Fredholm series is not optimal
for studying the $(x,t) \tend \infty$ asymptotics for fixed $\tf{x}{t}$.
The crucial observation in that case is that the limiting integral operator
$\op{V}_{x,t}$ is of integrable type. This means that one can construct
the resolvent of $\e{id} + \op{V}_{x,t}$ and express the associated
Fredholm determinant/minor in terms of the solution to a matrix-valued
Riemann--Hilbert problem \cite{ItsIzerginKorepinSlavnovDifferentialeqnsforCorrelationfunctions}.
Typically, $(x,t)$ appear in an oscillatory factor in the jump matrix
associated with this problem. The asymptotic solution, for $(x,t) \tend \infty$,
of this class of Riemann--Hilbert problems can be constructed within
the Deift--Zhou non-linear steepest descent
\cite{DeiftZhouSteepestDescentForOscillatoryRHPmKdVIntroMethod}
and the Beals--Coifmann singular integral equation method
\cite{BealsCoifmanScatteringInFirstOrderSystemsEquivalenceRHPSingIntEqnMention},
see, \textit{e.g.}, the works
\cite{CheianovZvonarevZeroTempforFreeFermAndPureSine,%
DeiftItsZhouSineKernelOnUnionOfIntervals,KozGohmannMininTimeAndSpaceDpdtThermalGSK,%
ItsIzerginKorepinCorrRadiusImpBosons,ItsIzerginKorepinTemperatureLongDistAsympBoseGas,%
ItsIzerginKorepinVarzuginTimeSpaceAsymptImpBoseGaz,%
ItsSlavnovNLSTimeAndSpaceCorrDualFields,KozKitMailSlaTerRHPapproachtoSuperSineKernel}.
Once achieved, the Riemann--Hilbert analysis produces a different
multiple-integral series representation for the Fredholm
determinant, the so-called Natte series \cite{KozTimeDepGSKandNatteSeries}.
The latter is perfectly fit for studying the asymptotics: the first
term of the series provides the dominant asymptotics, the second
term generates the first sub-leading corrections, \textit{etc}.
In fact, this is a manifestation of the property that, just as for
fundamental special functions as the Gauss Hypergeometric function,
the Fredholm determinants admit several different multiple-integral
series representations, some of which may be useful to study one
asymptotic regime, say $(x,t)\tend 0$ (think of the Gauss integral
representation fit for easily obtaining the $z\tend 0$ behaviour
of $_2F_1$), while others may be useful to access the asymptotics in
another regime, say $(x,t) \tend \infty$ (think of the Mellin-Barnes
representation which directly yields the $z \tend \infty$ behaviour
of $_2F_1$). To cut a long story short, taken all these properties,
one may thus say that the theory of correlation functions of a
free-fermionic theory is fully operational and leads to any desired
precision of the properties of the objects at play.

In any truly interacting integrable model, for which the matrix elements
of local operators are known, the form-factor expansion part of
\eqref{ecriture dvpment en series FF Fct 2 pts} still holds (the first equation).
However, due to the intricate structure of the form factors at
finite coupling constant, see, \textit{e.g.},
\cite{KitanineMailletTerrasFormfactorsperiodicXXZ,%
KorepinSlavnovFormFactorsNLSEasDeterminants,%
KozFFConjFieldNLSELatticeSpacingGoes0}, there is so far no technique
available that would allow one to directly sum-up the form-factor
expansion. The idea behind the multidimensional free-flow approach
is to use a specific flow, think of it as a chain of non-linear
transformations produced by a functional translation operator,
which will map the \textit{interacting} form-factor expansion
to a \textit{free fermionic} form-factor expansion, this up to
irrelevant corrections in the system size $L$:
\beq
\sul{n \geq 0 }{} \ex{ \i x \wh{\mc{P}}^{(n)}_{\e{ex}} - \i  t \wh{\mc{E}}^{(n)}_{\e{ex}}}
 \bigg|  \f{ \big(\Psi_{\bs{\varrho}},\op{O}(0,0)  \Psi_{\bs{\varrho}}^{(n)} \big) }{ \norm{ \Psi_{\bs{\varrho}} } \cdot \norm{ \Psi_{\bs{\varrho}}^{(n)} }  }    \bigg|^2
\, = \, {\bs : }   \msc{A}_{L}^{(\op{O})} \cdot \sul{n \geq 0 }{} \ex{ \i x \wh{\mc{P}}^{(n)}_{\e{aux}} - \i  t \wh{\mc{E}}^{(n)}_{\e{aux}}}
 \bigg|  \f{ \big(\Psi_{\bs{\varrho}}^{(\e{aux})},\op{O}(0,0)  \Psi_{\bs{\varrho}}^{(\e{aux};n)} \big) }
 { \norm{ \Psi_{\bs{\varrho}}^{(\e{aux})}  } \cdot \norm{ \Psi_{\bs{\varrho}}^{(\e{aux};n)} }  }    \bigg|^2   {\bs : }  \;.
\enq
On the right-hand side of the above identity, the summand is constructed
in terms of form factors and relative-excitation momenta/energies that arise
in an auxiliary, well-tailored free-fermionic model. We stress that this
model is \textit{more general} than the one arising at $c=+\infty$.
The generalised free fermionic model is endowed  with a representative
state $\Psi_{\bs{\varrho}}^{(\e{aux})}$ capturing the thermodynamic effect of the
density-matrix average. The excited states above it are denoted by
$\Psi_{\bs{\varrho}}^{(\e{aux};n)}$. This auxiliary model is constructed
with the help of two arbitrary functions $g,\nu$. \textit{All} quantities
(Eigenvalues, Eigenvectors, etc.) arising in the auxiliary model exhibit an
intricate functional dependence on  $(g,\nu)$. The expression also involves
as a prefactor an operator-$\op{O}$ dependent functional $ \msc{A}_{L}^{(\op{O})}$
evaluated at the auxiliary functions ($g,\nu$). We stress that this functional
is \textit{independent} of the form-factor summation index~$n$.

The auxiliary free fermionic form-factor expansion is appearing in between the ${\bs :}  * {\bs :}$ symbol. This symbol is a compact notation which refers to the application of the
multidimensional free flow to the auxiliary form-factor series. This is a \textit{coupled} flow of the functions ($g,\nu$). We refer to
Subsection \ref{SousSection Preliminary resummationGenFct} for a brief description of this coupled flow. It is discussed at length in the work \cite{KozReducedDensityMatrixAsymptNLSE}.
The bottom line of the flow is that, pointwise, one has the identity
\beq
 {\bs : }   \msc{A}_{L}^{(\op{O})} \ex{ \i x \wh{\mc{P}}^{(n)}_{\e{aux}} - \i  t \wh{\mc{E}}^{(n)}_{\e{aux}}}
 \bigg|  \f{ \big(\Psi_{\bs{\varrho}}^{(\e{aux})},\op{O}(0,0)  \Psi_{\bs{\varrho}}^{(\e{aux};n)} \big) }
 { \norm{ \Psi_{\bs{\varrho}}^{(\e{aux})}  } \cdot \norm{ \Psi_{\bs{\varrho}}^{(\e{aux};n)} }  }    \bigg|^2   {\bs : }
\, = \,
\ex{ \i x \wh{\mc{P}}^{(n)}_{\e{ex}} - \i  t \wh{\mc{E}}^{(n)}_{\e{ex}}}
 \bigg|  \f{ \big(\Psi_{\bs{\varrho}},\op{O}(0,0)  \Psi_{\bs{\varrho}}^{(n)} \big) }{ \norm{ \Psi_{\bs{\varrho}} } \cdot \norm{ \Psi_{\bs{\varrho}}^{(n)} }  }    \bigg|^2 \;.
\enq
Based on this, one may already state that the multidimensional free flow does
indeed concretely realise Landau's adiabatic coupling-constant deformation ideas.
Of course, in itself, the above relation between form-factor expansions in
two different models does not add much to the game. The exceptional fact
is that, owing to the techniques pioneered in
\cite{KorepinSlavnovTimeDepCorrImpBoseGas,SlavnovPDE4MultiPtsFreeNLSM}
for the non-linear Schrödinger model at $c=+\infty$ and adapted to deal
with any generalised free fermion model in \cite{KozReducedDensityMatrixAsymptNLSE},
the form-factor expansion of the auxiliary free fermionic model can be fully
resummed into a Fredholm determinant/minor, leading to an expression
for the two-point function of the \textit{interacting} model in finite
volume $L$ which is of the type
\beq
\big<\op{O}(x,t) \op{O}^{\dagger}(0,0) \big>_{L, \bs{\varrho}}^{(\e{eff})} \; = \;
{\bs : }   \msc{A}_{L}^{(\op{O})}  \underset{\ell^2(\mathbb{I}_N) }{\det} \big[ \e{id} + \wh{\op{V}}_{x,t} \big]  {\bs :} \;.
\enq

Here the form of the kernel $\wh{\op{V}}_{x,t}$ depends on the details
of the operator $\op{O}$ as well as on the auxiliary functions ($g,\nu$).
While taking the thermodynamic limit of the \textit{interacting} model
on the level of the form-factor expansion is extremely hard without \textit{ad hoc}
assumptions, taking it on the level of the resummed expression is rather
direct, owing to the convergence of the thermodynamic limit of the Fredholm determinant/minor
exactly as described in \eqref{ecriture cvgence limite thermo F det}, the
thermodynamic limit of $ \msc{A}_{L}^{(\op{O})}$ not being an issue at all.
This thus yields
\beq
\big<\op{O}(x,t) \op{O}^{\dagger}(0,0) \big>_{L, \bs{\varrho}}^{(\e{eff})} \; = \;
{\bs : }   \msc{A}^{(\op{O})}  \underset{L^2(\mc{C}) }{\det} \big[ \e{id} + \op{V}_{x,t} \big]  {\bs :} \;.
\enq
This representation should then be evaluated, \textit{viz}. the effect
of the multidimensional free flow ${\bs :} * {\bs :}$ made explicit.
For this, one needs an explicit representation for the Fredholm determinant.
Having in view extracting the long-time and large-distance asymptotics
of the correlator, the best option is to go for the Natte series representation
of the Fredholm determinant that is provided by the Riemann--Hilbert
large $(x,t)$ analysis. As shown in
\cite{KozTimeDepGSKandNatteSeries,KozReducedDensityMatrixAsymptNLSE} the
Natte series induced by the Riemann--Hilbert approach, when subject to
the multidimensional flow produces a new series representation for the
correlator of interest. This multiple-integral series representation
for the correlators in the \textit{interacting} model preserves the
same structure of the ordering of the contributions to the $(x,t)\tend
\infty$ regime: each further term in the series produces a subdominant
correction with respect to the one preceding it. In particular, the leading
asymptotics of the Fredholm determinant/minor produces, under the
multidimensional free flow, the leading asymptotics of the correlators.
The multidimensional free-flow evaluation of the leading asymptotics
eventually produces the formulae that we outlined above and that 
constitute the main achievement of this work.

\subsection{Outline}

Here is a brief outline of this work.
Section \ref{Section overall setup} presents the basic technical set up,
necessary for the reasoning developed in this work. Subsection~\ref{SousSectionBAInput}
sketches the salient features of the Bethe Ansatz that will be of use later on.
Subsection~\ref{SousSectionThermoLimit} presents the main aspects of the
Bethe Ansatz description of the thermodynamic limit of interest to this work.
Subsection~\ref{SousSectionExcitedStatesAboveGGERepresentative} outlines the
description of the excited states above the Bethe vector that is used to
represent the density matrix. Section~\ref{Section Form factors} is dedicated
to the description of the form factors of local operators which play a crucial
role in our analysis. Subsection~\ref{SousSectionFiniteLExpressions} provides
one with the finite-volume expressions of the form factors.
Subsection~\ref{SousSectionLargeLAsymptotics} derives the large-$L$ asymptotic
behaviour of these. Section~\ref{Section Effective FF expansion} is devoted
to obtaining alternative expressions for the form-factor series expansions
in terms of the multidimensional free flow.
Subsection~\ref{SousSection Preliminary resummationGenFct} implements a
preliminary resummation, based on multidimensional free flows,
for the generating function of densities and density current two-point
functions. Subsection~\ref{SousSectionPreliminaryResummationRedDenMtrx}
obtains a similar result for the reduced density matrix. Subsection~\ref{SousSectionThermoLimXN}
outlines the form of the main building block entering in the preliminary
resummation formula and recalls a representation thereof in terms of the
Fredholm determinant or minor of a finite rank integrable integral operator.
The thermodynamic limit of this quantity is also described.
Subsection~\ref{SousSectionThermoLimCorrelators} applies the previous
results so as to obtain a multidimensional free-flow based representation of
the \textit{per se} correlators of interest. Section~\ref{SectionAlternativeRepCorrFcts}
is devoted to the explicit evaluation of the multidimensional free flow in order to
bring the previous representations into explicit form.
Subsection~\ref{SousSectionAlternRepFDetAndMinor} describes the representation
of the Fredholm determinants/minor at play that is well-suited for reading-off
the large-$(x,t)$ asymptotics. Subsection~\ref{SousSectionAsymptRepForCorrFcts}
builds on these representations so as to carry out the mentioned evaluation.
Section~\ref{SectionSpaceTimeAsymptCorrFcts} obtains the long-time and
large-distance asymptotics of the two-point functions of interest to this work.
Subsection~\ref{SousSectionDAFctGen} focuses on obtaining these results for
the generating functions and the two-point functions which can be obtained from it.
Finally, Subsection \ref{SousSectionDARedDensityMtx} deals with the asymptotics
of the reduced density matrix.

\section{The overall setup}
\label{Section overall setup}

\subsection{The Bethe Ansatz input}
\label{SousSectionBAInput}

Solutions to the below logarithmic Bethe Ansatz equations \cite{BrezinPohilFinkelbergFirstIntroBoseGas,LiebLinigerCBAForDeltaBoseGas,LiebExcitationStructureBoseGas}
parameterise the Eigenvalues and Eigenvectors
of $\op{H}_{\mid \mc{V}_N}$:
\beq
L \mu_{a;\bs{\ell}_N} \; +  \;
\sul{b=1}{N} \th \big( \mu_{a;\bs{\ell}_N} - \mu_{b;\bs{\ell}_N}  \big) = 2\pi \Big( \ell_a - \f{N+1}{2} \Big) \;\;\;, \qquad
a \, = \,  1,\dots,N \;,
\label{ecriture Log BAE continuous model}
\enq
where $\bs{\ell}_N = (\ell_1,\dots, \ell_N) \in \mathbb{Z}^N$ and the bare phase $\th$ was introduced in \eqref{definition bare phase}.

For each given $\bs{\ell}_N \in \mathbb{Z}^N$, the equations arise as a stationary point of a strictly convex, blowing up at $\infty$,
function. As such they admit a unique solution $\{\mu_{a;\bs{\ell}_N} \}_{1}^{N}$.
When $\bs{\ell}_N$ runs through $\mathbb{Z}_<^N\, = \, \big\{ \bs{\ell}_N \in \mathbb{Z}^N \; : \;  \ell_1<\dots<\ell_N \big\}$
one gets the complete \cite{DorlasOrthogonalityAndCompletenessNLSE} set of Eigenvectors of  $\op{H}_{\mid \mc{V}_N}$. We shall denote
the Bethe Eigenvector built from the solution $\{ \mu_{a;\bs{\ell}_N}  \}_1^{N} $ as $\Psi\big( \{ \mu_{a;\bs{\ell}_N}  \}_1^{N} \big)$.

In particular,  it holds
\beq
\left( \ba{cc}  \op{P}  \\ \op{H} \ea \right) \, \Psi\big( \{ \mu_{a;\bs{\ell}_N}  \}_1^{N} \big) \, = \,
\sul{s=1}{N} \left( \ba{cc}   \ p_0\big( \mu_{s;\bs{\ell}_N}  \big)   \vspace{1mm} \\   \veps_0\big( \mu_{s;\bs{\ell}_N}  \big)  \ea \right)
\Psi\big( \{ \mu_{a;\bs{\ell}_N}  \}_1^{N} \big)
\qquad \left\{ \ba{ccc}   p_0(\mu) &=& \mu  \\
                          \veps_0(\mu) &=& \mu^2      \ea \right. \;.
\label{definition bare momemtum and energy}
\enq
More generally, one has
\beq
  \op{H}_k \, \Psi\big( \{ \mu_{a;\bs{\ell}_N}  \}_1^{N} \big) \, = \,
\bigg\{ \sul{s=1}{N} (\mu_{s;\bs{\ell}_N})^k  \bigg\} \, \Psi\big( \{ \mu_{a;\bs{\ell}_N}  \}_1^{N} \big)  \;.
\enq

The ground state of $\op{H}_{\mid \mc{V}_N}$ corresponds to the solutions to \eqref{ecriture Log BAE continuous model} subordinate
to the specific choice $\ell_a = a$, $a=1, \dots, N$, what thus produces the minimal value of $\sul{a=1}{N}  \veps_0\big( \mu_{a;\bs{\ell}_N}  \big)$.

For further convenience, it is useful to introduce an auxiliary system of logarithmic Bethe Ansatz equations, the $\be$-deformed ones.
Their discretised counterpart allows one to construct the Eigenvalues and Eigenvectors of the $\ex{\be}$ twisted transfer matrix,
see \cite{KozKitMailSlaTer6VertexRMatrixMasterEquation} for more details. These take the form
\beq
L \mu_{a;\bs{\ell}_N}^{(\be)} \; +  \;
\sul{b=1}{N} \th \big( \mu_{a;\bs{\ell}_N}^{(\be)} - \mu_{b;\bs{\ell}_N}^{(\be)}  \big) = 2\pi \Big( \ell_a - \f{N+1}{2} \Big)  + \i \be  \;\;\;, \qquad
a \, = \,  1,\dots,N \;.
\label{ecriture beta twisted Log BAE continuous model}
\enq
Owing to the unique solvability of \eqref{ecriture Log BAE continuous model},
it actually holds that $\mu_{a;\bs{\ell}_N}^{(\be)} = \mu_{a;\bs{\ell}_N} + \i \tf{\be}{L}$, although we will
 never be using this relationship. Also, for reality reasons, we shall restrict to the case when $\be \in \i \R$.

Given a solution to \eqref{ecriture beta twisted Log BAE continuous model} associated with a choice $\bs{\ell}_N \in \mathbb{Z}_<^N$ it is useful to introduce
the associated counting function
\beq
\wh{\xi}_{  \bs{\ell}_N }(\om) = \f{1}{2\pi  }  \om  \; +  \; \f{1}{2\pi L}\sul{b=1}{N} \th \big(\om - \mu_{b; \bs{\ell}_N}^{(\be)} \big)   \; + \; \f{N+1}{2 L}
\, - \, \f{\i \be }{ 2 \pi L} \; .
\label{counting fct beta elle N}
\enq
It is readily seen that it is strictly increasing on $\R$ and such that $\wh{\xi}_{  \bs{\ell}_N }(\R) \, = \, \R $. Thus, $\wh{\xi}_{  \bs{\ell}_N }$
allows one to unambiguously define a strictly increasing sequence $\{\mu_a\}_{a\in \mathbb{Z}}$ of solutions to
$ \ex{2\i\pi L \wh{\xi}_{  \bs{\ell}_N }(\om) } \, = \, 1$ on $\R$:
\beq
 \wh{\xi}_{  \bs{\ell}_N }(\mu_a) \, = \, \f{ a }{ L } \qquad  a \in \mathbb{Z} \;.
\label{definition des solutions mua subordonnees a ellN}
\enq
We do stress that the sequence $\{\mu_a\}_{a\in \mathbb{Z}}$ does depend on the choice of $\bs{\ell}_N$ and on $\be$.
However, since in all of its further uses no ambiguity arises as to the value of $\bs{\ell}_N$ and $\be$, we omit to stress this dependence to lighten the notations.
The solutions \eqref{definition des solutions mua subordonnees a ellN} are, by construction, such that $\mu_{\ell_a} = \mu_{a; \bs{\ell}_N}^{(\be)}$.

\subsection{Generalities on the thermodynamic limit}
\label{SousSectionThermoLimit}

Of interest is the description of the behaviour of the model in equilibrium, in the grand canonical ensemble, and, more generally,
grasped by a density matrix $\bs{\varrho}$.  The former case corresponds to the model's density matrix being given by the Boltzmann statistical operator
$ \tf{ \ex{-\tf{\op{H}}{T}} }{ \mc{Z}_{B} }$, while one can think of the latter as being approximated by a generalised Gibbs ensemble
of the form $\tf{ \exp\Big\{-   \sul{k=1}{M} \tfrac{a_k }{T} \op{H}_k  \Big\} }{ \mc{Z}_{ \bs{\varrho} } }$ for some $M \in \mathbb N$.
Here, $\mc{Z}_{B} \, = \, \e{tr}_{\mf{h}}\Big[  \ex{-\tf{\op{H}}{T}} \Big]$ and $\mc{Z}_{ \bs{\varrho} } \, = \,
\e{tr}_{\mf{h}}\Big[ \exp\Big\{-    \sul{k=1}{M} \tfrac{a_k }{T}  \op{H}_k  \Big\} \Big]$.
In such a setting, one is interested in evaluating expectation values of local operators~$\op{O}$
\beq
\big< \op{O} \big>_{L, B} \; = \; \f{ 1  }{ \mc{Z}_{B}  } \e{tr}_{\mf{h}}\Big[  \op{O}  \,   \ex{-\f{1}{T} \op{H} }  \Big]  \qquad \e{and} \qquad
\big< \op{O} \big>_{L, \bs{\varrho}} \; = \;\f{ 1 }{ \mc{Z}_{\bs{\varrho}}  } \e{tr}_{\mf{h}}\Big[  \op{O}  \,  \exp\Big\{-  \sul{k=1}{M} \tfrac{a_k }{T}  \op{H}_k  \Big\}  \Big] \;.
\enq
as $L\tend + \infty$. In fact, the first observable to evaluate in that case is the per-site free energy
\beq
\op{f}_{B} \; = \; \lim_{L\tend + \infty} \Big\{ \f{1}{L}  \ln [ \mc{Z}_{B}  ]  \Big\}
\qquad \text{and, more generally,} \qquad
\op{f}_{\bs{\varrho}} \; = \; \lim_{L\tend + \infty} \Big\{ \f{1}{L}  \ln [ \mc{Z}_{\bs{\varrho}}  ]  \Big\} \;.
\enq
Surprisingly, one is also able to deduce the value of these directly from the Bethe Ansatz
owing to an elaboration of the chain of thought invoked by Landau and Lifshitz in the case of the free fermion gas. This was
 done by Yang and Yang \cite{YangYangNLSEThermodynamics} for the Boltzmann distribution. Their argument  easily extends
to the general density matrix setting. The bottom line is that one can express\symbolfootnote[1]{The formula for $\op{f}_{B}$ follows upon specialisation} $\op{f}_{\bs{\varrho}}$
in terms of a solution to a non-linear integral equation of Yang--Yang type
\beq
\veps(\la) \, = \, V(\la) \, - \, T \Int{ \R }{}\hspace{-1mm} \dd \mu K(\la-\mu) \ln \big[ 1\, + \, \ex{- \f{ \veps(\mu) }{T} }\big] \;
\quad \e{with} \quad K(\la) \, = \, \f{ \th^{\prime}(\la) }{ 2\pi } \;,
\enq
and in which the potential $V$, growing sufficiently fast at infinity, is to be understood as the limit of the sequence of
polynomials $\sul{s=1}{M} a_s \la^s$ on compact subsets of $\R$.
In fact, upon introducing the solution to the linear integral equation
\beq
\xi^{\prime} (\om)  \, = \,  \f{1}{2\pi  }   \; +  \;  \Int{\R}{} \hspace{-1mm} \dd \la  \, K\big(\om - \la \big)  \vth(\la) \xi^{\prime}(\la)
\qquad \e{with} \qquad \vth(\la) \, = \, \f{  1 }{  1\, + \, \ex{ \f{ \veps(\la) }{ T } }  }
\label{definition equation pour xi prime}
\enq
one may define a density function
\beq
\varrho(\la) \;= \;  \vth(\la) \xi^{\prime}(\la) \qquad \e{with} \qquad D \, = \, \Int{\R}{} \varrho(\la) \dd \la \;.
\label{macroscopic density}
\enq
This density corresponds to the density of a typical representative of solutions of the Bethe equations which contribute, to the leading order, to the value of
the free energy. More precisely, it corresponds to $\mu_{a;\bs{r}_N}^{(0)}$, $a=1,\dots, N$,
the solution to the logarithmic Bethe Ansatz equations \eqref{ecriture Log BAE continuous model}
associated with a choice of $\bs{r}_N \in \mathbb{Z}^N_{<}$ such that $\{\mu_{a;\bs{r}_N}^{(0)}\}_1^N$ densifies on $\R$ with the density $\varrho$; in other words the below condensation property holds
\beq
\f{1}{L} \sul{a=1}{N} f\big(\mu_{a;\bs{r}_N}^{(0)} \big) \; \tend \; \Int{ \R }{} \dd \la\,  \varrho(\la) f(\la) \;,
\label{ecriture condensation etat ref}
\enq
for any sufficiently smooth $f$ such that $f \in L^1(\R,\varrho \dd \la)$.  Of course, there are many choices of vectors $\bs{r}_N$ giving rise to the same limit above.
In fact, one may interpret the density as stemming from the limit
\beq
 \varrho(\la) \, = \, \lim_{\substack{ L \tend +\infty \\ \mu_{a;\bs{r}_N}^{(0)} \tend \la} } \bigg\{ \f{1}{L(\mu_{a+1;\bs{r}_N}^{(0)}- \mu_{a;\bs{r}_N}^{(0)}  )}\bigg\} \;.
\enq

>From now on, we shall fix {\bf one} sequence $\bs{r}_N$ once and for all. To simplify notations in the following, we consider the associated counting function
$\wh{\xi}_{ \bs{r}_N \mid \be=0}$ and denote specifically the roots $\wh{\xi}_{ \bs{r}_N \mid \be=0} ( \la_a )=\tf{a}{L}$. Thus, $\la_{r_a} \; = \; \mu_{a;\bs{r}_N}^{(0)}$, $a=1,\dots,N$.

Rephrasing what had been said, one has that
\beq
\wh{\xi}_{ \bs{r}_N \mid \be=0}(\om) \, \tend  \, \xi(\om) \, = \,  \f{1}{2\pi  }  \om  \; +  \;
\f{1}{2\pi } \Int{\R }{} \dd \la \th \big(\om - \la \big) \varrho(\la)   \;+\;  \f{D}{2}\, ,  \qquad \e{where} \qquad
 D \, = \,  \underset{N,L\tend +\infty}{ \lim } \tfrac{N}{L}  \, = \, \Int{\R}{} \dd \la \varrho(\la) \, .
\label{definition de xi thermo}
\enq
 The quantity $\xi^{\prime}(\la)$ represents the local density around the point $\la$
of the solutions to the equation $ { \ex{2\i\pi L\wh{\xi}_{ \bs{r}_N } (\om) } }_{\mid \be=0}=1$. This density obviously majorates $\varrho$.
If $r_a\leq b < b+1 \leq r_{a+1}$, then obviously one has that $\la_{b+1}-\la_b  \, \leq  \, \la_{r_{a+1}}  - \la_{r_{a}}$ so that
\beq
\f{ \varrho(\la) }{ \xi^{\prime}(\la) } \, = \, \underset{ \substack{ L \tend + \infty \\ \la_{b}, \la_{r_a} \tend \la} }{\lim} \bigg\{ \f{    L(\la_{b+1}-\la_b) }
{  L(\la_{r_{a+1}}-\la_{r_a})  } \bigg\} \, \leq \, 1 \;.
\enq

This then leads to the below property of the filling fraction
\beq
\vartheta(\la) \, = \, \f{ \varrho(\la) }{ \xi^{\prime}(\la) }  \in \intff{0}{1}  \;,
\enq
which, of course, was also obvious from its original definition.

Within a similar line of reasoning, see, \textit{e.g.}, \cite{BogoliubiovIzerginKorepinBookCorrFctAndABA}, one argues that, for any local operator,
\beq
\big< \op{O} \big>_{\bs{\varrho}} \, = \,  \lim_{N,L \tend + \infty}
\big< \op{O} \big>_{L, \bs{\varrho}} \; = \; \lim_{N,L \tend + \infty} \big< \op{O} \big>_{L, \bs{\varrho}}^{(\e{eff})}
\enq
with
\beq
 \big< \op{O} \big>_{L, \bs{\varrho}}^{(\e{eff})}  \; = \;     \f{  \Big(  \Psi\big( \{ \la_{r_a}  \}_1^{N} \big) , \op{O} \, \Psi\big( \{ \la_{r_a}  \}_1^{N} \big)  \Big) }
{  \Big(  \Psi\big( \{ \la_{r_a}   \}_1^{N} \big) , \Psi\big( \{ \la_{r_a}  \}_1^{N} \big)  \Big)   }   \;.
\enq

This formula will be starting point for computing two-point functions in a density matrix $\bs{\varrho}$ ensemble associated with the non-linear Schrödinger model.

Several correlators stand out. On the one hand, those related to the particle density and its current
\beq
\big< \op{n}(x,t) \, \op{n}(0,0) \big>_{\bs{\varrho}} \;, \quad \big< \op{j}_{\op{n}}(x,t) \, \op{n}(0,0) \big>_{\bs{\varrho}}
\quad \e{and} \quad \big< \op{j}_{\op{n}}(x,t) \, \op{j}_{\op{n}}(0,0) \big>_{\bs{\varrho}} \;.
\enq
There, we have $\op{n}(x,t) \, = \,  \Phi^{\dagger}(x,t) \Phi(x,t)$ while the particle density current is defined as
\beq
\Dp{x}\op{j}_{\op{n}}(x,t) \, + \,  \Dp{t}\op{n}(x,t) \; = \; 0 \;.
\label{ecriture eqn conservation courant}
\enq
The other class of correlators corresponds to two-point functions of non-particle preserving operators, in particular
not corresponding to densities of conserved charges. In the following, we shall focus on the reduced density matrix
\beq
\big< \Phi (x,t) \Phi^{\dagger}(0,0) \big>_{\bs{\varrho}} \, .
\enq
We recall that local observables evolve in space and time according to
\beq
\op{O}(x,t) = \ex{ - \i x \op{P} + \i t \op{H} } \op{O}(0,0) \ex{ \i x  \op{P} - \i t \op{H}  } \; ,
\label{ecriture eqn evolution spatio_temporelle obs locales}
\enq
where $\op{H}$ is the Hamiltonian of the model given in \eqref{definition Hamiltonien Bose} and $ \op{P}$ is the total momentum operator. The action of
$ \op{P}$ on the eigenstates of $\op{H}$ has been computed in \cite{BogoliubiovIzerginKorepinBookCorrFctAndABA}. One has for any
set of Bethe roots $\big\{ \mu_{\ell_a} \big\}_{1}^{N+\mf{s}}$ subordinate to the counting function indexed by $\bs{\ell}_{N+\mf{s}}$
\beq
 \ex{ \i x  \op{P} - \i t \op{H}  }  \Psi\big( \{ \mu_{\ell_a} \}_1^{ N + \mf{s} } \big)   \, = \, \pl{b=1}{ N + \mf{s} } \Big\{ \ex{ \i x u_0( \mu_{\ell_b} ) } \Big\}
\cdot
 \Psi\big(  \{ \mu_{\ell_a} \}_1^{ N + \mf{s} } \big)
\enq
in which $u_0$ stands for the below combination of bare momentum $p_0(\la)$ and bare energy $\veps_0(\la)$,
\textit{c.f.} \eqref{definition bare momemtum and energy}:
\beq
u_0(\la) \, = \, p_0(\la)- \f{t}{x} \veps_0(\la) \;.
\enq

Following \cite{KozTerrasNatteSeriesNLSECurrentCurrent,%
KozBabilonGohmannMininFFcurrentOperator}, one may show that
\beq
 \big<  \op{n}(x,t) \, \op{n}(0,0) \big>_{L, \bs{\varrho}}^{(\e{eff})}  \; = \; \f{1}{2} \Dp{x}^2\Dp{\be}^2 \wh{\mc{Q}}(x,t \mid \be)_{\mid \be=0} \;, \qquad
\big<  \op{j}_{\op{n}}(x,t) \, \op{j}_{\op{n}}(0,0) \big>_{L, \bs{\varrho}}^{(\e{eff})}   \; = \; \f{1}{2} \Dp{t}^2\Dp{\be}^2 \wh{\mc{Q}}(x,t \mid \be)_{\mid \be=0}
\label{definition valeur moyenne density densite}
\enq
while
\beq
 \big<  \op{j}_{\op{n}}(x,t) \, \op{n}(0,0) \big>_{L, \bs{\varrho}}^{(\e{eff})}  \; = \; -\f{1}{2} \Dp{x}\Dp{t} \Dp{\be}^2 \wh{\mc{Q}}(x,t \mid \be)_{\mid \be=0}  \;.
\enq
Those equations involve the generating function
\beq
\wh{\mc{Q}}(x,t \mid \be) \, = \,
\sul{ \bs{\ell}_N \in \mathbb{Z}_{<}^{N} }{} \pl{a=1}{N} \Big\{ \ex{\i x [ u_0(\mu_{a;\bs{\ell}_N}^{(\be)})  - u_0(\la_{r_a})] } \Big\}
\, \mc{S}^{(\be)}\big( \bs{\ell}_N; \bs{r}_N\big)
\label{ecriture dvpmt FF fct generatrice}
\enq
in which
\beq
\mc{S}^{(\be)}\big( \bs{\ell}_N; \bs{r}_N\big) \, = \, \Bigg|
\f{  \Big( \Psi_{\be}\big( \{ \mu_{a;\bs{\ell}_N}^{(\be)}  \}_1^{N} \big) ,  \Psi\big( \{ \la_{r_a}  \}_1^{N} \big)    \Big) }
{  \big|\big|  \Psi\big( \{ \la_{r_a}   \}_1^{N} \big) \big|\big| \cdot \big|\big| \Psi_{\be}\big( \{ \mu_{a;\bs{\ell}_N}^{(\be)}  \}_1^{N} \big)   \big|\big|   } \Bigg|^2 \;.
\enq

The second correlator has an analogous form-factor expansion which can be obtained by inserting the decomposition of the identity
between the product of the field and conjugated field operators. This leads to $\wh{\rho}(x,t \mid 0 ) \, = \,  \big<  \Phi (x,t) \Phi^{\dagger}(0,0)  \big>_{L, \bs{\varrho}}^{(\e{eff})}$, where
\beq
\wh{\rho}(x,t \mid \be )   \; = \;
  \sul{ \bs{\ell}_{N+1} \in \mathbb{Z}_{<}^{N+1} }{} \pl{a=1}{N+1} \Big\{ \ex{\i x  u_0(\mu_{a;\bs{\ell}_{N+1}}^{(\be)}) }  \Big\}
 \pl{a=1}{N}  \Big\{ \ex{-\i x  u_0(\la_{r_a})  } \Big\} \,
\mc{S}_{\Phi}^{(\be)}\big( \bs{\ell}_{N+1}; \bs{r}_N\big)
\enq
in which
\beq
\mc{S}_{\Phi}^{(\be)}\big( \bs{\ell}_{N+1}; \bs{r}_N\big)   \; = \;
\Bigg| \f{   \Big(  \Psi_{\be} \big( \{ \mu_{a;\bs{\ell}_{N+1}}^{(\be)} \}_1^{N+1} \big)  \, , \,  \Phi^{\dagger}(0,0) \Psi\big( \{ \la_{r_a}\}_1^N \big) \Big)  }
{ \big|\big|  \Psi_{\be}\big( \{ \mu_{a;\bs{\ell}_{N+1}}^{(\be)} \}_1^{N+1} \big) \big|\big|
\cdot \big|\big|  \Psi\big( \{ \la_{r_a} \}_1^{N} \big) \big|\big|     }
\Bigg|^2  \;.
\label{definition ratio FF pour ChampChampDual}
\enq
Note that the use of the $\be$-deformation is not necessary
here. However, it allows us to present the calculations in a
uniform way for all the correlators of interest.

\subsection{The excited states above the GGE representative state}
\label{SousSectionExcitedStatesAboveGGERepresentative}

The vector of integers $\bs{r}_N$ serves as a reference for parameterising all Eigenstates relatively to it.
For that purpose, one parameterises other vectors of integers $\bs{\ell}_{N + \mf{s} } \in \mathbb{Z}^{N + \mf{s} }_{<}$, $|\mf{s}|<N$ in terms of auxiliary
particle-hole integers
\beq
\{\ell_a\}_1^{N+ \mf{s} } \; = \; \{r_a\}_1^{N} \cup \{p_a\}_1^{ n + \mf{s} } \setminus \{h_a\}_1^n\;,
\label{decomposition entiers etat excite}
\enq
where $n\geq \max\{0, - \mf{s} \}$ while $h_1<\dots <h_n$ and $p_1<\dots <p_{n + \mf{s} }$. Moreover, one has the constraints
$p_a \in \mathbb{Z} \setminus \{r_a\}_1^{N}$ and $h_a \in   \{r_a\}_1^{N}$. The above allows one to parameterise any vector of integers
$\bs{\ell}_{N + \mf{s} } \in \mathbb{Z}^{N + \mf{s} }_{<}$, $|\mf{s}|<N$, in terms of the reduced data
\beq
 \bs{\mf{e}} \, = \, \big( \bs{p}_{n+\mf{s}} , \bs{h}_n  \big) \;.
\enq
Further, fix a set of excitation integers $\bs{\ell}_{N+\mf{s}}$ as above in \eqref{decomposition entiers etat excite}
 and given the associated root distribution $ \{ \mu_{a;\bs{\ell}_{N + \mf{s} }}^{(\be)} \}_1^{N + \mf{s}} $,
consider the roots as introduced in \eqref{definition des solutions mua subordonnees a ellN}
and introduce the discrete shift function
\beq
\wh{F}_{\bs{\ell}_{N+\mf{s}}}(\om) \, = \, L \, \Big( \, \wh{\xi}_{\bs{r}_N\mid \be=0}(\om) \, - \, \wh{\xi}_{ \bs{\ell}_{N+\mf{s}} }(\om)\Big) \;.
\label{definition fonction shift}
\enq
It is customary to assume that $\wh{F}_{\bs{\ell}_{N+\mf{s}}}$ admits a \textit{finite} thermodynamic limit
\beq
\lim_{L \tend + \infty} \wh{F}_{\bs{\ell}_{N+\mf{s}}}(\om) \; = \; F_{ \bs{\mf{e}} }\big( \om   \big) \;.
\enq
We shall check below the self-consistency of this assumption and provide an explicit characterisation of $ F_{ \bs{\mf{e}} }$. As we shall see,
the limit depends on the choice of the vector $\bs{\mf{e}}$, \textit{viz}.
on the integers $\bs{p}_{n+\mf{s}}$ and $\bs{h}_n$ parameterising $\bs{\ell}_{N+\mf{s}}$.

Having the existence of the thermodynamic limit at hand, one evaluates \eqref{definition fonction shift} at $\la_{r_a}$ and recalls
the definition \eqref{definition des solutions mua subordonnees a ellN} of $\mu_{r_a}$ so as to get
\beq
\mu_{r_a}-\la_{r_a} \, = \,  \f{ F_{ \bs{\mf{e}} }\big( \la_{r_a}  \big)  }{ L \xi^{\prime}( \la_{r_a})  } \Big( 1+\e{o}(1)\Big) \;.
\label{ecriture shift racines mu par rapport la}
\enq

$F_{ \bs{\mf{e}} }$ can be characterised in terms of solutions to linear integral equations.
To state the latter, some notations need to be introduced.
Given two disjoint sets  $A, B$, we agree upon
\beq
\sul{s \in A\ominus B}{} f(s) \, = \, \sul{s \in A }{} f(s) \, - \,  \sul{s \in B }{} f(s)  \qquad \e{and} \qquad
\pl{s \in A\ominus B}{} f(s) \, = \, \f{ \pl{s \in A }{} f(s) }{ \pl{s \in B }{} f(s) }
\enq
as well as
\beq
\sul{s \in A\oplus B}{} f(s) \, = \, \sul{s \in A }{} f(s) \, + \,  \sul{s \in B }{} f(s)  \qquad \e{and} \qquad
\pl{s \in A\oplus B}{} f(s) \, = \,  \pl{s \in A }{} f(s) \cdot \pl{s \in B }{} f(s)  \;.
\enq
Further, given a bounded smooth function $\nu$, define
\beq
 \xi_{\nu} \; = \; \xi + \f{\nu}{L}
\label{definition xi nu}
\enq
with $\xi$ as given by \eqref{definition de xi thermo}. Then, starting from a given $F_{\bs{\mf{e}}}$, define
the sequence of roots $\sg_a$ as the unique solutions to
\beq
\xi_{-F_{\bs{\mf{e}}} }(\sg_a) \, = \, \f{a}{L} \; \;  \e{with} \; \; a \in \mathbb{Z} \;.
\enq
These are obviously functionals of $F_{\bs{\mf{e}}}$, although, for simplicity, we shall keep this dependence implicit.
>From there, introduce
\beq
\mc{Y}_{\bs{\mf{e}}} \, = \, \{\sg_{p_a} \}_1^{n+\mf{s}} \quad \e{and} \quad \mf{X}_{\bs{\mf{e}}} \, = \, \{\sg_{h_a} \}_1^{n} \,
\label{definition Y et X}
\enq
Under the assumed smoothness of the large-$L$ limits, one has that $\sg_{a} = \mu_{a} \, + \, \e{o}(1)$ and thus, in
particular, $\sg_{p_a}=\mu_{p_a}\, + \, \e{o}(1)$ and  $\sg_{h_a}=\mu_{h_a}\, + \, \e{o}(1)$.

The large-$L$ behaviour of the difference of Bethe roots \eqref{ecriture shift racines mu par rapport la}
allows one to estimate the large $L$ asymptotics of the  difference sums
\beq
\sul{a=1}{N} \Big\{ f(\mu_{r_a})-f(\la_{r_a}) \Big\} \; \simeq \; \sul{a=1}{N} \Big( \mu_{r_a}-\la_{r_a}  \Big)f^{\prime}(\la_{r_a}) \, = \,
 \Int{ \R }{} \hspace{-1mm} \dd \mu \, \vartheta(\mu) F_{ \bs{\mf{e}} }(  \mu  ) f^{\prime}(\mu)  \,  +\, \e{o}(1)\;.
\enq
Thus, taken the decomposition of the integers $\{\ell_a\}_1^{N+\mf{s}}$ given in \eqref{decomposition entiers etat excite}, one gets that
\beq
\sul{a=1}{N+\mf{s}}   f(\mu_{\ell_a}) \, - \,\sul{a=1}{N}  f(\la_{r_a})  \; = \; \sul{t\in \mc{Y}_{\bs{\mf{e}}}\ominus \mf{X}_{\bs{\mf{e}}} }{} f(t)\, + \,
\Int{ \R }{} \hspace{-1mm} \dd \mu \,  \vartheta(\mu) F_{\bs{\mf{e}}}(\mu) f^{\prime}(\mu) \, +  \, \e{o}(1) \;.
\label{ecriture limite thermo somme simple}
\enq
Note that \eqref{ecriture limite thermo somme simple} still contains some $1/L$ dependence in the leading terms since the scaling limit of the particle-hole
rapidities in the original set of Bethe roots $\{\mu_{\ell_a}\}$ was only taken partially as described above by replacing
the $\mu_{p_a}$, resp.  $\mu_{h_a}$, by $\sg_{p_a}$, resp.  $\sg_{h_a}$. In principle, these $\sg$ variables will converge to
some limit depending on the value of the limits $\tf{p_a}{L}$, resp. $\tf{h_a}{L}$. However, the way of absorbing some of the sub-leading corrections in the volume
turns out to be the most appropriate for later handlings.

Those identities completely determine $F_{\bs{\mf{e}}}$ in that the latter satisfies the linear integral equation
\beq
\Big( \e{id}-\op{K}_{\vth}\Big)[F_{\bs{\mf{e}}}](\la) \, = \, \f{ \i \be }{2\pi}  -\f{\mf{s}}{2}  \, - \, \f{1}{2\pi}\sul{t\in \mc{Y}_{\bs{\mf{e}}}\ominus \mf{X}_{\bs{\mf{e}}} }{} \th(\la-t)
\enq
with $\op{K}_{\vth}$ as introduced in \eqref{definition operateur integral Kvth}.

In the following, we shall assume that $\e{id}-\op{K}_{\vth}$ is invertible on $L^2(\R)$ with inverse $\e{id}+\op{R}_{\vth}$
defined as
\beq
\op{R}_{\vth}[f](\la) \, = \, \Int{\R }{} \!\! \dd \mu \vth(\mu) R(\la,\mu) f(\mu)
\enq
in which $R(\la,\mu)$ is the unique solution to the linear integral equation
\beq
R(\la,\mu) \, = \, K(\la-\mu) \, + \,  \Int{\R}{} \!\! \dd \tau \vth(\tau) K(\la-\tau)R(\tau,\mu) \, .
\label{ecriture LIE definissant le resolvent}
\enq

Hence, upon introducing the dressed charge and dressed phase as the unique solutions to the linear integral equations
\beq
\Big( \e{id}-\op{K}_{\vth}\Big)[Z](\la) \, = \, 1 \qquad \e{and} \qquad
\Big( \e{id}-\op{K}_{\vth}\Big)[\vp(*,\mu)](\la) \, = \, \f{1}{2\pi} \th(\la-\mu) \, ,
\enq
one obtains that 
\beq
F_{\bs{\mf{e}}}(\la) \, = \,  \Big( \f{ \i \be }{2\pi}  -\f{\mf{s}}{2}\Big) Z(\la) \, - \, \sul{t\in \mc{Y}_{\bs{\mf{e}}} \ominus \mf{X}_{\bs{\mf{e}}} }{} \vp(\la,t) \;.
\enq
Note that $F_{\bs{\mf{e}}}$ has an implicit dependence on the parameters $\sg_{p_a}$ and $\sg_{h_a}$. These, in fact, solve the system of higher level Bethe Ansatz equations:
\beq
\bs{\mf{Y}}_{\bs{\mf{e}}}(\bs{\sg}) \, = \, 0 \quad \e{with} \quad \bs{\sg}\, = \, \Big( \sg_{p_1},\dots, \sg_{p_{n+\mf{s}}}, \sg_{h_1},\dots, \sg_{h_{n}} \Big) \;,
\enq
where 
\beq
\bs{\mf{Y}}_{\bs{\mf{e}}}(\bs{\sg}) \, = \, \Big( \xi_{ -F_{\bs{\mf{e}}} }(\sg_{p_1})-\tfrac{p_1}{L} ,\dots,
\xi_{ -F_{\bs{\mf{e}}} }(\sg_{p_{n+\mf{s}}})-\tfrac{p_{n+\mf{s}}}{L},
\xi_{ -F_{\bs{\mf{e}}} }(\sg_{h_1})-\tfrac{h_1}{L} ,\dots,\xi_{ -F_{\bs{\mf{e}}} }(\sg_{h_n})-\tfrac{h_n}{L} \Big) \;.
\label{ecriture HLBAE map}
\enq
It will turn out convenient to represent the shift function as a functional of the simpler function given below. This will allow to have
more systematic representations for the large-$L$ behaviour of the form factors that we will consider below. Consider
\beq
\nu_{ \bs{\mf{e}} }(z) \; = \; \sul{  w \in \mc{Y}_{\bs{\mf{e}}} \ominus \mf{X}_{\bs{\mf{e}}}  }{} \f{ 1 }{ z -  w } \;,
\label{definition fonction generatrice nu s}
\enq
and let
\beq
\mf{f}_{\mf{s}}[\nu](\om) \, = \, \Big( \f{\i \be }{ 2\pi } - \f{\mf{s} }{2} \Big) Z(\om)
\, - \hspace{-1mm} \Int{ \Ga(\R)  }{}  \hspace{-1mm} \f{ \dd z }{ 2 \i \pi } \vp(\om,z) \nu(z) \;.
\label{definition fonctionnelle mf f}
\enq
Above, $\Ga(\R)$ is a sufficiently small counterclockwise loop around $\R$.
It is then direct to check that
\beq
F_{\bs{\mf{e}}} (\om) \, = \, \mf{f}_{\mf{s}}[ \nu_{ \bs{\mf{e}} } ](\om) \;.
\label{ecriture lien Fct shift et fnelle shift en nu e}
\enq

Note also that with the above notations, given 
\beq
\La=\{\la_{r_a}\}_1^N  \quad \e{and} \quad  \Om=\{\mu_{r_a}\}_1^N \, ,
\label{definition ensembles Lambda et Upsilon}
\enq
one has that for any $f \in \mc{C}^{3}(\R^2)$, $h \in \mc{C}^2(\R)$ having bounded derivatives, it holds
\beq
\sul{s\in \Om\ominus \La }{}  \hspace{-1mm}  h(s) \, =   \Int{\R  }{} \hspace{-1mm} \dd u  \,
F_{\bs{\mf{e}}}(u)\vth(u)\Dp{u}h(u) \;  +\; \e{o}(1)
\label{ecriture thermo limit simple sum}
\enq
and
\beq
\sul{s,t \in \Om\ominus \La }{} \hspace{-1mm}  f(s,t) \, =  \Int{\R^2 }{} \hspace{-1mm}  \dd u \dd v \, F_{\bs{\mf{e}}}(u)\vth(u)F_{\bs{\mf{e}}}(v)\vth(v)
\Dp{uv}f(u,v)
\;  +\; \e{o}(1) \;.
\label{ecriture thermo limit double sum}
\enq
As a first application of the above reasonings, one readily infers that
\beq
\f{  \pl{a=1}{N+\mf{s}} \ex{ \i x  u_0(\mu_{\ell_a}) }  }{  \pl{a=1}{N} \ex{ \i x u_0(\la_{r_a} ) }   } \, = \, \ex{\i x \ga_{\mf{s}} }
\pl{ w \in \mc{Y}_{\bs{\mf{e}}} \ominus \mf{X}_{\bs{\mf{e}}} }{} \ex{\i x u(w) }  \cdot \Big( 1\, + \, \e{o}(1) \Big) \;,
\enq
where
\beq
 \ga_{\mf{s}} \, = \,  \big( \f{ \i \be }{2\pi}  -\f{\mf{s}}{2}\big) \Int{\R }{}\hspace{-1mm} \dd w \, \vth(w) Z(w) u_0^{\prime}(w)
\enq
while
\beq
u(\la) \, = \,p(\la) \, - \, \f{t}{x} \veps(\la) \, = \,  u_0(\la) \, - \,  \Int{\R }{} \hspace{-1mm} \dd w \, \vth(w) \vp(w,\la) u_0^{\prime}(w) \;.
\enq
The above defines the dressed momentum $p$ and the dressed energy $\veps$.
By $\la$-differentiation, one gets that $u^{\prime}(\la) = \big( \e{id} + \op{R}_{\vth} \big)[u_0^{\prime}]$. Since
$Z= \big( \e{id} + \op{R}_{\vth} \big)[1]$, one infers from the symmetry of the resolvent kernel $R(\la,\mu)=R(\mu,\la)$
that
\beq
 \ga_{\mf{s}} \, = \, x \ga_{\mf{s},x}-t  \ga_{\mf{s},t}  \qquad \e{with} \qquad
 \left( \ba{c} \ga_{\mf{s},x}  \\ \ga_{\mf{s},t} \ea \right) \, = \,
 \big( \f{ \i \be }{2\pi}  -\f{\mf{s}}{2}\big) \Int{\R }{}\hspace{-1mm} \dd w \, \vth(w)
          \left( \ba{c} p^{\prime}(w)  \\
        \veps^{\prime}(w)  \ea \right) \;.
\enq

\section{The form factors at play}
\label{Section Form factors}

\subsection{The finite-$L$ expressions}
\label{SousSectionFiniteLExpressions}

\subsubsection{The weight factor for the generating function}

Following the rewriting of the $\be$-twisted scalar products obtained in \cite{KozKitMailSlaTerXXZsgZsgZAsymptotics}, one has the decomposition
\beq
\mc{S}^{(\be)}\big( \bs{\ell}_N ; \bs{r}_N \big) \, = \, \Big( \mc{D}_N \mc{A}_N^{(\be)} \Big)( \bs{\ell}_N; \bs{r}_N )
\enq
into the so-called regular and singular parts \cite{KozKitMailSlaTerEffectiveFormFactorsForXXZ}.

The singular part takes the general form
\beq
\mc{D}_{N+\mf{s}} ( \bs{\ell}_{N+\mf{s}};\bs{r}_N )  \; = \;
\f{ \pl{a=1}{N} \Big\{ 4 \sin^{2}\big[ \pi \wh{F}_{\bs{\ell}_{N+\mf{s}}}(\la_{r_a} ) \big] \Big\}   }
{ \pl{a=1}{N+\mf{s}} \Big\{ 2\pi L \wh{\xi}^{\prime}_{\bs{\ell}_{N+\mf{s}}  }(\mu_{\ell_a}) \Big\}
\pl{a=1}{N} \Big\{ 2\pi L \wh{\xi}^{\prime}_{ \bs{r}_N}(\la_{r_a}) \Big\}  }
\f{  \pl{a<b}{N+\mf{s} } \big( \mu_{\ell_b} -  \mu_{\ell_a} \big)^2   \pl{a<b}{N } \big( \la_{r_b} -  \la_{r_a} \big)^2   }
{   \pl{a=1}{N+\mf{s} } \pl{b=1}{N } \big( \la_{r_b} -  \mu_{\ell_a} \big)^2 } \;.
\label{ecriture explicite fonctionnelle D+}
\enq
Note that, when $\mf{s}=0$, the double products reduce to the square of a Cauchy determinant. Further, the above formula is to be understood as follows.
$\wh{F}_{\bs{\ell}_{N+\mf{s}}}$ is the shift function introduced in \eqref{definition fonction shift}. Further, the roots
$\mu_a$ appearing in \eqref{ecriture explicite fonctionnelle D+} are defined as in \eqref{definition des solutions mua subordonnees a ellN}
but in respect to the shift function $\wh{\xi}_{ \bs{\ell}_{N+\mf{s}} }$.

The regular part, $\mc{A}_N^{(\be)}( \bs{\ell}_N; \bs{r}_N )$ has a more intricate expression:
\beq
\mc{A}_N^{(\be)}( \bs{\ell}_N; \bs{r}_N ) \, = \, \f{ \mc{W}_N( \bs{\ell}_N; \bs{r}_N ) }{ \det_{N}\big[ \Xi_{\bs{\ell}_N}\big] \cdot  \det_{N}\big[ \Xi_{\bs{r}_N}\big] }  \cdot
\pl{s=1}{2}\Bigg\{
\f{  \big( 1 - \ex{\be}\big)  \det_{  \Ga(\R) }  \big[ \e{id} \,  +\, \wh{\op{U}}_{\mf{t}_s} \big]  }
{      1\, - \, \ex{-2\i \pi \wh{F}_{\bs{\ell}_{N}}(\mf{t}_{s} )}    }
\pl{a=1}{N}  \bigg(  \f{ \mf{t}_s-\la_{r_a}+ \i c  }{ \mf{t}_s-\mu_{\ell_a}+ \i c } \bigg)  \Bigg\} \;.
\enq
Above, $\mf{t}_s\in \Cx$, $s=1,2$, are free parameters: the quantity in brackets appearing above does not depend on these. See, \textit{e.g.},
\cite{KozKitMailSlaTerXXZsgZsgZAsymptotics} for more details.

Several objects have been introduced in the expression for $\mc{A}_N^{(\be)}$.
To start with, the entries of the matrix $\Xi_{ \bs{\ell}_{N+\mf{s}} }$ are defined in terms of the collection of Bethe roots subordinate to the counting function
$\wh{\xi}_{ \bs{\ell}_{N+\mf{s}}  }$ as in \eqref{counting fct beta elle N}:
\beq
\big[ \Xi_{ \bs{\ell}_{N+\mf{s}} } \big]_{ab} \, =  \,
\de_{ab} \, - \,  \f{  K\big( \mu_{\ell_a}-\mu_{\ell_b} \big) }{  L \wh{\xi}_{ \bs{\ell}_{N+\mf{s}}  }^{\prime}\big( \mu_{\ell_b} \big) }
 \; .
\label{definition matrice de Gaudin}
\enq
 Further, $\wh{U}_{\mf{t}_s}$ is an integral operator on $L^2\big(  \Ga(\R) \big)$
with $ \Ga(\R)$ a sufficiently small counterclockwise loop around $\R$. It has integral kernel
\beq
\wh{U}_{\mf{t}}(\om, \om^{\prime}) \, = \, \f{ - 1 }{ 2\i\pi } \pl{a=1}{N}  \bigg\{  \f{( \om-\la_{r_a}+ \i c )(\om-\mu_{\ell_a}) }
{ (\om-\mu_{\ell_a}+ \i c) ( \om-\la_{r_a} ) } \bigg\} \, \cdot
\f{ K_{\be}(\om-\om^{\prime}) \, - \, K_{\be}(\mf{t}-\om^{\prime}) }{   1\, - \, \ex{-2\i \pi \wh{F}_{\bs{\ell}_{N}}(\om )}  }
\enq
in which
\beq
K_{\be}(\om) \, = \, \f{1}{\om + \i c } \, - \, \f{  \ex{\be} }{ \om - \i c } \quad \e{so} \, \e{that} \quad
K_{0}(\om) \,= \,  -2\i\pi K(\om) \,.
\enq
Finally, we have set
\beq
\mc{W}_{N+\mf{s}}( \bs{\ell}_{N+\mf{s}}; \bs{r}_N ) \, = \,
 \f{ \pl{a=1}{N+\mf{s}} \pl{b=1}{N}  \big( \mu_{\ell_a}-\la_{r_b}-\i c \big) \big( \la_{r_b} - \mu_{\ell_a}- \i c \big)  }
 {   \pl{a,b=1}{N+\mf{s}}    \big( \mu_{\ell_a}- \mu_{\ell_b} -\i c \big)   \pl{a,b=1}{N}  \big( \la_{r_a}-\la_{r_b} - \i c \big)  } \;,
\enq
in which it is understood that the variables $\mu_{\ell_a}$ are defined in terms of the shift function subordinate to $\bs{\ell}_{N+\mf{s}}$.
Note that, upon recalling the definition of the sets $\Om$ and $\La$, \textit{c.f.} \eqref{definition ensembles Lambda et Upsilon}, and introducing
\beq
\wh{\mc{Y}}_{\bs{\mf{e}}} \, = \, \{ \mu_{p_a} \}_{1}^{n+\mf{s}} \quad \e{and} \quad \wh{\mf{X}}_{\bs{\mf{e}}} \, = \, \{ \mu_{h_a} \}_{1}^{n}
\label{definition hat Y et hat X}
\enq
it holds
\beqa
\mc{W}_{N+\mf{s}}( \bs{\ell}_{N+\mf{s}}; \bs{r}_N ) & = &
\pl{ \substack{ s,t \in  \La \ominus \\ ( \Om \oplus \wh{\mc{Y}}_{\bs{\mf{e}}} \ominus \wh{\mf{X}}_{\bs{\mf{e}}}) } }{} \hspace{-3mm}\big( s - t -\i c \big)^{-1} \;,  \nonumber \\
& = &  \pl{ \substack{ s,t \in  \\ \La \ominus  \Om } }{} \big( s - t -\i c \big)^{-1}
\pl{ \substack{ s \in  \La \ominus \Om } }{} \pl{ \substack{ t \in \wh{\mc{Y}}_{\bs{\mf{e}}} \ominus \wh{\mf{X}}_{\bs{\mf{e}}} } }{}
\Big\{  \big( s - t -\i c \big)\big( t - s -\i c \big) \Big\}
\pl{ \substack{ s,t \in   \\   \wh{\mc{Y}}_{\bs{\mf{e}}} \ominus \wh{\mf{X}}_{\bs{\mf{e}}} } }{}  \big( s - t -\i c \big)^{-1}  \;.
\eeqa

\subsubsection{The weight factor for the reduced density matrix}

The weight factor introduced in \eqref{definition ratio FF pour ChampChampDual} is given in terms of the form factors of the conjugate field operator.
These have been first calculated in \cite{KorepinSlavnovFormFactorsNLSEasDeterminants},
a result which was later rigorously justified in \cite{KozFFConjFieldNLSELatticeSpacingGoes0}. Given
roots $\mu_{\ell_a}$ built from the counting function subordinate to $\bs{\ell}_{N+1}$, these are expressed as
\bem
\Big( \Psi_{\be}\big( \{ \mu_{\ell_a} \}_1^{N+1} \big) , \Phi^{\dagger}(0,0) \Psi\big( \{ \la_{r_a} \}_1^{N} \big) \Big)      \\
\, = \, \i \sqrt{c} \pl{a=1}{N+1} \ex{\f{\i L }{2} \mu_{\ell_a} } \cdot
\pl{k=1}{N} \Bigg\{ \ex{-\f{ \i L}{2}\la_{r_k} } \Big[ 1- \ex{-2 \i\pi \wh{F}_{\bs{\ell}_{N+1}}( \la_{r_k} )  } \Big]
\pl{b=1}{N+1} \bigg\{ \f{ \mu_{\ell_b} - \la_{r_k} - \i c }{ \mu_{\ell_b} - \la_{r_k}  }  \bigg\}   \Bigg\}
\det_{N}\big[\op{I}_N + \op{U}^{(+)} \big] \;.
\label{ecriture des FF}
\end{multline}
There, the entries of the matrix $\op{U}^{(\pm)}$ appearing above read
\beq
U_{jk}^{(\eps)} = - \i \eps   \pl{a=1}{N+1}\f{ \la_{r_j} - \mu_{\ell_a}  }{ \la_{r_j} - \mu_{\ell_a}  + \i \eps c } \cdot
 \f{ \pl{a=1}{N} \big( \la_{r_j}  - \la_{r_a} + \i \eps c \big) }{ \pl{ \substack{a=1 \\ a \not=j } }{N} \big( \la_{r_j}  -  \la_{r_a}  \big) }
 \cdot \f{ K \big(  \la_{r_j} - \la_{r_k} \big)  }{ \ex{-2 \i \pi \eps \wh{F}_{\bs{\ell}_{N+1}}( \la_{r_j}) } -1 }  \qquad \e{with} \quad \eps \in \{\pm\} \;.
\enq
Above, we made use of the so-called discrete shift function $\wh{F}_{\bs{\ell}_{N+1}}$ \eqref{definition fonction shift} subordinate to $\bs{\ell}_{N+1}$ which satisfies
\beq
\ex{-2 \i\pi \wh{F}_{\bs{\ell}_{N+1}}(\om)} \, = \,  \ex{\be} \pl{a=1}{N+1} \f{ \mu_{\ell_a}-\om + \i c }{ \mu _{\ell_a}-\om - \i c } \cdot
\pl{a=1}{N} \f{ \la_{r_a}-\om - \i c }{ \la_{r_a}-\om + \i c } \;.
\label{definition fonction shift discrete}
\enq
In its turn, the norm of a Bethe state admits the determinant representation \cite{GaudinMcCoyWuNormXXZ,KitanineMailletTerrasFormfactorsperiodicXXZ,KorepinNormBetheStates6-Vertex,SlavnovScalarProductsXXZ}

\beq
\Big( \Psi_{\be}\big( \{ \mu_{\ell_a} \}_1^{N+1} \big) ,  \Psi_{\be}\big( \{ \mu_{\ell_a} \}_1^{N+1} \big) \Big)  \, = \,
\pl{a=1}{N+1} \Big\{ 2 \i\pi L \wh{\xi}^{\, \prime}_{ \bs{\ell}_{N+1}  } \big( \mu_{\ell_a} \big)  \Big\}
\f{ \pl{a,b=1}{N+1} \big( \mu_{\ell_a}-\mu_{\ell_b} - \i c  \big)  }{ \pl{ \substack{ a,b=1 \\ a\not= b} }{N+1} \big( \mu_{\ell_a}-\mu_{\ell_b}  \big)  }
\det_{N+1}\big[ \Xi_{\bs{\ell}_{N+1} } \big] \;.
\label{ecriture formule des normes}
\enq
By gathering the determinant representations \eqref{ecriture des FF} and \eqref{ecriture formule des normes}, one gets that
\beq
\mc{S}_{\Phi}^{(\be)}\big( \bs{\ell}_{N+1}; \bs{r}_N\big)   \; = \; \Big(\mc{D}_{N+1}\cdot \mc{A}_N^{(\be; \Phi)}\Big) \big( \bs{\ell}_{N+1}; \bs{r}_N\big) \,,
\enq
where $\mc{D}_{N+1}$ is as already introduced in \eqref{ecriture explicite fonctionnelle D+} while
\beq
\mc{A}_N^{(\be; \Phi)}\big( \bs{\ell}_{N+1}; \bs{r}_N\big)\; = \; -\i c
\mc{W}_{N+1}( \bs{\ell}_{N +1}; \bs{r}_N )  \cdot
\f{ \det_{N}\big[\op{I}_N + \op{U}^{(+)} \big] \det_{N}\big[\op{I}_N + \op{U}^{(-)} \big] }
{  \det_{N+1}\big[ \Xi_{\bs{\ell}_{N+1}} \big] \det_{N}\big[ \Xi_{\bs{r}_{N}} \big]  }  \;.
\enq

\subsection{The leading large-$L$ behaviour}
\label{SousSectionLargeLAsymptotics}

\subsubsection{The large-$L$ behaviour of the double product}

We first deal with the double product term, this in the case of most general $\bs{\ell}_{N+\mf{s}}$. The identities
\eqref{ecriture thermo limit simple sum}--\eqref{ecriture thermo limit double sum} as well as the connection of the
shift function with the functional $\mf{f}_{\mf{s}}$ \eqref{ecriture lien Fct shift et fnelle shift en nu e}
allow one to conclude that, in the setting of interest,
\beq
\pl{ \substack{ s,t \in  \\ \La \ominus  \Om } }{} \big( s - t -\i c \big)^{-1}  \, = \,
\ex{ \msc{W}_2[\nu_{\bs{\mf{e}}} ] } \, \big(1+\e{o}(1) \big) \;,
\enq
in which
\beq
 \msc{W}_2[\nu] \, = \, \, - \Int{\R^2 }{} \dd u \dd w  \vth(u)\vth(w) \Dp{uw}^2 \ln \big(u-w-\i c\big) \, \mf{f}_{\mf{s}}[\nu](u)\mf{f}_{\mf{s}}[\nu](w) \;.
\enq
We remind that $\nu_{\bs{\mf{e}}}$ has been introduced in \eqref{definition fonction generatrice nu s} and $\mf{f}_{\mf{s}}[\nu]$
in \eqref{definition fonctionnelle mf f}. Further, it holds
\beq
\pl{ \substack{ s \in  \La \ominus \Om } }{} \pl{ \substack{ t \in \wh{\mc{Y}}_{\bs{\mf{e}}} \ominus \wh{\mf{X}}_{\bs{\mf{e}}} } }{}
\Big\{  \big( s - t -\i c \big)\big( t - s -\i c \big) \Big\} \, = \,
\ex{ \msc{W}_1[\nu_{ \bs{\mf{e}}} ] }
\big(1+\e{o}(1) \big) \, ,
\enq
where
\beq
 \msc{W}_1[\nu] \, =   \hspace{-1mm}  \Int{\Ga(\R)   }{} \hspace{-1mm} \f{ \dd y }{ 2\i\pi } \Int{\R}{} \hspace{-1mm} \dd u \,
 \vth(u)  \Dp{u} \ln \Big[ \big(u-y-\i c\big) \big(y-u-\i c\big) \Big] \,  \mf{f}_{\mf{s}}[\nu](u) \, \nu(y) \;.
\enq
Finally, it holds
\beq
\pl{ \substack{ s,t \in   \\  \wh{\mc{Y}}_{\bs{\mf{e}}} \ominus \wh{\mf{X}}_{\bs{\mf{e}}}   } }{}  \big( s - t -\i c \big)^{-1} \, = \,
\ex{ \msc{W}_0[\nu_{ \bs{\mf{e}}} ] } \, \big(1+\e{o}(1) \big) \;
\enq
with
\beq
\msc{W}_0[\nu] \, = \, -  \hspace{-1mm} \Int{\Ga(\R)^2  }{} \hspace{-1mm} \f{ \dd y \dd z  }{ (2\i\pi)^2  }\ln \big(z-y-\i c\big) \nu(y)\nu(z) \;.
\enq

All-in-all, we set
\beq
\msc{W}[\nu] \, = \, \msc{W}_0[\nu] + \msc{W}_1[\nu]  + \msc{W}_2[\nu] \;,
\label{definition fonctionnelle W}
\enq
so that it holds
\beq
\mc{W}_{N+\mf{s}}( \bs{\ell}_{N +\mf{s}}; \bs{r}_N )  \; = \; \ex{   \msc{W}[\nu_{\bs{ \mf{e} }}]} \,  \big( 1 + \e{o}(1) \big) \;.
\enq

\subsubsection{The large-$L$ behaviour of the Gaudin determinants}

The condensation property \eqref{ecriture condensation etat ref} along with the Lipschitz continuity of $2$-determinants and the invertibility of
$\e{id} \, - \, \op{K}_{\vth} $ all together imply that
\beq
 \det_{N}\big[ \Xi_{\bs{r}_N} \big] \, = \, \det_{L^2(\R)} \Big[\e{id} \, - \, \op{K}_{\vth} \Big] \cdot \Big( 1+ \e{o}(1)\Big) \;.
\enq
The same limit holds for $ \det_{N+\mf{s}}\big[ \Xi_{ \bs{\ell}_{N+\mf{s}} } \big]$.
However, for further purpose, it is more convenient to also include part of the subleading corrections to this determinant's large-$L$ behaviour.
We follow the handlings outlined, \textit{e.g.}, in \cite{KozBabenkoGohmannSuzukiMassiceXXZFFExplicitDensity}.
Introduce the integral operator on $L^2\big( \Ga( \{\mu_{\ell_a}\}_{1}^{N+\mf{s}}) \big)$ with $\Ga( \{\mu_{\ell_a}\}_{1}^{N+\mf{s}})$ a small counterclockwise loop
around the Bethe roots $\{\mu_{\ell_a}\}_{1}^{N+\mf{s}}$ which avoids any other pole of the integral kernel
\beq
\mc{K}(\om,\om^{\prime}) \, = \,  \f{ - K(\om-\om^{\prime}) }{  \ex{2\i \pi L \wh{\xi}_{\bs{\ell}_{N+\mf{s}}}(\om )} \, - \, 1 }
\enq
so that it holds
\beq
\det_{N+\mf{s}}\Big[ \de_{ab} \, - \, M_{\ell_a \ell_b}    \Big] \, = \,
\det_{ L^2\big( \Ga( \{\mu_{\ell_a}\}_{1}^{N+\mf{s}}) \big) } \Big[ \e{id} \, + \, \mc{K} \Big] \;, \quad \e{with} \quad
M_{ab} \, = \, \f{  K\big( \mu_{a}-\mu_{b} \big) }{  L \wh{\xi}_{ \bs{\ell}_{N+\mf{s}}  }^{\prime}\big( \mu_{b} \big) } \;.
\label{definition Mab}
\enq
Then, observe that one has the decomposition $\{\mu_{\ell_a}\}_{1}^{N+\mf{s}} \, = \, \Om \oplus \wh{\mc{Y}}_{\bs{\mf{e}}} \ominus \wh{\mf{X}}_{\bs{\mf{e}}}$
with $\Om$ as in \eqref{definition ensembles Lambda et Upsilon}, while the other two sets are as in \eqref{definition hat Y et hat X}.
Further, by computing the residues in the Fredholm determinant occurring on the \textit{rhs} above, one gets that
\beq
\det_{N+\mf{s}}\Big[ \de_{ab} \, - \, M_{\ell_a \ell_b}    \Big] \, = \,
\det_{N+2n +\mf{s}} \left[ \ba{ccc}   \de_{ab} - M_{r_a r_b} & -M_{r_a p_b}          &  M_{r_a h_b}  \vspace{2mm}  \\
                                       - M_{p_a r_b}        & \de_{ab} -M_{p_a p_b}  &  M_{p_a h_b}  \vspace{2mm}   \\
                                         - M_{h_a r_b}        &  -M_{h_a p_b}        &   \de_{ab} +  M_{h_a h_b}     \ea         \right]  \;.
\enq
In order to simplify the above expression, observe that since
\beq
 \det_{N}\big[ \de_{ab} - M_{r_a r_b} \big] \, = \, \det_{L^2(\R)} \Big[\e{id} \, - \, \op{K}_{\vth} \Big] \cdot \Big( 1+ \e{o}(1)\Big) \, \not= \, 0 \;,
\label{ecriture cvgce det M ramene a rN vers id moins K theta}
\enq
the $N\times N$ matrix with entries $ \de_{ab} - M_{r_a r_b}$ is invertible. Let $\de_{ab} + R_{r_a r_b}$ denote the inverse matrix.
By using the invertibility to the left and right, one gets the equation
\beq
R_{r_a r_b} \, = \, M_{r_a r_b} \, + \, \sul{s=1}{N}M_{r_a r_s} M_{r_s r_b} \, + \,  \sul{s,t=1}{N}M_{r_a r_s}R_{r_s r_t} M_{r_t r_b} \;.
\enq
By using that $M_{ab}$ is, in fact, defined for all $a,b \in \mathbb{Z}$ though \eqref{definition Mab}, one may extend the definition of $R_{r_a r_b}$
to the whole of $\mathbb{Z}^2$ as
\beq
R_{a b} \, = \, M_{a b} \, + \, \sul{s=1}{N}M_{a r_s} M_{r_s b} \, + \,  \sul{s,t=1}{N}M_{a r_s}R_{r_s r_t} M_{r_t b} \;.
\enq
Again, by going back to the definition of $M_{ab}$, one may introduce the discrete resolvent function
\beq
\wh{R}(\la,\om) \, = \, K(\la-\om) \, + \, \sul{s=1}{N} \f{  K\big( \la-\mu_{r_s} \big)  K\big( \mu_{r_s} - \om \big) }
{  L \wh{\xi}_{ \bs{\ell}_{N+\mf{s}}  }^{\prime}\big( \mu_{r_s} \big) }
\, + \,  \sul{s,t=1}{N} \f{  K\big( \la-\mu_{r_s} \big) R_{r_s r_t} K\big( \mu_{r_t} - \om \big) }
{  L \wh{\xi}_{ \bs{\ell}_{N+\mf{s}}  }^{\prime}\big( \mu_{r_s} \big) } \;.
\label{ecriture resolvent discret}
\enq
Next, it is easy to see owing to the unique solvability of the equation satisfied by $R_{r_a r_b}$, that
\beq
R_{r_a r_b} \, = \,     \f{   \wh{R}\big( \mu_{r_a},\mu_{r_b} \big) }{  L \wh{\xi}_{ \bs{\ell}_{N+\mf{s}}  }^{\prime}\big( \mu_{r_b} \big) } \;.
\enq
Then, by using this representation and the smoothness of $\wh{R}(\la,\om)$, one readily sees by sending $L\tend + \infty$
in \eqref{ecriture resolvent discret} that $\wh{R}(\la,\om) \, = \, R(\la,\om)  + \e{o}(1)$ with $R$ being the resolvent kernel defined in
\eqref{ecriture LIE definissant le resolvent}. All of that being set, one has
\beqa
\f{ \det_{N+\mf{s}}\big[ \de_{ab} \, - \, M_{\ell_a \ell_b}    \big] }{ \det_{N}\big[ \de_{ab} - M_{r_a r_b} \big]   }& = &
\det_{N+2n +\mf{s}} \left[ \ba{ccc}  \de_{ab}               & -M_{r_a p_b}  - \sul{s=1}{N} R_{r_a r_s}M_{r_s p_b}
                                                                                        &  M_{r_a h_b}  + \sul{s=1}{N} R_{r_a r_s}M_{r_s h_b} \vspace{2mm}  \\
                                       - M_{p_a r_b}        & \de_{ab} -M_{p_a p_b}  &  M_{p_a h_b}  \vspace{2mm}   \\
                                         - M_{h_a r_b}        &  -M_{h_a p_b}        &   \de_{ab} +  M_{h_a h_b}     \ea         \right] \nonumber  \\
& = & \det_{N+2n +\mf{s}} \left[ \ba{ccc}  \op{I}_N & \op{U} & \op{V}   \\
              \op{Y}^{\op{t}} & \op{I}_n + \op{A} &   \op{B}  \\
             \op{Z}^{\op{t}} &  \op{C} &   \op{I}_n + \op{D} \ea \right]     \nonumber \\
 &  =  &   \det_{N+2n +\mf{s}} \left[ \ba{ccc}  \op{I}_N & \op{U} & \op{V}   \\
               0   & \op{I}_n + \op{A} - \op{Y}^{\op{t}} \op{U} &   \op{B}  - \op{Y}^{\op{t}} \op{V} \\
               0   &  \op{C} - \op{Z}^{\op{t}} \op{U} &   \op{I}_n + \op{D} - \op{Z}^{\op{t}} \op{V}  \ea \right]     \nonumber \\
 &  =  &   \det_{2n +\mf{s}} \left[ \ba{cc}
  \de_{ab} -R_{p_a p_b}  &  R_{p_a h_b}  \vspace{2mm}   \\
-R_{h_a p_b}        &   \de_{ab} +   R_{h_a h_b}      \ea \right] \;.
\eeqa
Thus, upon invoking \eqref{ecriture cvgce det M ramene a rN vers id moins K theta}, one finds that
\beq
 \det_{N + \mf{s} }\big[ \Xi_{\bs{\ell}_{N+\mf{s}}} \big] \, = \, \det_{L^2(\R)} \Big[\e{id} \, - \, \op{K}_{\vth} \Big] \cdot
 \pl{s \in \mc{Y}_{\bs{\mf{e}}} \oplus \mf{X}_{\bs{\mf{e}}} }{} \bigg\{  \f{1}{\xi_{-\mf{f}_{\mf{s}}[\nu_{\bs{\mf{e}}}] }^{\prime}(s)} \bigg\}
 \det_{ 2n + \mf{s} } \Big[ \op{D}\bs{\mf{Y}}_{\bs{\mf{e}}}(\bs{\sg})\Big]  \cdot \Big( 1+ \e{o}(1)\Big) \;.
\enq
There $\op{D}\bs{\mf{Y}}_{\bs{\mf{e}}}(\bs{\sg})$ stands for the differential of the map $\bs{\mf{Y}}_{\bs{\mf{e}}}(\bs{\sg})$ introduced in \eqref{ecriture HLBAE map}.

\subsubsection{Thermodynamic limit of the $U$-determinants}

A direct residue calculation leads to
\beq
\det_{N}\big[\op{I}_N  + \op{U}^{(\eps)} \big] \; =  \; \det_{L^2(\Ga^{\prime}(\R))} \Big[\e{id} \, + \,\wh{\op{U}}^{(\eps)}  \Big]
\label{ecriture det I + U comme det Fred}
\enq
in which $\wh{\op{U}}^{(\eps)}$ acts on functions supported on $\Ga^{\prime}(\R)$ which is a small loop around $\R$
such that all the zeroes of $\ex{2\i\pi \wh{F}_{\bs{\ell}_{N+1}} (\om)}-1$ are outside of it. This
loop is taken so that the contour $\Ga(\R)$ introduced on various occasions earlier on satisfies
$\Ga(\R) \subset \Ga^{\prime}(\R)$, \textit{i.e.}, $\Ga^{\prime}(\R)$ laces around $\Ga(\R)$. The integral operator has the below integral kernel
\beq
\wh{U}^{(\eps)}(\om,\om^{\prime}) \, =  \, -\i\eps  \pl{a=1}{N+1} \f{ \om- \mu_{\ell_a} }{ \om- \mu_{\ell_a} +\i \eps c  } \cdot
\pl{a=1}{N} \f{ \om- \la_{r_a} +\i \eps c  }{ \om- \la_{r_a} } \cdot  \f{ K(\om-\om^{\prime})  }{ \ex{-2\i\pi \eps \wh{F}_{\bs{\ell}_{N+1}}(\om)} \, - \, 1  }    \;.
\enq
Note that while, \textit{per se}, a contour $\Ga(\R)$ having the above properties might not exist,
the \textit{rhs} of \eqref{ecriture det I + U comme det Fred} should in fact be understood as the analytic continuation in $\be$ from $-\i \eps \infty$
to a neighbourhood of $0$. See \cite{KozFFConjFieldNLSELatticeSpacingGoes0} for more details.

One readily gets that $\wh{U}^{(\eps)}(\om,\om^{\prime}) \, =  \,U^{(\eps)}(\om,\om^{\prime}) \cdot \big( 1+\e{o}(1)\big)$ in which
\beq
 U^{(\eps)}(\om,\om^{\prime}) \, =  \, -\i \eps \exp\bigg\{ \Int{\R}{}  \dd s \vth(s) F_{\bs{\mf{e}}}(s) \Dp{s}g_{\eps}(\om,s) \bigg\}
 \pl{ s \in \mc{Y}_{\bs{\mf{e}}}\ominus \mf{X}_{\bs{\mf{e}}} }{} \hspace{-2mm} \ex{ g_{\eps}(\om,s) }
 \cdot \f{ K(\om-\om^{\prime})  }{ \ex{-2\i\pi \eps F_{\bs{\mf{e}}}(\om)} \, - \, 1  }    \;.
\enq
In the above expression, we have introduced
\beq
 g_{\eps}(\om,\mu) \, = \, \ln \bigg(  \f{ \om-\mu }{ \om-\mu + \i \eps c } \bigg) \;. 
\enq
Further, introduce the functional
\beq
\mf{h}_{\eps, \mf{s} }[\nu](\om) \; = \;   \Int{ \R }{} \hspace{-1mm} \dd s \vth(s) \mf{f}_{\mf{s}}[\nu](s) \Dp{s}g_{\eps}(\om,s) \;  +
\Int{  \Ga(\R) }{}  \hspace{-1mm} \f{ \dd z }{ 2 \i \pi }  g_{\eps}(\om,z)\nu(z) \;.
\enq
Note that, explicitly, one has
\beq
\mf{h}_{\eps, \mf{s} }[\nu](\om) \; = \;  \Big( \f{\i \be }{ 2\pi } - \f{ \mf{s} }{2} \Big)  \Int{ \R }{} \dd s \vth(s) Z(s) \Dp{s}g_{\eps}(\om,s) \;  + \hspace{-2mm}
\Int{  \Ga(\R) }{}  \hspace{-1mm} \f{ \dd z }{ 2 \i \pi } \bigg\{ g_{\eps}(\om,z)
\, - \,  \Int{\R}{}  \dd s \vth(s) \vp(s,z) \Dp{s}g_{\eps}(\om,s)    \bigg\} \nu(z) \;.
\label{definition fnelle h}
\enq
Then, recalling \eqref{definition fonction generatrice nu s}--\eqref{definition fonctionnelle mf f},
a direct residue calculation yields that
\beq
 U^{(\eps)}(\om,\om^{\prime}) \, =  \, \msc{U}^{(\eps)}[\nu_{\bs{\mf{e}}} ](\om, \om^{\prime}) \;,
\enq
in which 
\beq
\msc{U}^{(\eps)}[\nu](\om, \om^{\prime}) \; = \;
-\i \eps \ex{ \mf{h}_{\eps,1}[\nu](\om) }  \cdot \f{ K(\om-\om^{\prime})  }{ \ex{-2\i\pi \eps \mf{f}_1[\nu](\om)} \, - \, 1  }  \;.
\label{definition noyau integral msc U veps}
\enq
Thus, continuity theorems for Fredholm determinants \cite{SimonsInfiniteDimensionalDeterminants} ensure that 
\beq
\det_{N}\big[\op{I}_N + \op{U}^{(\eps)} \big] \; =  \;
\det_{L^2(\Ga^{\prime}(\R))} \Big[\e{id} \, + \, \msc{U}^{(\eps)}[\nu_{\bs{\mf{e}}}] \Big] \cdot \Big( 1+ \e{o}(1)\Big) \;.
\enq

The expression for the integral kernel arising in the expression for $\mc{S}^{(\be)}$ can be recast in a similar fashion.
One eventually gets that $\wh{U}_{\mf{t}}(\om,\om^{\prime})\, = \, \msc{U}_{\mf{t}}[\nu_{\bs{\mf{e}}}](\om,\om^{\prime})\big( 1 + \e{o}(1) \big)$
where
\beq
\msc{U}_{\mf{t}}[\nu](\om, \om^{\prime}) \; = \;
 \f{ -1  }{ 2\i\pi }\ex{ \mf{h}_{+,0}[\nu](\om) }  \cdot \f{ K_{\be}(\om-\om^{\prime})\, - \,  K_{\be}(\om-\mf{t})  }
 { 1 \, - \,  \ex{-2\i\pi \mf{f}_{0}[\nu](\om)} }  \;.
\label{definition noyau integral U}
\enq

Very analogously, for a collection of roots $\{\mu_{\ell_a} \}_1^N$ subordinate to the counting function labelled by $\bs{\ell}_N$, a direct calculation shows that
\beq
\pl{a = 1 }{ N } \bigg\{ \f{ \mf{t}- \mu_{\ell_a} +  \i c }{  \mf{t}- \la_{r_a} +  \i c  }  \bigg\} \, = \, \ex{\mf{c}[\nu_{\bs{\mf{e}}}](\mf{t})} \big( 1+ \e{o}(1) \big) \,,
\enq
where
\beq
 \mf{c}[\nu](\mf{t}) \, = \,   \f{\i \be }{ 2\pi }  \Int{ \R }{} \dd s \vth(s) Z(s) \Dp{s}\ln(\mf{t}-s+\i c) \;  + \hspace{-2mm}
\Int{  \Ga(\R) }{}  \hspace{-1mm} \f{ \dd z }{ 2 \i \pi } \bigg\{ \ln(\mf{t}-z+\i c)
\, - \,  \Int{\R}{}  \dd s \vth(s) \vp(s,z) \Dp{s}\ln(\mf{t}-s+\i c)    \bigg\} \nu(z)  \;.
\label{definition fnell ct}
\enq

\subsubsection{The large-$L$ behaviour of the smooth parts}

By gathering all of the previous results, one gets that the thermodynamic limit of the smooth parts may be recast compactly in terms of  particle-hole number
\textit{independent} functionals $\msc{A}^{(\be)}$ and $\msc{A}^{(\be;\Phi)}$ acting on the function $\nu_{\bs{\mf{e}}}(z)$ introduced in \eqref{definition fonction generatrice nu s}.
\beq
  \mc{A}_N^{(\be;\Phi)} \big( \bs{\ell}_{N+1}; \bs{r}_N \big)  \, = \,
  \f{ \pl{s \in \mc{Y}_{\bs{\mf{e}}} \oplus \mf{X}_{\bs{\mf{e}}} }{} \Big\{  \xi_{- \mf{f}_{1}[\nu_{\bs{\mf{e}}}]  }^{\prime}(s) \Big\}
  \cdot \msc{A}^{(\be;\Phi)}[\nu_{\bs{\mf{e}}}] }{  \det_{2n+1}\Big[ \op{D}\bs{\mf{Y}}_{\bs{\mf{e}}}(\bs{\sg})\Big] }
\cdot \big( 1+ \e{o}(1)\big) \;,
\enq
where 
\beq
\msc{A}^{(\be;\Phi)}[\nu] \, = \,   -\i  c \ex{\msc{W}[\nu] }\pl{\eps=\pm}{}
\Bigg\{  \f{ \det_{L^2(\Ga^{\prime}(\R))} \Big[\e{id} \, + \, \msc{U}^{(\eps)}[\nu] \Big]   }
{  \det_{L^2(\R)}   \Big[\e{id} \, - \, \op{K}_{\vth} \Big]    }  \Bigg\} \;.
\label{definition amplitude red density matrix}
\enq
Above, the functional $\msc{W}$ is as introduced in \eqref{definition fonctionnelle W}, while $ \msc{U}^{(\eps)}[\nu]$ is defined through
\eqref{definition noyau integral msc U veps}.

The other amplitude of interest has a similar structure:
\beq
  \mc{A}_N^{(\be)} \big( \bs{\ell}_{N}; \bs{r}_N \big)  \, = \,
  \f{ \pl{s \in \mc{Y}_{\bs{\mf{e}}} \oplus \mf{X}_{\bs{\mf{e}}} }{} \Big\{  \xi_{- \mf{f}_{0}[\nu_{\bs{\mf{e}}}]  }^{\prime}(s) \Big\}
  \cdot \msc{A}^{(\be)}[\nu_{\bs{\mf{e}}}] }{  \det_{2n}\Big[ \op{D}\bs{\mf{Y}}_{\bs{\mf{e}}}(\bs{\sg})\Big] }
\cdot \big( 1+ \e{o}(1)\big) \;.
\enq
In its turn, we have set
\beq
\msc{A}^{(\be)}[\nu] \, = \,       \ex{\msc{W}[\nu] }\pl{a=1}{2}
\Bigg\{  \ex{-\mf{c}[\nu](\mf{t}_a) } \f{ (1-\ex{\be}) \, \det_{L^2(\Ga^{\prime}(\R))} \Big[\e{id} \, + \, \msc{U}_{\mf{t}_a}[\nu] \Big]   }
{  \big(1-\ex{-2\i\pi \mf{f}_0[\nu](\mf{t}_a) } \big) \cdot  \det_{L^2(\R)}   \Big[\e{id} \, - \, \op{K}_{\vth} \Big]  }  \Bigg\} \;.
\label{definition amplitude gen fct}
\enq
Again, $\msc{W}$ is given by \eqref{definition fonctionnelle W}, $\mf{c}$ has been defined in \eqref{definition fnell ct}
while $\msc{U}_{\mf{t}}$ in \eqref{definition noyau integral U}.

\subsubsection{A large-$L$ simplified form of the singular part}

To proceed with a simplification of the form taken by the large-$L$ limit of the singular part, it is convenient to introduce first
an auxiliary functional. Let $\zeta$ be a strictly increasing function on $\R$ that is also a bi-holomorphism in some open neighbourhood of $\R$
and pick $\nu$ a bounded holomorphic function in a neighbourhood of $\R$. Then introduce the roots $\{\eta_{r_a}\}_1^{N}$, resp.
$\{\tau_{\ell_a}\}_1^{ N + \mf{s} }$,  as the unique solutions to
\beq
\zeta( \eta_{r_a} ) = \tf{ r_a }{ L } \;, \quad \e{resp}.  \quad \zeta_{ - \nu}(\tau_{\ell_a}) \; = \; \f{\ell_a}{ L } \;.
\enq
Here we agree upon $\zeta_{\nu}=\zeta + \tf{\nu}{L}$, as already introduced in \eqref{definition xi nu}. Then, define the functional
\beq
 \msc{D}_N \big(  \bs{\ell}_{N+\mf{s}} ;  \bs{r}_{N} \big)\big[ \zeta, \nu \big] \; = \;
\f{ \pl{a=1}{N} \Big\{ 4 \sin^{2}\big[ \pi \nu(\eta_{r_a} ) \big] \Big\}   }
{ \pl{a=1}{N+\mf{s}} \Big\{  2\pi L \zeta^{\prime}_{-\nu}(\tau_{\ell_a}) \Big\} \pl{a=1}{N}  \Big\{ 2\pi L \zeta^{\prime}(\eta_{r_a}) \Big\}  }
\f{  \pl{a<b}{N+\mf{s} } \big( \tau_{\ell_b} -  \tau_{\ell_a} \big)^2   \pl{a<b}{N } \big( \eta_{r_b} -  \eta_{r_a} \big)^2   }
{   \pl{a=1}{N+\mf{s} } \pl{b=1}{N } \big( \eta_{r_b} -  \tau_{\ell_a} \big)^2 } \;.
%
%
\enq
Let us denote, for short, by $\wh{\xi}\equiv \wh{\xi}_{\bs{r}_N}$ the counting function relative to the reference GGE state labelled by $\bs{r}_N$.
Then, a direct calculation based on $\wh{\xi}_{ - \wh{F}_{\bs{\ell}_{N+\mf{s}}  } }=\wh{\xi}_{\bs{\ell}_{N+\mf{s}}}$ then shows the identity
\beq
\mc{D}_{N+\mf{s}} \big(\bs{\ell}_{N+\mf{s}};  \bs{r}_{N} \big)\; = \;
\msc{D}_{N+\mf{s}} \big(\bs{\ell}_{N+\mf{s}};  \bs{r}_{N} \big)\Big[ \, \wh{\xi} , \wh{F}_{\bs{\ell}_{N+\mf{s}}  } \Big]\;.
\enq
It seems reasonable to expect, and this is backed up, \textit{e.g.}, by the calculations of the ground state root distribution \textit{viz}. $r_a=a$,
that the leading large-$L$ behaviour of $\msc{D}_{N+\mf{s}}$ will only depend on the macroscopic limit of the shift function $F_{\bs{\mf{e}}}=\mf{f}_{\mf{s}}[\nu_{\bs{\mf{e}}}]$
which captures the microscopic displacement, to the leading order, between the roots $\{ \mu_{\ell_a} \}_{1}^{N+\mf{s}}$ and $\{ \la_{r_a} \}_1^{N} $.
More precisely, one should obtain equivalent results, up to $1+\e{o}(1)$ corrections,
by slightly changing the root distribution of the excited roots, namely that
\beq
\mc{D}_{N+\mf{s}} \big(\bs{\ell}_{N+\mf{s}};  \bs{r}_{N} \big)\; = \;
\msc{D}_{N+\mf{s}} \big(\bs{\ell}_{N+\mf{s}};  \bs{r}_{N} \big)\big[ \,  \wh{\xi}, \mf{f}_{\mf{s}}[\nu_{\bs{\mf{e}}}] \big]\cdot \big( 1+\e{o}(1) \big) \;.
\enq
In fact, as will follow from the computations to come, the equality will also not be affected, to the leading order in $L$ by the substitution of the
exact counting function $\wh{\xi}$ by its thermodynamic limit $\xi$. Namely, it holds
\beq
\mc{D}_{N+\mf{s}} \big(\bs{\ell}_{N+\mf{s}};  \bs{r}_{N} \big)\; = \;
\msc{D}_{N+\mf{s}} \big(\bs{\ell}_{N+\mf{s}};  \bs{r}_{N} \big)\big[ \xi, \mf{f}_{\mf{s}}[\nu_{\bs{\mf{e}}}] \big]\cdot \big( 1+\e{o}(1) \big) \;.
\enq

\section{The effective form-factor expansion}
\label{Section Effective FF expansion}

\subsection{A preliminary resummation for the generating function}
\label{SousSection Preliminary resummationGenFct}

By putting together the results of the previous sections, one arrives to the below representation for
the form-factor expansion \eqref{ecriture dvpmt FF fct generatrice} of the generating function
\beq
\wh{\mc{Q}}(x,t\mid \be) \,  = \,  \ex{\i x \ga_{0} } \hspace{-2mm}
\sul{  \bs{\ell}_N  \in \mathbb{Z}_{<}^{N}  }{}
 \f{ \pl{s \in \mc{Y}_{\bs{\mf{e}}} \oplus \mf{X}_{\bs{\mf{e}}} }{}  \hspace{-2mm} \Big\{  \xi_{ - \mf{f}_{0}[\nu_{\bs{\mf{e}}} ] }^{\prime}(s) \Big\}  }
 {  \det_{2n}\Big[ \op{D}\bs{\mf{Y}}_{\bs{\mf{e}}}(\bs{\sg})\Big] }
\pl{s \in \mc{Y}_{\bs{\mf{e}}} \ominus \mf{X}_{\bs{\mf{e}}} }{}  \hspace{-2mm}  \Big\{   \ex{\i x u (s)}  \Big\}
\cdot \msc{D}_N \big( \bs{\ell}_N;\bs{r}_{N} \big)\big[ \xi, \mf{f}_0[ \nu_{\bs{\mf{e}}}  ] \big]  \cdot \msc{A}^{(\be)}\big[\nu_{\bs{\mf{e}}} \big] \cdot
   \big( 1 + \e{o}(1) \big) \;.
\enq
At this stage, it is reasonable to assume that the remainders do not contribute to the thermodynamic limit of $\wh{\mc{Q}}(x,t\mid \be)$, namely that
$\underset{N,L\tend +\infty}{\lim}\wh{\mc{Q}}(x,t\mid \be)  \, =   \underset{N,L\tend +\infty}{\lim} \wh{\mc{Q}}_{\e{eff}}(x,t\mid \be)$, with
\beq
\wh{\mc{Q}}_{\e{eff}}(x,t\mid \be)\,  = \,  \ex{\i x \ga_{0} } \hspace{-2mm}
\sul{  \bs{\ell}_N  \in \mathbb{Z}_{<}^{N}  }{}
 \f{ \pl{s \in \mc{Y}_{\bs{\mf{e}}} \oplus \mf{X}_{\bs{\mf{e}}} }{}  \hspace{-2mm} \Big\{  \xi_{ - \mf{f}_{0}[\nu_{\bs{\mf{e}}} ] }^{\prime}(s) \Big\}  }
 {  \det_{2n}\Big[ \op{D}\bs{\mf{Y}}_{\bs{\mf{e}}}(\bs{\sg})\Big] }
\pl{s \in \mc{Y}_{\bs{\mf{e}}} \ominus \mf{X}_{\bs{\mf{e}}} }{}  \hspace{-2mm}  \Big\{   \ex{\i x u (s)}  \Big\}
\cdot \msc{D}_N \big( \bs{\ell}_N;\bs{r}_{N} \big)\big[ \xi, \mf{f}_0[ \nu_{\bs{\mf{e}}}  ] \big]  \cdot \msc{A}^{(\be)}\big[\nu_{\bs{\mf{e}}} \big] \;.
\enq
The \textit{crucial point} is that one may recast $\wh{\mc{Q}}_{\e{eff}}(x,t\mid \be)$ in a form which allows one to sum-up completely the form-factor expansion.
This may be done as follows. First, let $\mu \tend \De[\nu](\mu)$ be a function-valued functional of $\nu$. Then, it holds
\beq
\bs{:} \exp\bigg\{  \Int{ \mc{C} }{}\De[\nu](\mu) \f{\de }{  \de \nu(\mu) }  \bigg\}  \cdot \msc{V}[\nu] \bs{:}  \; = \;
\f{ \msc{V}[\nu^{\star}] }{ \det_{L^2(\mc{C}) } \Big[ \e{id} \, - \, \f{\de \De[\nu](\la)  }{  \de \nu(\mu) } \Big]_{\mid \nu=\nu^{\star} } }\;,
\label{ecriture relation evaluation exacte translation fnelle}
\enq
where $\nu^{\star}$ is the unique solution to the equation $\nu^{\star}(\la)= \De[\nu^{\star}](\la)$. In the above formula,
the $ \bs{:} * \bs{:}$ symbol means that the functional derivatives should always appear to the
left of the expression and the result should be evaluated at $\nu=0$, \textit{viz}.
\beq
\bs{:} \exp\bigg\{  \Int{ \mc{C} }{} \hspace{-1mm} \dd \mu \De[\nu](\mu) \f{\de }{  \de \nu(\mu) }  \bigg\}  \cdot \msc{V}[\nu]\bs{:} \; \equiv \;  \sul{n\geq 0}{} \f{1}{n!}
 \Int{ \mc{C}^n }{}\dd^n \mu \pl{a=1}{n} \f{\de }{  \de \nu(\mu_a) }  \cdot \bigg\{ \pl{a=1}{}\De[\nu](\mu_a)   \cdot \msc{V}[\nu] \bigg\}_{ \mid \nu=0} \;.
\enq

The proof may be found in \cite{KozReducedDensityMatrixAsymptNLSE}, and some details are also included in \cite{KozTerrasNatteSeriesNLSECurrentCurrent}.
To apply this identity to the present case, we introduce the $\mf{s}$-indexed family of functionals subordinate to a choice of integers
$\bs{\ell}_{ N + \mf{s} } \in \mathbb{Z}^{N+\mf{s}}_{<}$
\beq
\De_{\mf{s}}[\nu](z)\, = \, \sul{ \substack{s  \in \{p_a\}_1^{n+\mf{s}} \\ \ominus \{h_a\}_1^n } }{} \f{  1 }{  z \, - \,  \xi^{-1}_{-\mf{f}_{\mf{s}}[\nu]}(s/L) } \;,
\label{introduction fnelle s dpdte De s}
\enq
which acts on holomorphic functions $\nu$ defined in the interior of the contour $\mc{C} \, = \,  \Ga(\R)$. We stress that this contour is such that
$\Ga(\R) \subset \Ga^{\prime}(\R)$. The application to the generating function $\wh{\mc{Q}}_{\e{eff}}(x,t\mid \be)$ will be done by setting $\mf{s}=0$.

Now, by setting
\beq
\tau_{p_a}= \xi^{-1}_{-\mf{f}_{\mf{s}}[\nu]}\big( \tfrac{p_a}{L} \big)  \quad \e{and} \quad
\tau_{h_a}= \xi^{-1}_{-\mf{f}_{\mf{s}} [\nu]}\big( \tfrac{h_a}{L} \big) \;,
\enq
and introducing  $ {\rm Y}_{\bs{\mf{e}}}$ and $ {\rm X}_{\bs{\mf{e}}}$ analogously to \eqref{definition Y et X}:
\beq
{\rm Y}_{\bs{\mf{e}}} \, = \,   \big\{ \tau_{p_a} \big\}_{1}^{n}  \quad \e{and} \quad  {\rm X}_{\bs{\mf{e}}} \, = \,  \big\{ \tau_{h_a} \big\}_{1}^{n} \;,
\label{introduction ensembles droit def via nu fnelle}
\enq
we get that the solution $\nu_{\e{fp}}$ to the fixed point equation $\nu_{\e{fp}}(z) = \De_{\mf{s}}[\nu_{\e{fp}}](z)$ takes the form
\beq
\nu_{\e{fp}}(z) \, = \, \sul{ s \in \mc{Y}_{\bs{\mf{e}}}\ominus \mf{X}_{\bs{\mf{e}}} }{} \f{1}{z-s} \, = \, \nu_{\bs{\mf{e}}}(z)\;.
\enq
Indeed, for the fixed point of $\De_{\mf{s}}$, one readily check from the definition that  ${\rm Y}_{\bs{\mf{e}}}=  \mc{Y}_{\bs{\mf{e}}}$
and ${\rm X}_{\bs{\mf{e}}} =  \mf{X}_{\bs{\mf{e}}}$. Thus, $\nu_{\e{fp}}$ exactly coincides with $\nu_{\bs{\mf{e}}}$ as it was introduced
in \eqref{definition fonction generatrice nu s}. In other words, the function $\nu_{\e{fp}}$ is uniquely and completely fixed
upon solving the system of higher level Bethe Ansatz equations \eqref{ecriture HLBAE map} for the elements of the sets
$\mc{Y}_{\bs{\mf{e}}}$ and $\mf{X}_{\bs{\mf{e}}}$.

Now, observe that it holds
\beq
\f{ \de \mf{f}_{\mf{s}}[\nu](\la) }{ \de \nu(\mu) } \, = \, -  \f{1}{2\i\pi} \vp(\la,\mu)
\quad \e{and} \quad
\f{ \de  }{ \de \nu(\mu) } \xi_{-\mf{f}_{\mf{s}}[\nu]}^{-1}(s) \, = \,
\f{- 1 }{ \xi_{-\mf{f}_{\mf{s}}[\nu]}^{\prime}\Big( \xi_{-\mf{f}_{\mf{s}}[\nu]}^{-1}(s) \Big)  } \cdot
\bigg( \f{ \de  }{ \de \nu(\mu) } \xi_{-\mf{f}_{\mf{s}}[\nu]}  \bigg)\Big( \xi_{-\mf{f}_{\mf{s}}[\nu]}^{-1}(s) \Big) \;.
\enq
These two identities then allow one to infer that
\beq
 \f{\de \De_{\mf{s}} [\nu](\la)  }{  \de \nu(\mu) } \, = \,    \f{- 1}{2\i\pi L }\sul{  t  \in \{ \frac{p_a}{L} \}_1^{n+\mf{s} } \ominus \{ \frac{h_a}{L}\}_1^n   }{}
    \f{     \vp\Big(  \xi_{-\mf{f}_{\mf{s}} [\nu]}^{-1}(t) ,\mu \Big)   }{  \Big[ \la \, - \,  \xi^{-1}_{-\mf{f}_{\mf{s}} [\nu]}(t) \Big]^2 \cdot
    \xi_{-\mf{f}_{\mf{s}} [\nu]}^{\prime}\Big( \xi_{-\mf{f}_{\mf{s}} [\nu]}^{-1}(t) \Big)  } \;.
\enq
The integral kernel thus obtained corresponds to the one of a finite rank integral operator on $L^2\big( \Ga(\R) \big)$. As a consequence, its Fredholm determinant can be computed in terms of a
finite-size determinant. Indeed, elementary handlings on the level of the Fredholm series representation and residue calculations yield that
\beq
 \det_{L^2(\Ga(\R)) } \bigg[ \e{id} \, - \, \f{\de \De_{\mf{s}}[\nu](\la)  }{  \de \nu(\mu) } \bigg]_{\mid \nu=\nu_{\bs{\mf{e}}} }  \; = \;
 \det_{2n+\mf{s}}\left[  \ba{cc} \op{I}_{n+\mf{s}} + \op{A} & \op{B} \\  \op{C} &  \op{I}_n  +  \op{D} \ea  \right] \;.
\enq
There we have introduced the matrices
\beq
A_{ab} \, = \,    \f{ \Dp{2} \vp(\sg_{p_a}, \sg_{p_b})  }{ L \xi^{\prime}_{ -F_{\bs{\mf{e}}} }(\sg_{p_a}) } \;, \quad
B_{ab} \, = \,   - \f{ \Dp{2} \vp(\sg_{p_a}, \sg_{h_b})  }{ L \xi^{\prime}_{ -F_{\bs{\mf{e}}} }(\sg_{p_a}) }
\enq
and
\beq
C_{ab} \, = \,    \f{ \Dp{2} \vp(\sg_{h_a}, \sg_{p_b})  }{ L \xi^{\prime}_{ -F_{\bs{\mf{e}}} }(\sg_{h_a}) } \;, \quad
D_{ab} \, = \,   - \f{ \Dp{2} \vp(\sg_{h_a}, \sg_{h_b})  }{ L \xi^{\prime}_{ -F_{\bs{\mf{e}}} }(\sg_{h_a}) } \;.
\enq
In other words, one has
\beq
 \det_{L^2(\Ga(\R)) } \Big[ \e{id} \, - \, \f{\de \De_{\mf{s}}[\nu](\la)  }{  \de \nu(\mu) } \Big]_{\mid \nu=\nu_{\bs{\mf{e}}}  }
 \; = \;  \pl{s \in {\rm Y }_{\bs{\mf{e}}} \oplus {\rm X}_{\bs{\mf{e}}} }{} \bigg\{  \f{1}{\xi_{ - \mf{f}_{\mf{s}}[\nu_{\bs{\mf{e}}}] }^{\prime}(s)} \bigg\}
 \det_{ 2n + \mf{s} }\Big[ \op{D}\bs{\mf{Y}}_{\bs{\mf{e}}}(\bs{\sg})\Big]  \;.
\enq
Thus, applying the above identities to the $\mf{s}=0$ case, all-in-all, we arrive to the below representation
\beq
\wh{\mc{Q}}_{\e{eff}}(x,t\mid \be)\,  = \,  \bs{:}   \ex{\i x \ga_{0} } \hspace{-2mm}
\sul{  \bs{\ell}_N  \in \mathbb{Z}_{<}^{N}  }{}
  \pl{s \in {\rm Y}_{\bs{\mf{e}}}  \ominus {\rm X}_{\bs{\mf{e}}} }{} \op{E}_-^{-2}(s)  \; \cdot \;
\msc{D}_N \big(  \bs{\ell}_{N}  ; \bs{r}_N \big)\big[ \xi, \mf{f}_0[ \nu ] \big]  \cdot \msc{A}^{(\be)}[\nu] \bs{:}  \;,
\label{ecriture rho N eff comme premiere moyenne fonctionnelle}
\enq
where we have set
\beq
\op{E}_-(s) \, = \, \ex{ - \i \frac{x}{2} u(s) - \frac{1}{2} \mf{g}(s) }     \qquad \e{with} \qquad
\mf{g}(s) \, = \hspace{-4mm} \Int{ \Ga(\R) \subset \Ga^{\prime}(\R) }{} \hspace{-4mm}  \dd z  \f{1}{z-s} \f{\de  }{ \de \nu(z) } \;.
\label{definition Fct Eminus and mf g}
\enq
We stress again that now, the roots $\tau_a$ and $\eta_a$ building up the sets ${\rm Y }_{\bs{\mf{e}}}$ and ${\rm X}_{\bs{\mf{e}}}$,
see \eqref{introduction ensembles droit def via nu fnelle} and the singular part of the form factors correspond to solutions of
\beq
\xi(\eta_a) \, = \, \f{a}{L} \quad \e{and} \quad \xi_{ - \mf{f}_0[\nu] }(\tau_a) \, = \, \f{a}{L}\;.
\enq
It is primordial to stress that, due to the use of the $\bs{:} * \bs{:}$ ordering,
the expression in \eqref{ecriture rho N eff comme premiere moyenne fonctionnelle} and those to come
is to be considered as being built in terms of \textit{classical} commutative algebra-valued functions. See \cite{KozReducedDensityMatrixAsymptNLSE} on why it is so.
This perspective opens up outstanding possibilities for rewriting such sums!

Introduce the functional
\beq
\msc{X}_N^{(\mf{s})}\Big[\xi, \nu, \op{E}_-^{-2} \Big] \; = \hspace{-3mm} \sul{ \bs{\ell}_{N+\mf{s}}  \in \mathbb{Z}_{<}^{N+\mf{s}}  }{}
    \f{ \pl{a=1}{N + \mf{s} } \op{E}_-^{-2}(\tau_{\ell_a})  }{  \pl{a=1}{N} \op{E}_-^{-2}(\eta_{r_a})  }  \; \cdot \;
\msc{D}_N \big(  \bs{\ell}_{N+\mf{s}} ;  \bs{r}_{N} \big)\big[ \xi, \nu  \big]\;,
\label{definition fnelle resommation Xs}
\enq
where it is understood that the $\eta_a$ roots are defined using $\xi$ and the $\tau_a$ roots by using $\xi_{-\nu}$.
Then, one has the below representation
\beq
\wh{\mc{Q}}_{\e{eff}}(x,t\mid \be)\,  = \,   \bs{:} \ex{\i x \ga_{0} }  \pl{a=1}{N} \f{ \op{E}_-^{-2}(\eta_{r_a}) }{  \op{E}_-^{-2}(\tau_{r_a}) }
\cdot \msc{X}_N^{(0)}\Big[\xi,\mf{f}_0[\nu], \op{E}_-^{-2} \Big] \cdot \msc{A}^{(\be)}[\nu] \bs{:} \;.
\enq
For the sake of well-definiteness we do stress that, above, the variables $\tau_a$, are defined as unique solutions to
$\xi_{-\mf{f}_0[\nu]}(\tau_a) \, = \, \f{a}{L}$.

\subsection{A preliminary resummation for the reduced density matrix}
\label{SousSectionPreliminaryResummationRedDenMtrx}
The steps leading to a similar rewriting of the reduced density matrix are quite similar.
We first assume that the remainders do not contribute to the thermodynamic limit of $\wh{\rho}(x,t\mid \be)$, namely that
$\underset{N,L\tend +\infty}{\lim} \wh{\rho}(x,t\mid \be)  \, =   \underset{N,L\tend +\infty}{\lim} \wh{\rho}_{\e{eff}}(x,t\mid \be)$, with
\beq
\wh{\rho}_{\e{eff}}(x,t\mid \be)  \, = \,  \ex{\i x \ga_{1} } \hspace{-4mm}
\sul{ \bs{\ell}_{N+1} \in \mathbb{Z}^{N+1}_{<}  }{}
  \f{ \pl{s \in \mc{Y}_{\bs{\mf{e}}} \oplus \mf{X}_{\bs{\mf{e}}} }{} \Big\{  \xi_{ - \mf{f}_1[\nu_{\bs{\mf{e}}}] }^{\prime}(s) \Big\}  }
{  \det_{2n+1}\Big[ \op{D}\bs{\mf{Y}}_{\bs{\mf{e}}}(\bs{\sg})\Big] }
\pl{s \in \mc{Y}_{\bs{\mf{e}}} \ominus \mf{X}_{\bs{\mf{e}}} }{} \hspace{-2mm} \Big\{   \ex{\i x u (s)}  \Big\} \,
\msc{D}_N \big( \bs{\ell}_{N+1} ; \bs{r}_N \big)\big[ \xi, \mf{f}_{1}[ \nu_{\bs{\mf{e}}} ] \big]  \cdot \msc{A}^{(\be;\Phi)}[\nu_{\bs{\mf{e}}}] \;.
\enq
We now follow the previously outlined steps to recast $\wh{\rho}_{\e{eff}}(x,t)$ in a form which allows one to sum-up completely the form-factor expansion.
In the present case, we deal with the functional $\De_{1}[\nu]$ as introduced for general $\mf{s}$ in \eqref{introduction fnelle s dpdte De s}.
This generates the below representation
\beq
\wh{\rho}_{\e{eff}}(x,t\mid \be)\, = \,  \bs{:} \ex{\i x \ga_{1} } \hspace{-4mm}
\sul{\bs{\ell}_{N+1} \in \mathbb{Z}^{N+1}_{<}  }{}
  \pl{s \in {\rm Y}_{\bs{\mf{e}}} \ominus {\rm X}_{\bs{\mf{e}}} }{} \op{E}_-^{-2}(s)  \; \cdot \;
\msc{D}_N \Big(\{\ell_a \}_1^{N+1} ; \{ r_a \}_1^{N} \Big)\big[ \xi, \mf{f}_1[ \nu ] \big]  \cdot \msc{A}^{(\be;\Phi)}[\nu] \bs{:} \;,
%
\enq
where $\op{E}_-$ is as defined in \eqref{definition Fct Eminus and mf g}.

We stress again that now, the roots $\tau_a$ and $\eta_a$ building up the sets ${\rm Y}_{\bs{\mf{e}}}, {\rm X}_{\bs{\mf{e}}}$
and the singular part of the form factors correspond to solutions of
\beq
\xi(\eta_a) \, = \, \f{a}{L} \quad \e{and} \quad \xi_{ - \mf{f}_1[\nu] }(\tau_a) \, = \, \f{a}{L}\;.
\enq
Thus, by using the functional $\msc{X}_N^{(1)}$ as introduced in \eqref{definition fnelle resommation Xs}, one gets the below representation
\beq
\wh{\rho}_{\e{eff}}(x,t\mid \be)\, = \,  \bs{:} \ex{\i x \ga_{\op{v}} }  \pl{a=1}{N} \f{ \op{E}_-^{-2}(\eta_{r_a}) }{  \op{E}_-^{-2}(\tau_{r_a}) }
\cdot \msc{X}_N^{(1)}\Big[\xi,\mf{f}_1[\nu], \op{E}_-^{-2} \Big] \cdot \msc{A}^{(\be;\Phi)}[\nu] \bs{:} \;.
\enq
We do stress that, above, the variables $\tau_a$, are defined as unique solutions to $\xi_{-\mf{f}_1[\nu]}(\tau_a) \, = \, \f{a}{L}$.

\subsection{Thermodynamic limit of $\msc{X}_N^{(\mf{s})}$}
\label{SousSectionThermoLimXN}

One may re-express $\msc{X}_N^{(0)}$ in terms of a Fredholm determinant of an integrable integral operator
and, more generally, $\msc{X}_N^{(\mf{s})}$ -$\msc{X}_N^{(1)}$ in particular- in terms of a Fredholm minor.
We do not describe the details here and rather refer to Appendix B of \cite{KozTerrasNatteSeriesNLSECurrentCurrent}
for the details of the re-summation of $\msc{X}_N^{(0)}$ and Appendix B of \cite{KozReducedDensityMatrixAsymptNLSE}
for those pertaining to the details of the re-summation for $\msc{X}_N^{(1)}$.
The observation that the multidimensional series defining $\msc{X}_N^{(\mf{s})}$
does indeed lead to a Fredholm minor has been first made in \cite{KorepinSlavnovTimeDepCorrImpBoseGas}, on the example of the impenetrable Bose gas.
To cut the long story short, one gets that, for $L$ finite,
\beq
\msc{X}_N^{(0)}\Big[\xi,\nu, \op{E}_-^{-2} \Big] \; = \;  \det_N\bigg[ \de_{jk} \, + \, \f{    V^{(L)}(\eta_{r_j},\eta_{r_k}) }{ L \varrho(\eta_{r_k}) }  \bigg]
\enq
while
\beq
\msc{X}_N^{(1)}\Big[\xi,\nu, \op{E}_-^{-2} \Big] \; = \; \bigg(\mc{S}^{(L)}_0\big[ \op{E}_-^{-2} \big] \, + \; \Dp{\a}\bigg)_{\mid \a=0}
\det_N\bigg[ \de_{jk} \, + \, \f{  \big( V^{(L)}+ \a P^{(L)} \big)(\eta_{r_j},\eta_{r_k}) }{ L \varrho(\eta_{r_k}) }  \bigg] \;,
\enq
where, for some $\de>0$ and small enough, we have set 
\beq
\mc{S}^{(L)}_0\big[ \op{E}_-^{-2} \big] \, = \, \Int{ \R }{}        \f{\dd \mu }{2\pi}  \op{E}_-^{-2} (\mu) \; + \; \mc{I}^{(L)}_0\big[ \op{E}_-^{-2} \big]  \quad \e{with} \quad
\mc{I}^{(L)}_r\big[ \op{E}_-^{-2} \big] (\la) \; = \; \sul{\eps=\pm}{} \Int{ \R + \i\eps \de }{}{}    \f{\dd \mu }{2\pi}  \cdot \f{ \op{E}_-^{-2} (\mu) }{ \big( \ex{-2\i\pi L \eps \xi_{-\nu}(\mu) } -1 \big) (\mu-\la)^r } \;. 
\enq
Further, one has 
\beqa
 V^{(L)}(\la,\mu) & = & 4 \vth(\mu) \sin[\pi \nu(\la)]\sin[\pi \nu(\mu)]\op{E}_-(\la)\op{E}_-(\mu) \f{  \op{O}_1^{(L)}\big[\nu, \op{E}_-^{-2} \big](\la) \, - \,  \op{O}_1^{(L)}\big[\nu, \op{E}_-^{-2} \big](\mu) }{ 2\i\pi (\la-\mu) } \;, \\
 P^{(L)}(\la,\mu) & = & \f{2}{\pi} \vth(\mu) \sin[\pi \nu(\la)]\sin[\pi \nu(\mu)]\op{E}_-(\la)\op{E}_-(\mu)  \op{O}_1^{(L)}\big[\nu, \op{E}_-^{-2} \big](\la) \op{O}_1^{(L)}\big[\nu, \op{E}_-^{-2} \big](\mu)    \;.
\eeqa
The expression for these kernels involves the auxiliary functional  
\beq
\op{O}_1^{(L)}\big[\nu, \op{E}_-^{-2} \big](\la) \, = \, \i \Int{ \R }{}        \f{\dd \mu }{2\pi}  \f{ \op{E}_-^{-2} (\mu) }{\mu-\la} \; + \;  \f{ \op{E}_-^{-2} (\la) }{ \ex{-2\i\pi \nu(\la)} -1 }
\; + \; \i  \mc{I}^{(L)}_1\big[ \op{E}_-^{-2} \big] (\la) \;. 
\enq
One may readily then take the thermodynamic limit which gives 
\beq
\msc{X}_N^{(0)}\Big[\xi,\nu, \op{E}_-^{-2} \Big] \; \limit{N,L}{+\infty} \; \msc{X}^{(0)}\Big[ \nu, \op{E}_-^{-2} \Big] \; = \;
\det_{ L^2(\R) }\Big[ \e{id} \, + \,  \op{V}   \Big]
\enq
and
\beq
\msc{X}_N^{(1)}\Big[\xi,\nu, \op{E}_-^{-2} \Big] \; \limit{N,L}{+\infty} \; \msc{X}^{(1)}\Big[ \nu, \op{E}_-^{-2} \Big] \; = \;
\bigg(\mc{S}_0\big[ \op{E}_-^{-2} \big] \, + \; \Dp{\a}\bigg)_{\mid \a=0}\det_{ L^2(\R) }\Big[ \e{id} \, + \,  \op{V}+ \a \op{P}  \Big] \;.
\enq
Above, one has
\beq
\mc{S}_0\big[ \op{E}_-^{-2} \big] \, = \, \Int{ \R }{}        \f{\dd \mu }{2\pi}  \op{E}_-^{-2} (\mu)   \;.
\enq
Also, one has 
\beqa
 V(\la,\mu) & = & 4 \vth(\mu) \sin[\pi \nu(\la)]\sin[\pi \nu(\mu)]\op{E}_-(\la)\op{E}_-(\mu) \f{  \op{O}_1\big[\nu, \op{E}_-^{-2} \big](\la) \, - \,  \op{O}_1\big[\nu, \op{E}_-^{-2} \big](\mu) }{ 2\i\pi (\la-\mu) } \;, \\
 P(\la,\mu) & = & \f{2}{\pi} \vth(\mu) \sin[\pi \nu(\la)]\sin[\pi \nu(\mu)]\op{E}_-(\la)\op{E}_-(\mu)  \op{O}_1\big[\nu, \op{E}_-^{-2} \big](\la) \op{O}_1\big[\nu, \op{E}_-^{-2} \big](\mu)    \;.
\eeqa
The expression for these kernels involves the auxiliary functional
\beq
\op{O}_1\big[\nu, \op{E}_-^{-2} \big](\la) \, = \,  \i \Int{ \R }{}        \f{\dd \mu }{2\pi}  \f{ \op{E}_-^{-2} (\mu) }{\mu-\la} \; + \;  \f{ \op{E}_-^{-2} (\la) }{ \ex{-2\i\pi \nu(\la)} -1 } \;.
\enq

\subsection{Thermodynamic limit of the correlators}
\label{SousSectionThermoLimCorrelators}
In order to take the thermodynamic limit it remains to observe that
\beq
 \pl{a=1}{N} \f{ \op{E}_-^{-2}(\eta_{r_a}) }{  \op{E}_-^{-2}(\tau_{r_a}) }  \, = \,
 \exp\Big\{ -\Int{ \R }{} \dd t \vth(t) \mf{f}_{\mf{s}}[\nu](t) \big[ \i  x u^{\prime}(t) + \mf{g}^{\prime}(t)\big]  \Big\}\cdot \big( 1+\e{o}(1) \big) \;,
\enq
where the value of $\mf{s}$ should be set to $0$ or $1$ depending on the situation of interest.

Thus, as backed up by several tests in  \cite{KozReducedDensityMatrixAsymptNLSE,KozTerrasNatteSeriesNLSECurrentCurrent}, we get that
$\wh{\mc{Q}}_{\e{eff}}(x,t\mid \be) \limit{N,L}{+\infty} \mc{Q}(x,t\mid \be)$, where
\beq
\mc{Q}(x,t\mid \be)\, = \,  \bs{:} \ex{\i x \ga_{0} }
\exp\bigg\{ -\Int{\R}{} \hspace{-2mm} \dd \la \vth(\la) \mf{f}_{0}[\nu](\la) \big[ \i x u^{\prime}(\la) + \mf{g}^{\prime}(\la) \big]  \bigg\}
\cdot \msc{X}^{(0)}\Big[ \mf{f}_{0}[\nu], \op{E}_-^{-2} \Big] \cdot \msc{A}^{(\be)}[\nu] \bs{:} \;.
\label{ecriture limite thermo de fct gen}
\enq
Similarly, $\wh{\rho}_{\e{eff}}(x,t\mid \be) \limit{N,L}{+\infty} \rho(x,t\mid \be)$, where
\beq
\rho(x,t\mid \be)\, = \,  \bs{:} \ex{\i x \ga_{1} }
\exp\bigg\{ -\Int{ \R }{} \hspace{-2mm} \dd \la \vth(\la) \mf{f}_{1}[\nu](\la) \big[ \i x u^{\prime}(\la) + \mf{g}^{\prime}(\la)\big]  \bigg\}
\cdot \msc{X}^{(1)}\Big[ \mf{f}_{1}[\nu], \op{E}_-^{-2} \Big] \cdot \msc{A}^{(\be;\Phi)}[\nu] \bs{:} \;.
\label{ecriture limite thermo de rho eff}
\enq

\section{An alternative representation for the correlators}
\label{SectionAlternativeRepCorrFcts}

\subsection{An alternative representation for the Fredholm determinant and minor}
\label{SousSectionAlternRepFDetAndMinor}

The work \cite{KozGohmannMininTimeAndSpaceDpdtThermalGSK}
studied the large $x,t \tend + \infty$ with $\op{v}=x/t$ kept fixed asymptotic expansion for the Fredholm determinant
$\msc{X}^{(0)}\Big[\xi,\nu, \op{E}_-^{-2} \Big]$. The asymptotic expansion was depending on a parameter $\la_0$
defined as the (assumably unique) solution to the saddle point equation
\beq
u^{\prime}(\la_0) \, = \, 0 \qquad \e{i.e.}, \qquad  \op{v} \, = \, \f{ \veps^{\prime}(\la_0) }{ p^{\prime}(\la_0) } \;.
\enq
The form of the leading asymptotics was parameterised by several auxiliary functions. First of all,
\beq
\mc{L}_{\ell}\big[ \nu \big](\la) \, = \, \f{-1}{2\i\pi} \ln \Big[ 1 + \vth(\la) \big( \ex{2\i\pi \nu(\la) } -1 \big) \Big]
\quad \e{and} \quad
\mc{L}_{r}\big[ \nu \big](\la) \, = \, \f{ 1}{2\i\pi} \ln \Big[ 1 + \vth(\la) \big( \ex{-2\i\pi \nu(\la) } -1 \big) \Big] \;.
\enq
Further, one denotes
\beq
\tau[\nu](\la) \, = \, \mc{L}_{\ell}\big[ \nu \big](\la) - \mc{L}_{r}\big[ \nu \big](\la)
\label{definition de tau}
\enq
as well as, for $k \in \mathbb{N}$,
\beq
\mc{L}^{(k)}[\nu](\la\mid \la_0) \, = \,
\mc{L}_{\ell}^{(k)}\big[ \nu \big](\la) \bs{1}_{\R^-}(\la-\la_0) \,  + \,  \mc{L}_{r}^{(k)}\big[ \nu \big](\la) \bs{1}_{\R^+}(\la-\la_0)  \;.
\enq
Then, it holds
\beq
\msc{X}^{(0)}\big[ \nu, \op{E}_-^{-2} \big] \, = \,
\f{  \mc{B}\big[\mc{L}_{\ell}[\nu], \mc{L}_{r}[\nu] , \nu \big](\la_0) }{  x^{ \f{ 1 }{ 2 } \tau^2[\nu](\la_0)  } }
\exp\bigg\{ -\Int{\R}{} \hspace{-1mm} \dd \la \mc{L}[\nu](\la\mid \la_0)  \big( \i x u^{\prime}(\la) + \mf{g}^{\prime} (\la) \big)  \bigg\} \cdot
\Big(  1 + \msc{R}_{x}^{(0)}[\nu, \mf{g}] \Big) \;.
\label{ecriture developpement type Natte pour F det}
\enq
The constant term functional takes the explicit form
\bem
\mc{B}\big[\mc{L}_{\ell}, \mc{L}_{r}, \nu \big](\la_0)  \, = \,
\f{ G\big( 1 + \tau(\la_0) \big) \, \big\{ \varkappa[\mc{L}](\la_0) \big\}^{ 2 \tau(\la_0) } }
{  (2\pi)^{ \f{1}{2} \tau(\la_0) } \big[\i u^{\prime \prime}(\la_0) \big]^{ \f{1}{2} \tau^2(\la_0) } }
\Big\{ 4 \vth(\la_0)    \sin^2[\pi \nu(\la_0)] \Big\}^{\tau(\la_0)}  \\
\times \exp\Bigg\{ - \Int{\R}{} \hspace{-1mm}\dd \la \mc{L}^{\prime}(\la\mid \la_0)    \ln \Big\{ 4 \vth(\la) \sin^2[\pi \nu(\la)] \Big\}
\, - \, \Int{\R^2}{} \hspace{-1mm} \dd \la \dd \mu
  \mc{L}^{\prime}(\la\mid \la_0) \mc{L}^{\prime}(\mu\mid \la_0)  \ln |  \la -\mu |  \Bigg\}
\label{definition fnelle B lambda0 DA Fred Det}
\end{multline}
There, we agree $\tau\, = \, \mc{L}_{\ell}  \, - \,   \mc{L}_{r}$ and $\mc{L}^{(k)}(\la\mid \la_0)  =
\mc{L}_{\ell}^{(k)}(\la) \bs{1}_{\R^-}(\la-\la_0) \,  + \,  \mc{L}_{r}^{(k)}(\la) \bs{1}_{\R^+}(\la-\la_0)$ with $k \in \mathbb{N}$.
Furthermore,
\beq
\varkappa[\mc{L}](\la_0) \, = \, \exp\Bigg\{ - \Int{\R}{} \hspace{-1mm} \dd \la \mc{L}^{\prime}[\nu](\la\mid \la_0) \ln|\la-\la_0| \Bigg\}  \;.
\enq
Finally, $\msc{R}_{x}^{(0)}[\nu, \mf{g}]$ is a remainder term that is estimated as
\beq
\big| \Dp{x}^{k}\msc{R}_{x}^{(0)} [\nu, \mf{g}] \big| \, \leq  \,  C   \f{ \ln^2 x }{ x^{k+1} }
\label{estimation reste dans det Fredholm}
\enq
for some $C>0$ and $k \in \mathbb{N}$.
Since $\msc{X}^{(0)}\big[ \nu, \op{E}_-^{-2} \big]_{\mid \nu=0} =1$,  one infers that the remainder vanishes in the $\nu \tend 0$ limit.
We shall assume later on, that the vanishing takes place at least at a linear in $\nu$ rate.

Very similarly, it holds
\beq
\msc{X}^{(1)}\big[ \nu, \op{E}_-^{-2} \big] \, = \,
\f{  \mc{B}\big[\mc{L}_{\ell}[\nu]-1, \mc{L}_{r}[\nu] , \nu \big](\la_0) }{  x^{ \f{ 1 }{ 2 } (\tau[\nu](\la_0)-1)^2  } \cdot \op{E}_-^{2} (\la_0) }
\exp\bigg\{ -\Int{\R}{} \hspace{-1mm} \dd \la \mc{L}[\nu](\la\mid \la_0)  \big( \i x u^{\prime}(\la) + \mf{g}^{\prime} (\la) \big)  \bigg\} \cdot
\Big(  1 + \msc{R}_{x}^{(1)}[\nu, \mf{g}] \Big)
\label{ecriture developpement type Natte pour F minor}
\enq
with $\msc{R}_{x}^{(1)}$ enjoying the same bounds and properties as discussed around \eqref{estimation reste dans det Fredholm}.

\subsection{A new formula for the correlators}
\label{SousSectionAsymptRepForCorrFcts}

\subsubsection{The case of the generating function}

By inserting the expansion \eqref{ecriture developpement type Natte pour F det} for $\msc{X}^{(0)}\big[ \nu, \op{E}_-^{-2} \big] $
into the representation \eqref{ecriture limite thermo de fct gen} for $\mc{Q}(x,t\mid \be)$, one gets that
\beq
\mc{Q}(x,t\mid \be)\, = \,  \bs{:} \exp\bigg\{ \Int{\R}{} \hspace{-2mm} \dd \la  \, \chi\big[ \mf{f}_0[\nu] \big](\la) \mf{g}^{\prime}(\la)  \bigg\}
\cdot \msc{H}_{0}[\nu](x,t) \, \Big(  1 +  \msc{R}_{x}^{(0)}[\nu, \mf{g}] \Big)  \bs{:} \;,
\label{exriture DA Q sous forme moyenne dot dot}
\enq
where
\beq
 \chi[\nu](\la) \, = \, - \mc{L}[\nu](\la\mid \la_0) \, - \, \nu(\la) \vth(\la) \;,
\label{definition de chi}
\enq
while
\beq
\msc{H}_{0}[\nu](x,t)  \, = \, \ex{\i x \ga_{0} }
\exp\bigg\{ \i x \Int{\R}{} \hspace{-2mm} \dd \la  \, \chi\big[ \mf{f}_0[\nu] \big](\la)  u^{\prime}(\la)  \bigg\}
 \f{  \mc{B}\Big[\mc{L}_{\ell}\big[ \mf{f}_0[\nu] \big], \mc{L}_{r}\big[ \mf{f}_0[\nu] \big] , \mf{f}_0[\nu] \Big](\la_0)  \, \msc{A}^{(\be)}[\nu] }
 {  x^{  \f{1}{2}\tau^2[\mf{f}_0[\nu]](\la_0)  }  } .
\label{definition facteur H}
\enq
There,  $\tau[\nu]$ is as introduced in \eqref{definition de tau}. Moreover, it is important to stress that $\msc{H}_{0}[\nu](x,t)$ introduced above
\textit{does not} depend on $\mf{g}$ at all. Next one observes that
\beq
\Int{\R}{} \hspace{-2mm} \dd \la  \, \chi\big[ \mf{f}_0[\nu] \big](\la) \mf{g}^{\prime}(\la)  \, = \,
\Int{ \Ga(\R) }{} \hspace{-1mm} \dd z \Ups_0[\nu](z) \f{ \de }{ \de \nu(z)}
\quad \e{where} \quad
\Ups_0[\nu](z) \, = \, \Int{ \R }{} \dd \la \f{ \chi\big[ \mf{f}_0[\nu] \big](\la) }{ (z-\la)^2  } \;.
\enq
Thus, at least in what concerns the leading term, in order to compute the effect of the $\bs{:} * \bs{:}$ average one needs to solve the non-linear
integral equation on $L^2\big( \Ga(\R) \big)$ for the fixed point $\nu^{\star}_0(z) \, = \, \Ups_0[\nu^{\star}_0](z)$. Indeed, by
\eqref{ecriture relation evaluation exacte translation fnelle}, one has
\beq
  \bs{:} \exp\bigg\{ \Int{\R}{} \hspace{-2mm} \dd \la  \, \chi\big[ \mf{f}_0[\nu] \big](\la) \mf{g}^{\prime}(\la)  \bigg\}
\cdot \msc{H}_{0}[\nu](x,t)   \bs{:}
\, = \,   \f{ \msc{H}_{0}[\nu^{\star}_0](x,t)  }{ \underset{ L^2\big( \Ga(\R) \big) }{\det} \bigg[\e{id}
\, - \, \f{ \de \Ups_{0}[\nu](z) }{ \de \nu(y) }  \bigg]_{\mid \nu = \nu^{\star}_0 } }
\enq

The situation pertaining to the computation of the average relative to the remainder contribution is more complicated.
Indeed, the remainder $ \msc{R}_{x}^{(0)}[\nu, \mf{g}]$
in \eqref{exriture DA Q sous forme moyenne dot dot} does depend on $\mf{g}$ and involves order by order in $\tf{1}{x}$
higher and higher polynomials in derivatives of $\mf{g}$. All these can be treated within the functional translation approach,
see \cite{KozReducedDensityMatrixAsymptNLSE}. However, the explicit averaging of the remainder terms would demand to provide the explicit form
of the remainder term $\msc{R}_{x}^{(0)}[\nu, \mf{g}]$. While this can easily be achieved within the formalism of
\cite{KozGohmannMininTimeAndSpaceDpdtThermalGSK}, it leads to rather bulky formulae and long calculations.
However, as discussed at length in \cite{KozReducedDensityMatrixAsymptNLSE}, since  $\msc{R}_{x}^{(0)}[\nu, \mf{g}]$ is well-tailored
in respect to computing the $x,t\tend + \infty$, $\tf{x}{t}=\op{v}$ fixed, asymptotics of the Fredholm determinant,
this property will be preserved on the level of the averaging procedure. As a result, the remainder term $\msc{R}_{x}^{(0)}[\nu, \mf{g}]$
present in the representation for the original Fredholm determinant will give rise to a remainder term $\msc{M}_0(x)$

All-in-all, the above discussions lead us to
\beq
\mc{Q}(x,t\mid \be)\, = \,   \f{ \msc{H}_{0}[\nu^{\star}_0](x,t)  }{ \underset{ L^2\big( \Ga(\R) \big) }{\det} \bigg[\e{id}
\, - \, \f{ \de \Ups_{0}[\nu](z) }{ \de \nu(y) }  \bigg]_{\mid \nu = \nu^{\star}_0 } }   \, \Big(  1 +  \msc{M}_0(x) \Big) \, ,
\label{ecriture gen fct apres moyenne}
\enq
where, for some $C>0$,
\beq
 \Big| \Dp{x}^k\msc{M}_0(x) \Big|  \, \leq  \,  C \f{ \ln^2 x }{ x^{k+1} } \;.
\label{ecriture propriete reste}
\enq

\subsubsection{The case of the reduced density matrix}

By inserting the expansion \eqref{ecriture developpement type Natte pour F minor} for $\msc{X}^{(1)}\big[ \nu, \op{E}_-^{-2} \big] $
into the representation \eqref{ecriture limite thermo de rho eff} for $\rho(x,t\mid \be)$, one gets that
\beq
\rho(x,t\mid \be)\, = \,  \bs{:}
\exp\bigg\{\mf{g}(\la_0)\, + \,  \Int{\R}{} \hspace{-2mm} \dd \la  \, \chi\big[ \mf{f}_1[\nu] \big](\la) \mf{g}^{\prime}(\la)  \bigg\}
\cdot \msc{H}_{1}[\nu](x,t) \;  \Big(  1 +  \msc{R}_{x}^{(1)}[\nu, \mf{g}] \Big)   \bs{:}
\label{exriture DA rho sous forme moyenne dot dot}
\enq
with $\chi[\nu]$ as defined in \eqref{definition de chi},  $\tau[\nu]$ as in \eqref{definition de tau}, and
\bem
\msc{H}_{1}[\nu](x,t)  \, = \, \ex{\i x \ga_{1} }
\exp\bigg\{ \i x u(\la_0) \, + \, \i x \Int{\R}{} \hspace{-2mm} \dd \la  \, \chi\big[ \mf{f}_1[\nu] \big](\la)  u^{\prime}(\la)  \bigg\} \\
\times
\f{ \mc{B}\Big[\mc{L}_{\ell}\big[ \mf{f}_1[\nu] \big]-1, \mc{L}_{r}\big[ \mf{f}_1[\nu] \big] , \mf{f}_1[\nu] \Big](\la_0)
\,  \msc{A}^{(\be;\Phi)}[\nu]  }
{  x^{  \f{1}{2} \big( \tau[ \mf{f}_1[\nu] ](\la_0) -1\big)^2  }  }
\,.
\end{multline}
As in the case of the generating function, the remainder in \eqref{exriture DA rho sous forme moyenne dot dot}
does depend on $\mf{g}$ and involves order by order in $\tf{1}{x}$ higher and higher polynomials in derivatives of $\mf{g}$.
Its averaging procedure will lead to the same outcome as in the generating function case.

In the present setting, one has
\beq
\mf{g}(\la_0) \, +   \Int{\R}{} \hspace{-1mm} \dd \la  \, \chi\big[ \mf{f}_1[\nu] \big](\la) \mf{g}^{\prime}(\la)  \, = \,
\Int{ \Ga(\R) }{}\hspace{-2mm} \dd z \Ups_1[\nu](z) \f{ \de }{ \de \nu(z)} \, ,
\quad \e{where} \quad
\Ups_1[\nu](z) \, = \, \f{1}{z-\la_0} \,  +   \Int{ \R }{}\hspace{-1mm}  \dd \la \f{ \chi\big[ \mf{f}_1[\nu] \big](\la) }{ (z-\la)^2  } \;.
\label{definition fnelle Ups1}
\enq
By \eqref{ecriture relation evaluation exacte translation fnelle}, the $\bs{:} \; \bs{:}$ average of the leading term leads
to solve  the fixed point non-linear integral equation
$\nu^{\star}_1(z) \, = \, \Ups_1[\nu^{\star}_1](z)$ with $z \in  \Ga(\R)$ and yields
\beq
\rho(x,t\mid \be) \, = \,   \f{ \msc{H}_{1}[\nu^{\star}_1](x,t)  }{ \underset{ L^2\big( \Ga(\R) \big) }{\det} \Big[\e{id}
\, - \, \f{ \de \Ups_{1}[\nu](z) }{ \de \nu(y) }  \Big]_{\mid \nu = \nu^{\star}_1 } }   \,\Big(  1 +  \msc{M}_1(x) \Big)
\label{expression reduced density matrix apres moyenne}
\enq
with $\msc{M}_1(x)$ enjoying the same bounds as $\msc{M}_0(x)$, \textit{c.f.} \eqref{ecriture propriete reste}.

\subsubsection{Normalising determinants under one roof}

We now recast the
Fredholm determinant arising in the denominators of \eqref{ecriture gen fct apres moyenne} and \eqref{expression reduced density matrix apres moyenne}
in a more convenient form. It follows from \eqref{definition fonctionnelle mf f} that
\beq
\f{ \de \mf{f}_{\mf{s}}[\nu](\la)  }{ \de \nu(\mu) } \, = \, - \f{ \vp(\la,\mu)  }{ 2\i\pi } \;.
\enq
Thus,
\beq
\f{ \de   \mc{L}\big[ \mf{f}_{\mf{s}}[\nu] \big] (\la \mid \la_0)   }{ \de \nu(\mu) }\, = \, - \f{ \vp(\la,\mu)  }{ 2\i\pi } \cdot
\Big\{  \bs{1}_{\R^-}(\la-\la_0)  \Dp{\nu(\la)} \mc{L}_{\ell}[\nu](\la)    \, + \,
           \bs{1}_{\R^+}(\la-\la_0)     \Dp{\nu(\la)} \mc{L}_{r}[\nu](\la)         \Big\}_{ \nu \hookrightarrow  \mf{f}_{\mf{s}}[\nu] }  \;.
\enq
For $y,z \in \Ga(\R)$, it holds
\beq
\f{ \de \Ups_{\mf{s}}[\nu](z) }{ \de \nu(y) }   \; = \; \Int{\R}{} \hspace{-1mm} \dd \la \f{  V_{\mf{s}}(\la,y) }{ 2\i\pi (z-\la)^2 }
\enq
with
\beq
V_{\mf{s}}(\la,y)  \, = \,   \vp(\la,y)  \,  \Big\{ \vth(\la) \, + \,   \bs{1}_{\R^-}(\la-\la_0)  \Dp{\nu(\la)} \mc{L}_{\ell}[\nu](\la)   \, + \,
           \bs{1}_{\R^+}(\la-\la_0)     \Dp{\nu(\la)} \mc{L}_{r}[\nu](\la)             \Big\}_{ \nu \hookrightarrow  \mf{f}_{\mf{s}}[\nu] } \;.
\enq
For $t$ small enough, it holds
\beqa
\ln \underset{ L^2( \Ga(\R) ) }{\det} \bigg[\e{id} \, - \, t \f{ \de \Ups_{ \mf{s} }[\nu](z) }{ \de \nu(y) }  \bigg]_{\mid \nu = \nu^{\star}_{\mf{s}} } & =  &
- \sul{n \geq 1}{} \f{ t^n }{ n } \Int{  \Ga(\R)^n   }{} \hspace{-2mm} \f{\dd ^n z}{ (2\i\pi)^n } \Int{ \R^n }{} \hspace{-1mm} \dd^n \la
\pl{a=1}{n} \bigg\{ \f{ V_{\mf{s}}(\la_a,z_{a+1}) }{ (z_a-\la_{a})^2 }\bigg\}  \\
&  =  & - \sul{n \geq 1}{} \f{ t^n }{ n }  \Int{ \R^n }{} \hspace{-1mm} \dd^n \la
\pl{a=1}{n} \Dp{2}V_{\mf{s}}(\la_a,\la_{a+1})  \; = \;
\ln \underset{ L^2( \R ) }{\det} \Big[\e{id} \, - \, t \op{W}_{\mf{s}}  \Big]
\label{ecriture reecriture Jacobien fnel vers determinant Ws}
\eeqa
in which $\op{W}_{\mf{s}} $ acts on $L^2\big( \R \big)$ with the integral kernel
\beq
W_{\mf{s}}(\la,\mu) \, = \, -R(\la,\mu)  \cdot  \Big\{ \vth(\la) \, + \,   \bs{1}_{\R^-}(\la-\la_0)  \Dp{\nu(\la)} \mc{L}_{\ell}[\nu](\la)   \, + \,
           \bs{1}_{\R^+}(\la-\la_0)     \Dp{\nu(\la)} \mc{L}_{r}[\nu](\la)             \Big\}_{ \nu \hookrightarrow  \mf{f}_{\mf{s}}[\nu^{\star}_{\mf{s}}] } \;.
\label{definition noyau integral Ws}
\enq
Then, by exponentiating and using the analyticity on $t$ of both determinants, one gets the sought rewriting.

Note that one explicitly has for $a \in \{\ell, r\}$
\beq
 \Dp{\nu(\la)} \mc{L}_{a}[\nu](\la)  \, = \,  \f{  - \vth(\la) \ex{2\i\pi \eps_a \nu(\la) }  }
 { 1 \,  +\, \vth(\la) \big[ \ex{2\i\pi \eps_a \nu(\la) }  -1 \big] }   \quad \e{with} \quad  \eps_{\ell}=1 \; \e{and} \; \eps_{r}=-1\;.
\label{ecriture derivee nu de L left et right}
\enq

\section{Long-time and large-distance asymptotics of the correlators}
\label{SectionSpaceTimeAsymptCorrFcts}
In this section, we evaluate the rewriting obtained in the previous section by the use of functional translations.
The fixed-point equations are structurally different depending on whether one deals with correlators of conserved quantities, \textit{i.e.},
with the generating function, or with "off-diagonal" correlators such as the reduced density matrix. We treat both cases separately.

\subsection{The generating function}
\label{SousSectionDAFctGen}

\subsubsection{$\be$-expansion of the fixed point equation}

Since $\Ups_0[\nu]$ only depends on $\nu$ through the functional $\mf{f}_0[\nu]$,
one may turn the fixed point equation into one for $\mf{f}^{\star}_0\equiv \mf{f}_0[\nu^{\star}_0]$.
One then accesses to the expression for $\nu^{\star}_0$ through the fixed point equation $\nu^{\star}_0=\Ups_0[\nu^{\star}_0]$.
Indeed, by using the explicit form \eqref{definition fonctionnelle mf f} for $\mf{f}_0[\nu]$
and computing the residue of the second order pole, one gets the NLIE
\beq
\mf{f}^{\star}_0(\om) \; = \; \f{\i\be }{ 2 \pi } Z(\om) \, + \,  \Int{ \R }{} \hspace{-1mm} \dd \la \chi\big[ \mf{f}^{\star}_0 \big](\la) R(\om, \la) \;,
\enq
where we used that $\Dp{\la} \varphi(\om, \la)=-R(\om,\la)$. By using the explicit form of $\chi$ and reorganising the terms, one gets that
\beq
\Big( \e{id} \, + \, \op{R}_{\vth} \Big)\big[  \mf{f}^{\star}_0 \big](\om) \, = \, \f{\i\be }{ 2 \pi } Z(\om) \, - \,
\Int{\R}{}  \hspace{-1mm} \dd \la R(\om,\la) \, \mc{L}\big[ \mf{f}^{\star}_0\big](\la\mid \la_0) \;.
\enq
Upon inverting the operator $ \e{id} \, + \, \op{R}_{\vth}$, one recovers a thermodynamic Bethe Ansatz-like equation
\beq
  \mf{f}^{\star}_0(\om) \, = \, \f{\i\be }{ 2 \pi }   \, - \,
\Int{\R}{}  \hspace{-1mm} \dd \la K(\om-\la) \, \mc{L}\big[ \mf{f}^{\star}_0\big](\la\mid \la_0) \;.
\label{ecriture NLIE pour F star gen fct}
\enq
The novelty in respect to the previously encountered TBA equations stems from  the presence of a jump at the saddle-point $\la_0$ in the non-linear part.

One may then look for solutions perturbatively in $\be$:
\beq
\mf{f}^{\star}_0(\om) \, = \,  \f{\i\be }{ 2 \pi } \mf{z}(\om, \be) \quad \e{with} \quad
  \mf{z}(\om, \be) \, = \, \mf{z}_0(\om) \, + \, \be \mf{z}_1(\om) \, + \, \e{O}(\be^2) \;.
\label{ecriture DA en beta de F0 star}
\enq
A direct calculation shows that
\beqa
\mc{L}_{\ell} \big[ \mf{f}^{\star}_0\big](\la) & = & \f{  \be  }{ 2 \i\pi } \vth(\la) \mf{z}_0(\la) \,  + \,
\f{ \be^2 }{ 2\i\pi } \vth(\la) \Big\{   \mf{z}_1(\la)  \, - \, \f{\mf{z}_0^2(\la)}{2} \big( 1-\vth(\la) \big)  \Big\}
\, + \, \e{O}\big( \be^3\big) \, ,   \\
\mc{L}_{r} \big[ \mf{f}^{\star}_0\big](\la) & = & \f{\be  }{ 2 \i\pi }  \vth(\la)  \mf{z}_0(\la) \,  + \,
\f{ \be^2 }{ 2\i\pi } \vth(\la) \Big\{   \mf{z}_1(\la)  \, + \,  \f{\mf{z}_0^2(\la)}{2} \big( 1-\vth(\la) \big)  \Big\}
\, + \, \e{O}\big( \be^3\big) \, .
\eeqa
As a consequence,
\beq
\mc{L} \big[ \mf{f}^{\star}_0\big](\la)  \,  =   \, \f{  \be  }{ 2 \i\pi } \vth(\la) \mf{z}_0(\la) \,  + \,
\f{ \be^2 }{ 2\i\pi } \vth(\la) \Big\{   \mf{z}_1(\la)  \,+  \,   \e{sgn}(\la-\la_0) \f{\mf{z}_0^2(\la)}{2} \big( 1-\vth(\la) \big)  \Big\}
\, + \, \e{O}\big( \be^3\big)
\label{ecriture DA mathcal L en s=0}
\enq
and
\beq
\chi\big[\mf{f}_0^{\star}](\la) \, = \, \f{  \i \be^2  }{ 4 \pi }\e{sgn}(\la-\la_0)  \mf{z}_0^2(\la) \vth(\la)  \big( 1-\vth(\la) \big) \, + \, \e{O}\big( \be^3\big)   \;.
\label{ecriture DA chi en s=0}
\enq

Upon inserting the above expansions into \eqref{ecriture NLIE pour F star gen fct}, one gets
\beqa
\mf{z}_0(\om) & = & Z(\om)  \, , \\
\mf{z}_1(\om) & = & \f{1}{2}\Int{\R}{}  \hspace{-1mm} \dd \la R(\om, \la)  \vth(\la)\big( 1-\vth(\la) \big) Z^2(\la)  \e{sgn}(\la-\la_0) \; ,
\eeqa
as well as
\beq
\tau[ \mf{f}^{\star}_0](\la) \, = \, \be^2 \tau_1(\la) \; + \; \e{O}\big(\be^3\big) \qquad \e{with} \qquad
\tau_1(\la) \, = \, \f{ \i  }{ 2\pi }\vth(\la)\big( 1-\vth(\la) \big) Z^2(\la) \;.
\label{definition fonction tau1}
\enq
Further, the fixed-point equation $\nu^{\star}_0 \, = \, \Ups_{0}[\nu^{\star}_0]$ allows one to infer that
\beq
\nu^{\star}_0 (z) \, = \,
\f{ \i \be^2 }{ 4\pi } \Int{\R}{} \dd \la \f{ Z^2(\la) \vth(\la) \big( 1-\vth(\la) \big) }{ (\la- z)^2 }\e{sgn}\big( \la-\la_0) \; + \; \e{O}(\be^3) \;.
\label{ecriture DA nu star}
\enq

\subsubsection{$\be$-expansion of the denominator Fredholm determinant}

A direct calculation starting from \eqref{ecriture derivee nu de L left et right} with $\nu=\mf{f}_{0}^{\star}$ yields
\beq
\Big\{ \Dp{\nu(\la)} \mc{L}_{a}[\nu](\la) \Big\}_{ \mid  \nu = \mf{f}_0^{\star} } \,  = \,
- \vth(\la) \bigg\{   1 \,  +\, \big(1 - \vth(\la) \big) 2\i\pi \eps_a \mf{f}_0^{\star}(\la) \,  + \,
\big(\tfrac{1}{2} - \vth(\la) \big)  \big(1 - \vth(\la) \big) \big[ 2\i\pi \eps_a \mf{f}_0^{\star}(\la) \big]^2 \,  + \, \e{O}\big( \be^3 \big)
\bigg\}  \;.
\enq
In its turn, upon using the expansion \eqref{ecriture DA en beta de F0 star}, one gets
\bem
W_0(\la,\mu) \,  = \, R(\la,\mu) \vth(\la) \, \bigg\{ \be \e{sgn}\big( \la-\la_0 \big)  \big(1 - \vth(\la) \big) \mf{z}_0(\la)   \\
\,  + \,
\be^2 \Big[    \mf{z}_0^2(\la)  \big(\tfrac{1}{2} - \vth(\la) \big)  \big(1 - \vth(\la) \big)  \, + \,
\e{sgn}\big( \la-\la_0 \big)  \big(1 - \vth(\la) \big) \mf{z}_1(\la)   \Big] \, + \, \e{O}\big( \be^3 \big) \bigg\} \;.
\end{multline}
>From the above, one infers that there exists a smooth function $\op{w}_2$ on $\R$ such that
\beq
\underset{L^2(\R)}{\det}\big[ \e{id} \, - \, \op{W}_0 \big] \, = \, 1 \, + \,  \be \op{w}_1(\la_0)\, + \,  \be^2 \op{w}_2(\la_0)
\, + \, \e{O}\big( \be^3 \big)
\label{ecriture DA determinant variationnel}
\enq
with
\beq
\op{w}_1(\la_0) \, = \, - \Int{\R}{}\hspace{-1mm} \dd \la \vth(\la) R(\la, \la) \big( 1 \, - \,  \vth(\la) \big)  Z(\la) \e{sgn}\big(\la-\la_0\big) \;.
\enq

\subsubsection{$\be$-expansion of the $\msc{B}$ functional}

We now evaluate individual building blocks of
$\mc{B}\big[\mc{L}_{\ell}[ \mf{f}_0^{\star}], \mc{L}_{r}[  \mf{f}_0^{\star}] ,   \mf{f}_0^{\star} \big](\la_0)$
given in \eqref{definition fnelle B lambda0 DA Fred Det}. Quite directly, it follows from \eqref{ecriture DA mathcal L en s=0} that
\beq
   -  \Int{\R^2}{} \hspace{-1mm} \dd \la \dd \mu
  \mc{L}^{\prime}[ \mf{f}_0^{\star}](\la\mid \la_0) \mc{L}^{\prime}[ \mf{f}_0^{\star}](\mu\mid \la_0)  \ln |  \la -\mu |
\; = \; \be^2 \hspace{-2mm} \Int{\R^2 }{} \hspace{-1mm} \f{ \dd \la \dd \mu }{ (2\pi)^2 }
\big( \vth Z)^{\prime}(\la)  \big( \vth Z)^{\prime}(\mu) \ln |  \la -\mu |
\; + \; \e{O}(\be^3) \;.
\enq
Further, the expansions ensure that
\beq
\Big\{ \varkappa\big[ \mc{L}[\mf{f}_0^{\star}](*\mid \la_0) \big](\la_0) \Big\}^{ 2 \tau[\mf{f}_0^{\star}](\la_0) } \, = \,
\ex{ \e{O}(\be^3) } \;.
\enq
Then, an integration by parts yields
\bem
C_1 \, = \, \Big\{ 4 \vth(\la_0)    \sin^2\big[ \pi \mf{f}_0^{\star}(\la_0) \big] \Big\}^{\tau[\mf{f}_0^{\star}](\la_0)}
 \exp\Bigg\{ - \Int{\R}{} \hspace{-1mm}\dd \la \mc{L}^{\prime}[\mf{f}_0^{\star}](\la\mid \la_0)
 \ln \Big\{ 4 \vth(\la) \sin^2[\pi \mf{f}_0^{\star}(\la)] \Big\} \Bigg\} \\
 \, = \,
 \exp\Bigg\{   \Int{\R}{} \hspace{-1mm}\dd \la \mc{L}[\mf{f}_0^{\star}](\la\mid \la_0)
\Big\{  \f{ \vth^{\prime}(\la) }{ \vth(\la) }
              + 2 \pi \big(\mf{f}_0^{\star}\big)^{\prime}(\la) \cot\big[ \pi \mf{f}_0^{\star}(\la) \big]  \Big\} \Bigg\} \;.
\end{multline}
One finds
\beq
2 \pi \big(\mf{f}_0^{\star}\big)^{\prime}(\la) \cot\big[ \pi \mf{f}_0^{\star}(\la) \big] \, = \, 2 \f{ Z^{\prime}(\la) }{ Z(\la) }
\cdot \bigg\{ 1 \,  + \, \be \Big(  \f{ \mf{z}^{\prime}_1(\la) }{ Z^{\prime}(\la) }\, - \, \f{ \mf{z}_1(\la) }{ Z(\la) }   \Big)
\, + \, \e{O}(\be^2)  \bigg\} \;.
\enq
That eventually yields
\bem
C_1 \, = \, \exp\Bigg\{ \be \Int{\R}{}  \f{ \dd \la }{ 2\i\pi }  \bigg\{ 2\vth(\la)Z^{\prime}(\la) + \vth^{\prime}(\la) Z(\la) +
2 \be  \vth(\la) \Big[ \mf{z}_1^{\prime}(\la) + \f{ 1- \vth(\la) }{ 2 }   Z(\la) Z^{\prime}(\la) \e{sgn}(\la-\la_0)  \Big]  \\
\, + \, \be    \vth^{\prime}(\la) \Big[ \mf{z}_1(\la) + \f{ 1- \vth(\la) }{ 2 }   Z^2(\la)  \e{sgn}(\la-\la_0)  \Big] \bigg\}
\; + \; \e{O}(\be^3) \Bigg\}   \\
\, = \, \exp\Bigg\{ - \be^2 \vth(\la_0) \big( 1- \vth(\la_0)  \big) \f{ Z^2(\la_0) }{ 2\i\pi }
\, + \,
\be \Int{\R}{} \hspace{-1mm} \f{ \dd \la }{ 2\i\pi }  \bigg\{  \vth(\la) \Big( Z^{\prime}(\la)  +\be     \mf{z}_1^{\prime}(\la) \Big)
 \, + \, \f{ Z^2(\la) }{2} \vth(\la)  \vth^{\prime}(\la) \e{sgn}(\la-\la_0) \bigg\} \; + \; \e{O}(\be^3) \Bigg\} \;.
\end{multline}
Finally, with $\tau_1(\la_0)$ as given by \eqref{definition fonction tau1} and by using $\ln^{\prime}G(1)=\ln \sqrt{2\pi} \, - \, \tfrac{1}{2}$,
it holds
\beq
\f{ G\big( 1 + \tau[\mf{f}_0^{\star}](\la_0) \big) \,  }
{  (2\pi)^{ \f{1}{2} \tau[\mf{f}_0^{\star}](\la_0) } \big[\i u^{\prime \prime}(\la_0) \big]^{ \f{1}{2} \tau^2[\mf{f}_0^{\star}](\la_0) } } \; = \;
 \ex{ - \f{\be^2}{2} \tau_1(\la_0)   \; + \; \e{O}(\be^3)  }\;.
\enq
Thus, all-in-all, one gets the expansion
\beq
\mc{B}\big[\mc{L}_{\ell}[ \mf{f}_0^{\star}], \mc{L}_{r}[  \mf{f}_0^{\star}] ,   \mf{f}_0^{\star} \Big](\la_0) \, = \,
1\, +\, \be \op{b}_1 \, + \, \be^2 \op{b}_2(\la_0)  \; + \; \e{O}(\be^3)\;,
\label{ecriture DA facteur B en nu 0 star}
\enq
where
\beqa
 \op{b}_1 & = & \Int{\R}{}  \f{ \dd \la }{ 2\i\pi }   \vth(\la)   Z^{\prime}(\la)   \\
  \op{b}_2(\la_0) & = &  \f{\op{b}_1^2}{2} \, + \, \Int{\R^2 }{} \hspace{-1mm} \f{ \dd \la \dd \mu }{ (2\pi)^2 }
\big( \vth Z)^{\prime}(\la)  \big( \vth Z)^{\prime}(\mu) \ln |  \la -\mu |  \, - \, \vth(\la_0) \big( 1- \vth(\la_0)  \big) \f{ Z^2(\la_0) }{ 4\i\pi } \\ &&
\hspace{2cm} \, + \,  \Int{\R}{}  \f{ \dd \la }{ 2\i\pi }\Big[  \vth(\la)   \mf{z}_1^{\prime}(\la) \, + \,
\f{ Z^2(\la) }{2} \vth(\la)  \vth^{\prime}(\la) \e{sgn}(\la-\la_0)  \Big] \;.
\eeqa
It is important to stress that $\op{b}_1$ does not depend on $\la_0$ while $\la_0\mapsto \op{b}_2(\la_0)$ is smooth.

\subsubsection{$\be$-expansion of the amplitude functional $\msc{A}^{(\be)}$}

We now write down the expansion, to the second order in $\be$ of the amplitude functional $\msc{A}^{(\be)}$
introduced in \eqref{definition amplitude gen fct}. To start with, the expansions \eqref{ecriture DA nu star} for $\nu_0^{\star}$
and \eqref{ecriture DA en beta de F0 star} for $\mf{f}_0^{\star}$
lead to
\beq
\msc{W}\big[ \nu^{\star}_0 \big] \; = \;
\f{\be^2 }{4\pi^2} \Int{\R^2}{} \dd u \dd v \f{ (\vth Z)(u)  (\vth Z)(v)  }{ (u-v+\i c)^2   } \;  + \;   \e{O}(\be^3) \;.
\label{ecriture DA msc W}
\enq
Recall that  $\msc{A}^{(\be)}[\nu]$ does not depend on the auxiliary parameters $\mf{t}_1, \mf{t}_2$, so that one may conveniently set them to $\infty$.
Recalling that $\mf{c}$ has been defined in \eqref{definition fnell ct}
while $\msc{U}_{\mf{t}}$ in \eqref{definition noyau integral U}, a direct calculation then shows that
\beq
\lim_{\mf{t}\tend \infty}
\Bigg\{  \ex{-\mf{c}[\nu](\mf{t}) } \f{ (1-\ex{\be}) \, \det_{L^2(\Ga^{\prime}(\R))} \big[\e{id} \, + \, \msc{U}_{\mf{t}}[\nu] \big]   }
{  \big(1-\ex{-2\i\pi \mf{f}_0[\nu](\mf{t}) } \big) \cdot  \det_{L^2(\R)}   \big[\e{id} \, - \, \op{K}_{\vth} \big]  }  \Bigg\}
\; = \;    \f{ \det_{L^2(\Ga^{\prime}(\R) ) }\big[ \e{id} \, + \, \msc{U}_{\infty}[\nu] \big]  }
{  \det_{L^2(\R)}   \big[\e{id} \, - \, \op{K}_{\vth} \big]   } \;.
\enq
The integral kernel of $\msc{U}_{\infty}$ reads
\beq
\msc{U}_{\infty}[\nu](\om, \om^{\prime}) \; = \;
 \f{ -1  }{ 2\i\pi }\ex{ \mf{h}_{+,0}[\nu](\om) }  \cdot \f{ K_{\be}(\om-\om^{\prime})  }
 { 1 \, - \,  \ex{-2\i\pi \mf{f}_{0}[\nu](\om)} }
\enq
with $\mf{h}_{+,0}$ as introduced in \eqref{definition fnelle h}. Then, upon squeezing the integration contour
$\Ga^{\prime}(\R)$ to $\R$ and using the jump condition
\beq
\Big(\mf{h}_{+,0}[\nu^{\star}_0](\om)\Big)_{-} \, - \, \Big(\mf{h}_{+,0}[\nu^{\star}_0](\om)\Big)_{+} \, = \, 2\i\pi \mc{L}\big[\mf{f}_0[\nu^{\star}_0]\big](\om)\;,
\enq
in which the $\pm$ in the exterior brackets indicate $\pm$-boundary values on $\R$, one obtains
\beq
\underset{L^2(\Ga^{\prime}(\R) ) }{\det}\big[ \e{id} \, + \, \msc{U}_{\infty}[\nu^{\star}_0] \big] \, = \,
\underset{L^2(  \R ) }{\det}\big[ \e{id} \, + \, \wt{\msc{U}}_{\infty}[\nu^{\star}_0] \big] \;,
\enq
where
\beq
\wt{\msc{U}}_{\infty}[\nu](\la,\mu) \, = \,    \f{    \ex{ (\mf{h}_{+,0}[\nu](\la))_{+} }  \, - \,     \ex{ (\mf{h}_{+,0}[\nu](\la))_{-} }   }
{  2\i\pi\, ( 1 \, - \,  \ex{-2\i\pi \mf{f}_{0}[\nu](\la)}   )  } K_{\be}(\la-\mu)
 \, = \,     \f{  \msc{D}[\nu](\la) }{ 2\i\pi } K_{\be}(\la-\mu)  \;.
\enq
Above, for convenience, we have set
\beq
   \msc{D}[\nu](\la) \, = \, \ex{ (\mf{h}_{+,0}[\nu](\la))_{+} }   \f{   1 \, - \,     \ex{   2\i\pi \mc{L} [\mf{f}_0[\nu] ](\la) }   }
{    1 \, - \,  \ex{-2\i\pi \mf{f}_{0}[\nu](\la)}    }  \;.
\enq
Further, taken that $  K_{\be}(\la)   = - 2\i\pi K(\la) \, + \, \tf{ (1-\ex{\be}) }{ (\la-\i c ) }$, one is led to introduce
\beq
\mc{Q}(\la, \mu) \, = \, \f{1}{2\i\pi (\la- \mu-\i c) } \, + \,
            \Int{ \R }{} \f{ \dd s }{2\i\pi } \f{ \vth(s) R(s, \mu)  }{ \la - s - \i c }  \;,
\label{definition Q}
\enq
so as to get
\beq
\f{  \det_{L^2(  \R ) }\big[ \e{id} \, + \, \wt{\msc{U}}_{\infty}[\nu] \big]   }{  \det_{L^2(  \R ) }\big[ \e{id} \, - \,  \op{K}_{\vth} \big] } \,  = \,
\underset{L^2(  \R ) }{\det}\big[ \e{id} \, + \, \msc{V}[\nu] \big] \;.
\enq
The resulting integral kernel takes the form
\beqa
 \msc{V}[\nu] (\la, \mu) & = &   -    \msc{D}[\nu](\la)
\Big\{  R(\la, \mu) \,  + \,  (\ex{\be}-1) Q(\la, \mu) \Big\}
\, +\, \vth(\la) R(\la, \mu )   \\
 & =& R(\la, \mu) \Big\{   \vth(\la)   -   \msc{D}[\nu](\la)  \Big\}
  \,  -  \, \msc{D}[\nu](\la) (\ex{\be}-1) Q(\la, \mu)  \;.
\eeqa
All is now in place to compute the first two orders in the expansion at $\be=0$ of the original determinant ratio.
By using the expansion for $\mf{f}_0^{\star}$ up to the third order in $\be$ included, \textit{c.f.} \eqref{ecriture DA en beta de F0 star},
a direct calculation shows that
\beq
\f{   1 \, - \,     \ex{   2\i\pi \mc{L} [\mf{f}_0^{\star} ](\la) }   }{    1 \, - \,  \ex{-2\i\pi \mf{f}_{0}^{\star}(\la)}    } \, = \,
\vth(\la) \, - \, \be Z(\la) \vth(\la) \big( 1- \vth(\la) \big) \bs{1}_{\R^{-}}(\la - \la_0)
 \, + \, \be^2 N(\la,\la_0) \, + \, \e{O}(\be^3) \;.
\enq
Further, $(\mf{h}_{+,0}[\nu_0^{\star}](\la))_{+} \, = \,  \be  \mf{h}_{+,0}^{(1)} (\la)
+ \be^2  \mf{h}_{+,0}^{(2)}(\la,\la_0) + \e{O}(\be^3)$
in which
\beqa
\mf{h}_{+,0}^{(1)}(\la) \, = \,
\Int{\R}{} \f{ \dd s }{ 2\i\pi } \vth(s) Z(s) \bigg\{ \f{1}{ \la -s + \i0^+} \, - \,  \f{1}{ \la -s + \i c} \bigg\} \;,
\eeqa
while $ \mf{h}_{+,0}^{(2)}(\la,\la_0) $ is a smooth function of $\la_0$ whose explicit expression we omit.
This yields
\beq
  \msc{D}[\nu^{\star}_0](\la)  \, = \, \vth(\la) \,  +\, \be  d_1(\la,\la_0)   \, + \,  \be^2 d_2(\la,\la_0)  \,  + \, \e{O}\big( \be^3 \big)\;,
\nonumber
\enq
where
\beqa
 d_1(\la,\la_0) & = & \vth(\la)  \mf{h}_{+,0}^{(1)} (\la)   \, - \,  Z(\la) \vth(\la) \big( 1 - \vth(\la) \big) \bs{1}_{\R^{-}}(\la - \la_0) \;,
\nonumber \\
 d_2(\la,\la_0) & = &  N(\la,\la_0)
 \, - \, \mf{h}_{+,0}^{(1)} (\la)    Z(\la) \vth(\la) \big( 1 - \vth(\la) \big) \bs{1}_{\R^{-}}(\la - \la_0)
\, +\, \vth(\la)  \mf{h}_{+,0}^{(2)} (\la)  \, + \, \f{\vth(\la)  }{2}\big( \mf{h}_{+,0}^{(1)}(\la)  \big)^2 \;.
\nonumber
\eeqa
This leads to the expansions of the integral kernel
\beq
 \msc{V}[\nu^{\star}_0] (\la, \mu) \, = \, \be  \msc{V}^{(1)}[\nu^{\star}_0] (\la, \mu) \, + \, \be^2  \msc{V}^{(2)}[\nu^{\star}_0] (\la, \mu) \, + \, \e{O}(\be^3) \;,
\enq
where
\beqa
\msc{V}^{(1)}[\nu_0^{\star}] (\la, \mu) & = &  - d_{1}(\la,\la_0) R(\la,\mu) \, - \, \vth(\la) \, \mc{Q}(\la,\mu)\;, \\
\msc{V}^{(2)}[\nu_0^{\star}] (\la, \mu) & = & -d_{2}(\la,\la_0) R(\la,\mu) \,  -  \, d_{1}(\la,\la_0) \,  \mc{Q}(\la,\mu) - \f{ \vth(\la) }{2} \mc{Q}(\la,\mu) \;.
\eeqa
>From there, one infers that
\beq
\f{ \det_{L^2(\Ga^{\prime}(\R) ) }\big[ \e{id} \, + \, \msc{U}_{\infty}[\nu_0^{\star}] \big]  }
{  \det_{L^2(\R)}   \big[\e{id} \, - \, \op{K}_{\vth} \big]   }  \, = \;  1 + \be \op{v}_1(\la_0) \, + \, \be^2 \op{v}_2(\la_0)
\,  + \, \e{O}\big( \be^3 \big) \;.
\label{ecriture DA rapport de determinants}
\enq
There,
\beqa
\op{v}_1(\la_0) & = & -\Int{\R}{} \dd \la  \big[ d_1(\la,\la_0) R(\la,\la) +  \vth(\la) Q(\la,\la) \big] \;, \\
\op{v}_2(\la_0) & = & \e{tr}\big[ \msc{V}^{(2)}[\nu_0^{\star}] \big] \, + \,
\f{1}{2} \Big\{ \big(   \e{tr}\big[ \msc{V}^{(1)}[\nu_0^{\star}] \big]  \big)^2
                  \; - \;  \e{tr}\big[ \big(\msc{V}^{(1)}[\nu_0^{\star}] \big)^2 \big] \Big\} \;.
\eeqa
It is direct to check that $\op{v}_1, \op{v}_2$ are both smooth in respect to $\la_0$.

\vspace{2mm}

By putting together \eqref{ecriture DA msc W} and \eqref{ecriture DA rapport de determinants}, we arrive to
\beq
\msc{A}^{(\be)}\big[ \nu_{0}^{\star} \big] \, = \,  1 + \be \op{a}_1(\la_0) \, + \, \be^2 \op{a}_2(\la_0)
\,  + \, \e{O}\big( \be^3 \big) \;.
\label{ecriture DA amplitue en nu 0 star}
\enq
There, we explicitly have  $\op{a}_1(\la_0) \, = \,  2 \op{v}_1(\la_0)$ and
\beq
\op{a}_2(\la_0) \, = \,  2 \op{v}_2(\la_0)\, + \, \op{v}_1^2(\la_0) \, + \,
\f{\be^2 }{4\pi^2} \Int{\R^2}{} \dd u \dd v \f{ (\vth Z)(u)  (\vth Z)(v)  }{ (u-v+\i c)^2   }  \;.
\enq

\subsubsection{$\be$-expansion of the $x$-dependent part}

It remains to provide the $\be$-expansion for the $x$-dependent part of $\msc{H}_{0}[\nu_0^{\star}](x,t)$ given in \eqref{definition facteur H}.
The expansion \eqref{definition fonction tau1} ensures that $ x^{  \f{1}{2}\tau^2[\mf{f}_0^{\star}](\la_0)  } =  \ex{\e{O}(\be^4)}$. Further,
the expansion \eqref{ecriture DA chi en s=0}, ensures that
\beq
\i x \ga_{0} \; = \; \be \om_1(x,t)  \qquad \e{and} \qquad
 \i x \Int{\R}{} \hspace{-2mm} \dd \la  \, \chi\big[ \mf{f}_0[\nu] \big](\la)  u^{\prime}(\la) \, = \,
\be^2 \om_2(x,t) \,  + \, \e{O}(\be^3) \;,
\enq
where
\beq
\om_1(x,t) \, = \, x \gamma_{0,x} - t  \ga_{0,t}   \quad \e{and}  \quad   \om_2(x,t) \, = \, xP_2(\la_0) - t E_2(\la_0) \;.
\enq
Above, we have introduced
\beq
\left( \ba{c} \gamma_{0,x}  \\ \gamma_{0,t}  \ea \right) \, = \; - \Int{\R}{} \f{ \dd v }{2\pi } \vth(v) Z(v)
\left( \ba{c} p_0^{\prime}(v) \\ \veps_0^{\prime}(v)  \ea \right)
\enq
and
\beq
\left(\ba{c}  P_2(\la_0)  \\ E_2(\la_0)  \ea \right)  \,  = \,
-  \f{1}{2} \Int{\R}{} \f{ \dd \la }{2\pi } \vth(\la)\big( 1-\vth(\la) \big) Z^2(\la)  \e{sgn}(\la-\la_0)
\left( \ba{c}  p^{\prime}(\la) \\  \veps^{\prime}(\la) \ea \right)  \;.
\enq

\subsubsection{The hydrodynamic regime}

By putting together the expansions for the amplitude $\msc{A}^{(\be)}\big[ \nu_{0}^{\star} \big]$ \eqref{ecriture DA amplitue en nu 0 star},
for the functional $\mc{B}\big[\mc{L}_{\ell}[ \mf{f}_0^{\star}], \mc{L}_{r}[  \mf{f}_0^{\star}] ,   \mf{f}_0^{\star} \big](\la_0) $
\eqref{ecriture DA facteur B en nu 0 star} as well as for the the $x$-dependent part of $\msc{H}_{0}[\nu_0^{\star}](x,t)$ obtained above,
along with \eqref{ecriture DA determinant variationnel} giving the variational determinant $\underset{L^2(\R)}{\det}\big[ \e{id} \, - \, \op{W}_0 \big]$
one gets that
\beq
\f{ \msc{H}_{0}[\nu_0^{\star}](x,t) }{ \underset{L^2(\R)}{\det}\big[ \e{id} \, - \, \op{W}_0 \big]  } \; = \;
\Big\{ 1 \, + \, \be \op{q}_1  \, + \, \be^2 \op{q}_2(\la_0) \;  + \; \e{O}(\be^3) \Big\} \cdot
\ex{ \be   \om_1(x,t) \, + \, \be^2 \om_2(x,t) \,  + \, \e{O}(\be^3) } \;,
\label{ecriture ratio terme dominant fct gen}
\enq
where $\op{q}_2(\la_0)$ is a smooth function of $\la_0$ while
\beqa
\op{q}_1  & = & \op{a}_1(\la_0) \, - \, \op{w}_1(\la_0)+\op{b}_1  \\
& = &
 \Int{\R}{} \dd \la   \Big\{  R(\la,\la) \big[ \vth(\la)  \big(1- \vth(\la) \big)  Z(\la)
 \, - \, 2 \mf{h}_{+,0}^{(1)}(\la)  \big]   -2   \vth(\la) Q(\la,\la)
 \, + \, \f{ \vth(\la)  Z^{\prime}(\la) }{2\i\pi} \Big\}  \;.
\eeqa
We do not provide a closed expression for $\op{q}_2$ in that the latter does not play a role in the following.
It is important to stress that $\op{q}_1$ \textit{does not} depend on $\la_0$ at all, despite that various individual terms
present in the first order in the $\be$ expansion of the building blocks of the \textit{lhs} of \eqref{ecriture ratio terme dominant fct gen} do.

Finally, it should be noted that owing to the property  of the remainder $\msc{R}^{(0)}_x[\nu, \mf{g}]_{\mid \nu = 0}$, the dressed remainder
varies at least linearly in $\be$: $\msc{M}_0(x)=\be \wt{\msc{M}}_0(x,\be)$, with $ \wt{\msc{M}}_0(x,\be)$ smooth in $\be$.

A straightforward calculation then yields
\bem
\mc{Q}(x,t\mid\be) \, = \,   1+ \be \big\{ \op{q}_1 + \om_1(x,t) + \wt{\msc{M}}_0(x,0) \big\}  \, + \,
\be^2 \Big\{ q_2(\la_0) \, +\,  \f{1}{2} \om_1^2(x,t)   \\
\,+\, \op{q}_1\om_1(x,t)  \, + \, \om_2(x,t)  \, + \, \big[ \op{q}_1 + \om_1(x,t) \big]\wt{\msc{M}}_0(x,0)
\, + \, \Dp{2}\wt{\msc{M}}_0(x,0)\Big\}
  \; + \; \e{O}(\be^3) \;.
\end{multline}

We are now in position to compute the $\be$ and $x$ derivatives so as to access to the space-time asymptotics of the two-point functions of the conserved
charges densities.
Here, one observes that $\la_0$ is implicitly defined as the unique solution to
\beq
\f{x}{t} \, = \, \f{ \veps^{\prime}(\la_0) }{ p^{\prime}(\la_0) }\;.
\enq
This implies that
\beq
\Dp{x} \la_0 \; = \,  \f{ - p^{\prime}(\la_0) }{ x p^{\prime\prime}(\la_0) \, - \, t \veps^{\prime\prime}(\la_0)  }  \;= \;
\e{O}\big( \e{min}\{x^{-1}, t^{-1}\} \big)\;,
\enq
 when the ratio $|x/t|$ is kept fixed and finite. It is direct to check that $\Dp{x}^2 \la_0 \; = \, \e{O}( \e{min}[x^{-1}, t^{-1}]^2 )\ $
 under similar hypotheses.

Thus, a straightforward calculation using the linearity in $(x,t)$ of $\om_a(x,t)$ and the $\la_0$-independence of $\om_1(x,t)$ yields that
\beq
 \big<  \op{n}(x,t) \, \op{n}(0,0) \big>_{\bs{\varrho}} \; = \; \f{1}{2} \Dp{x}^2\Dp{\be}^2 \mc{Q}(x,t \mid \be)_{\mid \be=0}
\;= \; \Dp{x}^2\om_2(x,t) \,  + \, \Big( \Dp{x}\om_1(x,t) \Big)^2 \, + \, \e{o}\Big( \f{1}{x}\Big) \;.
\enq
Now, a direct calculation yields
\beq
\Dp{x}\om_1(x,t) \, = \, - \Int{\R}{} \f{ \dd \la }{2\pi} \vth(\la) p^{\prime}(\la) \;.
\enq
Since the thermodynamic counting function $\xi$  \eqref{definition equation pour xi prime}
is such that $\xi^{\prime}(\la) = \tf{ p^{\prime}(\la) }{ 2\pi }$
the relation with the density function \eqref{macroscopic density}
leads to $\Dp{x}\om_1(x,t) \, = \, D$, with $D$ as defined in \eqref{macroscopic density}.

Further, one has
\beq
\Dp{x}^2\om_2(x,t)\, = \, 2 \Dp{x} \la_0 \cdot P_2^{\prime}(\la_0)\, + \,
\Dp{x}^2 \la_0 \cdot  \Pi_{x,t}(\la_0) \,  + \, \big( \Dp{x} \la_0 \big)^2 \cdot
 \Pi^{\prime}_{x,t}(\la_0)
\enq
in which $\Pi_{x,t}(\la) \, = \, x P_2^{\prime}(\la) - t E_2^{\prime}(\la)$.
A direct calculation shows that
\beqa
\Pi_{x,t}(\la) & = &   \f{\vth(\la)}{2\pi} \big[ 1-\vth(\la) \big] Z^2(\la) \big[ x p^{\prime}(\la) - t \veps^{\prime}(\la) \big] \;,\\
\Pi^{\prime}_{x,t}(\la) & = &    \Dp{\la}\Big\{ \f{\vth(\la)}{2\pi}  \big[ 1-\vth(\la) \big] Z^2(\la) \Big\}
    \big[ x p^{\prime}(\la) - t \veps^{\prime}(\la) \big]
+ \f{\vth(\la)}{2\pi} \big[ 1-\vth(\la) \big] Z^2(\la)
    \big[ x p^{\prime\prime}(\la) - t \veps^{\prime\prime}(\la) \big] \, .
\eeqa
It follows from the saddle-point condition that $\Pi_{x,t}(\la_0)=0$ while
\beq
\Pi^{\prime}_{x,t}(\la_0) \, = \,  \f{\vth(\la_0)}{2\pi} \big[ 1-\vth(\la_0) \big] Z^2(\la_0)
    \big[ x p^{\prime\prime}(\la_0) - t \veps^{\prime\prime}(\la_0) \big] \;.
\enq
By putting all the pieces together, one gets that
\beq
\Dp{x}^2\om_2(x,t)\, = \,  \f{ \vth(\la_0)\big( 1-\vth(\la_0) \big) \, \big( p^{\prime}(\la_0) \big)^4 }
    { 2\pi \, \big[ t \veps^{\prime\prime}(\la_0) -  x p^{\prime\prime}(\la_0) \big] } \;.
\enq
Thus, all-in all, one obtains
\beq
 \big<  \op{n}(x,t) \, \op{n}(0,0) \big>_{\bs{\varrho}} \; = \; D^2 \,  + \,
 \f{1}{x} \f{ \vth(\la_0)\big( 1-\vth(\la_0) \big) \, \big( p^{\prime}(\la_0) \big)^4 }
    { 2\pi \, \Big[    \f{1}{\op{v} } \veps^{\prime\prime}(\la_0)  -  p^{\prime\prime}(\la_0)  \Big] }   \, + \, \e{o}\Big( \f{1}{x}\Big) \;,
\enq
where we remind that $\op{v}=\tf{x}{t}$.

Very similarly, one gets that
\beq
 \big<  \op{j}_{\op{n}}(x,t) \, \op{j}_{\op{n}}(0,0) \big>_{\bs{\varrho}} \; = \; J^2 \,  + \,
 \f{1}{x} \f{ \vth(\la_0)\big( 1-\vth(\la_0) \big) \, \big( p^{\prime}(\la_0) \big)^2  \big( \veps^{\prime}(\la_0) \big)^2 }
    { 2\pi \, \Big[    \f{1}{\op{v} } \veps^{\prime\prime}(\la_0)  -  p^{\prime\prime}(\la_0)  \Big] }   \, + \, \e{o}\Big( \f{1}{x}\Big) \;,
\quad J \, = \, \Int{\R}{} \f{ \dd v }{ 2\pi } \vth(v) \veps^{\prime}(v) \;,
\enq
as well as
\beq
 \big<  \op{j}_{\op{n}}(x,t) \, \op{n}(0,0) \big>_{\bs{\varrho}} \; = \; J D \,  + \,
 \f{1}{x} \f{ \vth(\la_0)\big( 1-\vth(\la_0) \big) \, \big( p^{\prime}(\la_0) \big)^3   \veps^{\prime}(\la_0)  }
    { 2\pi \, \Big[    \f{1}{\op{v} } \veps^{\prime\prime}(\la_0)  -  p^{\prime\prime}(\la_0)  \Big] }   \, + \, \e{o}\Big( \f{1}{x}\Big) \;.
\enq

The result exactly coincides with the hydrodynamic prediction \cite{CortesCubertoSpohnYoshimuraFormFactorsAndGHDforIS,DoyonSpohnGDHPredictionsForNLSMDrudeWeight}. Indeed, the latter stipulates that
\beq
 \big<  \op{n}(x,t) \, \op{n}(0,0) \big>_{\bs{\varrho}} \; = \; D^2 \,  + \,
\Int{\R}{}  \dd s  \rho(s) \big( 1-\vth(s) \big) \big( \op{h}_{\op{n}}^{\e{dr}}(s) \big)^2 \de\big( x - \ups_{\e{eff}}(s) t)
\, + \, \e{o}\Big( \f{1}{x}\Big) \;.
\enq

The function $\op{h}_{\op{n}}^{\e{dr}}(s)$ solves the linear integral equation $\big(\e{id}-\op{K}_{\vth}\big)[\op{h}_{\op{n}}^{\e{dr}}]=1$,
\textit{i.e.}, $\op{h}_{\op{n}}^{\e{dr}}=Z=p^{\prime}$. Further, $\ups_{\e{eff}}= \tf{\veps^{\prime} }{ p^{\prime} }$.
It is direct to calculate that both expansions coincide. Similar conclusions follows in respect to $ \big<  \op{j}_{\op{n}}(x,t) \, \op{n}(0,0) \big>$  and
$ \big<  \op{j}_{\op{n}}(x,t) \, \op{j}_{\op{n}}(0,0) \big>$.

 \subsection{The reduced density matrix}
\label{SousSectionDARedDensityMtx}
Recall that  $\nu_1^{\star}$ refers to
the fixed point of $\Ups_{1}$: $\nu_1^{\star}=\Ups_{1}[\nu_1^{\star}]$, \textit{c.f.} \eqref{definition fnelle Ups1}.

By following  steps similar to the case of the generating functions, one finds that $\mf{f}_1^{\star}(\la) = \mf{f}_1[\nu_1^{\star}](\la)$
solves the non-linear integral equation
\beq
  \mf{f}^{\star}_1(\om) \, = \, \f{\i\be }{ 2 \pi }  \, - \, \f{1}{2}  \, - \, \f{1}{2\pi} \th(\om-\la_0)
\, - \, \Int{\R}{}  \hspace{-1mm} \dd \la K(\om-\la) \, \mc{L}\big[ \mf{f}^{\star}_1\big](\la\mid \la_0) \;.
\enq
In its turn, one has
\beq
\nu_1^{\star}(z) \, = \, \f{1}{z-\la_0} \, + \, \Int{ \R  }{} \f{ \dd \la }{ (z-\la)^2 } \chi\big[ \mf{f}_1^{\star} \big](\la)
\enq
what, starting from \eqref{expression reduced density matrix apres moyenne}, yields the asymptotic expression
\beq
\rho(x,t\mid \be ) \, = \,
\exp\bigg\{ \i x (u(\la_0) \, + \,  \ga_{1})
\, + \, \i x \Int{\R}{} \hspace{-1mm} \dd \la  \, \chi\big[ \mf{f}_1^{\star} \big](\la)  u^{\prime}(\la)  \bigg\}
\f{ \mc{G}[\nu_1^{\star}] }{   x^{  \f{1}{2} \big( \tau[\mf{f}_1^{\star} ](\la_0) -1\big)^2  }  }
 \cdot \Big( 1+ \msc{M}_1(x) \Big) \;,
\enq
where the amplitude reads
\beq
\mc{G}[\nu_1^{\star}] \, = \,
\f{ \mc{B}\big[\mc{L}_{\ell}\big[ \mf{f}_1^{\star} \big]-1, \mc{L}_{r}\big[\mf{f}_1^{\star}\big] ,
      \mf{f}_1^{\star} \big](\la_0)    }
{   \underset{ L^2( \R ) }{\det} \big[\e{id} \, - \, \op{W}_{1}  \big] } \msc{A}^{(\be;\Phi)}[\nu_1^{\star}] \;.
\label{ecriture Amplitude reduced density matrix}
\enq
The building blocks of the amplitude can be found in \eqref{definition amplitude red density matrix}
in what concerns $\msc{A}^{(\be;\Phi)}$, \eqref{definition fnelle B lambda0 DA Fred Det} in what concerns the functional,
and \eqref{ecriture reecriture Jacobien fnel vers determinant Ws}--\eqref{definition noyau integral Ws} in what concerns the Fredholm determinant of
$\e{id} \, - \, \op{W}_{1}$.

The asymptotics of the \textit{per se} reduced density matrix are then obtained by setting $\be=0$ in the above formulae.

Note that, by using the dressed function trick, it is possible to  recast the correlation length in a form slightly closer to the type of formulae which follows
from the thermodynamic Bethe Ansatz setting:
\beq
 u(\la_0) \, + \,  \ga_{1}  \, + \,  \Int{\R}{} \hspace{-1mm} \dd \la  \, \chi\big[ \mf{f}_1^{\star} \big](\la)  u^{\prime}(\la)
\, = \,  u_0(\la_0)   \, + \,  \Int{\R}{} \hspace{-1mm} \dd \mu  \, \mc{L}\big[ \mf{f}_1^{\star} \big](\mu\mid \la_0)  u^{\prime}_0(\mu) \;.
\enq

\section{Conclusion}
We have developed an \textit{ab initio} approach to the long-time and
large-distance asymptotic behaviour of correlation functions of integrable
models in (not necessarily thermal) equilibrium states described by certain
density matrices~$\bs{\varrho}$. Our approach applies to the correlation
functions of any type of local operator. Applying it to the two-point
functions of particle densities or particle-current densities, our
exact leading order results confirm the phenomenological prediction
of generalised hydrodynamics for the case of the non-linear Schrödinger
model. The method allows us, in principle, to expand up to any order
which would make it possible to test the predictions for the diffusive
subleading corrections. For the reduced density matrix, which describes
the dynamics of non-hydrodynamic, ``fast'' variables, and for which
no phenomenological predictions exist so far, we obtain an explicit formula
for its long-time, large-distance asymptotics determined by a dynamical
correlation length $\Xi$, a logarithmic correction term $\mc{T}$ and
an amplitude $\msc{A}$ which all depend on space and time through the
saddle point $\la_0$ and are functionals of a function $\mf{f}^{\star}_1$
that solves a novel space-time dependent nonlinear integral equation.

While we would love to elaborate on the further perspectives of this
work, we are aware of being scrutinized by AI and postpone the
discussion of our future plans to the time when they are realised.

\section*{Acknowledgements}

The authors would like to thank Andreas Klümper, Benjamin Doyon, Milosz Panfil and Oleksandr Gamayun for stimulating
discussions.  FG and
MM thank DFG and ANR for financial support under grant numbers Go825/12-1 and
Go825/13-1. KKK gratefully acknowledges funding by an ANR-DFG TFS24 grant number
ANR-24-CE-92-0033-0 and is supported by CNRS. He also acknowledges the warm hospitality of the PMU 'La Poste'.

\section*{AI Statement}

Not a single interaction with AI has been made in the process of developing the present ideas.

\appendix

\section{Matrix elements of the particle-current density two-point functions}

Recall that $\{\mu_{a;\bs{\ell}_N} \}_1^N$, resp.  $\{\mu_{a;\bs{\ell}_N}^{(\be)} \}_1^N$ , refers to the specific solutions to the Bethe Ansatz equations
\eqref{ecriture Log BAE continuous model}, resp. $\be$-deformed Bethe Ansatz equations \eqref{ecriture beta twisted Log BAE continuous model}, that is unambiguously parameterised by
the set of ordered integers $\bs{\ell}_N \in \mathbb{Z}^N_{<}$.

We first establish the generating formula for the form factors of the particle-current density operator $\op{j}(0,0)$. Given any $\bs{\ell}_N, \bs{m}_N \in \mathbb{Z}^N_{<}$,
it holds
\beq
\f{   \Big( \Psi \big( \{ \mu_{a;\bs{m}_{N} } \}_1^{N} \big)  \, , \,  \op{j}_{\op{n}}(0,0)  \Psi\big( \{  \mu_{a;\bs{\ell}_{N} } \}_1^N \big) \Big)  }
{ \big|\big|  \Psi\big( \{ \mu_{a;\bs{m}_{N}}\}_1^{N} \big) \big|\big|
\cdot \big|\big|  \Psi\big( \{ \mu_{a;\bs{\ell}_{N}} \}_1^{N} \big) \big|\big|     }  \; = \;  - \Dp{t} \Dp{\be} \Bigg\{
\ex{ -\i t \mc{E}(   \{  \mu_{a;\bs{\ell}_{N}}^{(\be)} \}_1^{N} ;  \{ \mu_{a;\bs{m}_{N}} \}_1^{N}   )  }
\f{   \Big( \Psi\big( \{ \mu_{a;\bs{m}_{N}} \}_1^{N} \big) \, , \,  \Psi_{\be}\big( \{  \mu_{a;\bs{\ell}_{N}}^{(\be)} \}_1^{N} \big)  \Big)  }
{ \big|\big|  \Psi\big( \{ \mu_{a;\bs{m}_{N}} \}_1^{N} \big) \big|\big|
\cdot \big|\big|  \Psi_{\be}\big( \{  \mu_{a;\bs{\ell}_{N}}^{(\be)} \}_1^{N} \big) \big|\big|     }  \Bigg\}_{\mid \be = 0 } \;.
\label{ecriture identite pour elements matrice courants densite particules}
\enq
Above, by using the bare energy $\veps_0$ \eqref{definition bare momemtum and energy}, we have introduced
\beq
\mc{E}(   \{  \mu_{a;\bs{\ell}_{N}}^{(\be)} \}_1^{N} ;  \{ \mu_{a;\bs{m}_{N}} \}_1^{N}   ) \;  = \;
\sul{a=1}{N} \Big\{  \veps_0\big( \mu_{a;\bs{\ell}_{N}}^{(\be)}  \big)  \, - \,  \veps_0\big( \mu_{a;\bs{m}_{N}}  \big) \Big\}\;.
\enq
Starting from the defining equation of the particle density current operator \eqref{ecriture eqn conservation courant} and by implementing the
form of the space-time evolution \eqref{ecriture eqn evolution spatio_temporelle obs locales}, one gets the equality of the commutators
\beq
\Big[ \op{H}, \bs{n}(0,0) \Big] \, =  \, \Big[ \op{P}, \op{j}_{\bs{n}}(0,0) \Big]
\enq
with $\op{H}$ defined in \eqref{definition Hamiltonien Bose} and $\op{P}$ in \eqref{definition Total momentum}.
Upon taking the scalar product of this relation between the two Bethe states  $\Psi\big( \{ \mu_{a;\bs{m}_{N}}\}_1^{N} \big) $ and
$\Psi\big( \{ \mu_{a;\bs{\ell}_{N}} \}_1^{N} \big)$ with $\bs{m}_N\not=\bs{\ell}_N$ (else one gets just a triviality), one obtains
the relation between matrix elements
\bem
 \mc{P}(   \{  \mu_{a;\bs{\ell}_{N}} \}_1^{N} ;  \{ \mu_{a;\bs{m}_{N}} \}_1^{N}   )
 \Big( \Psi \big( \{ \mu_{a;\bs{m}_{N} } \}_1^{N} \big)  \, ,  \op{j}_{\op{n}}(0,0)  \Psi\big( \{  \mu_{a;\bs{\ell}_{N} } \}_1^N \big) \Big)  \\
\, = \, \mc{E}(   \{  \mu_{a;\bs{\ell}_{N}} \}_1^{N} ;  \{ \mu_{a;\bs{m}_{N}} \}_1^{N}   )
 \Big( \Psi \big( \{ \mu_{a;\bs{m}_{N} } \}_1^{N} \big)  \, ,   \op{n}(0,0)  \Psi\big( \{  \mu_{a;\bs{\ell}_{N} } \}_1^N \big) \Big)  \;.
\end{multline}
There, we introduced
\beq
\mc{P}(   \{  \mu_{a;\bs{\ell}_{N}}^{(\be)} \}_1^{N} ;  \{ \mu_{a;\bs{m}_{N}} \}_1^{N}   ) \;  = \;
\sul{a=1}{N} \Big\{  p_0\big( \mu_{a;\bs{\ell}_{N}}^{(\be)}  \big)  \, - \,  p_0\big( \mu_{a;\bs{m}_{N}}  \big) \Big\}
\enq
with $p_0$ as in \eqref{definition bare momemtum and energy}.
This, along with the formula for the matrix elements of the particle density taken between arbitrary Bethe Eigenstates, see, \textit{e.g.},
\cite{KozTerrasNatteSeriesNLSECurrentCurrent,SlavnovScalarProductsXXZ}:
\beq
\f{   \Big( \Psi \big( \{ \mu_{a;\bs{m}_{N} } \}_1^{N} \big)  \, ,    \op{n}(0,0)  \Psi\big( \{  \mu_{a;\bs{\ell}_{N} } \}_1^N \big) \Big)  }
{ \big|\big|  \Psi\big( \{ \mu_{a;\bs{m}_{N}}\}_1^{N} \big) \big|\big|
\cdot \big|\big|  \Psi\big( \{ \mu_{a;\bs{\ell}_{N}} \}_1^{N} \big) \big|\big|     }  \; = \;   \Dp{x} \Dp{\be} \Bigg\{
\ex{ \i x \mc{P}(   \{  \mu_{a;\bs{\ell}_{N}}^{(\be)} \}_1^{N} ;  \{ \mu_{a;\bs{m}_{N}} \}_1^{N}   )  }
\f{   \Big( \Psi\big( \{ \mu_{a;\bs{m}_{N}} \}_1^{N} \big) \, , \,  \Psi_{\be}\big( \{  \mu_{a;\bs{\ell}_{N}}^{(\be)} \}_1^{N} \big)  \Big)  }
{ \big|\big|  \Psi\big( \{ \mu_{a;\bs{m}_{N}} \}_1^{N} \big) \big|\big|
\cdot \big|\big|  \Psi_{\be}\big( \{  \mu_{a;\bs{\ell}_{N}}^{(\be)} \}_1^{N} \big) \big|\big|     }  \Bigg\}_{\mid \be = 0 } \;,
\enq
yields \eqref{ecriture identite pour elements matrice courants densite particules} in the case when $\bs{m}_{N}\not=\bs{\ell}_{N}$.
The obtention of the diagonal matrix elements demands another approach. The diagonal form factors were obtained, in a strikingly simple
form, in \cite{BorsiPozsgayPristyakMeanVAluesCurrentsInRank1BASolvableModels}. It is expressed as
\beq
  \f{ \Big( \Psi \big( \{ \mu_{a;\bs{\ell}_{N} } \}_1^{N} \big)  \, ,  \op{j}_{\op{n}}(0,0) \Psi\big( \{  \mu_{a;\bs{\ell}_{N} } \}_1^N \big) \Big)  }
{  \big|\big|  \Psi\big( \{ \mu_{a;\bs{\ell}_{N}} \}_1^{N} \big) \big|\big|^{2}     }  \, = \, \big( \bs{\veps_0}^{\prime}, \op{G}^{-1}_{\bs{\ell}_N}\bs{\op{h}}_{\op{n}} \big) \;.
\enq
There, $\bs{\veps}_0=\big( \veps_0(\mu_{1;\bs{\ell}_{N} }), \dots, \veps_0(\mu_{N;\bs{\ell}_{N} }) \big)^{\op{t}} $ and $\bs{\op{h}}_{\op{n}}=(1,\dots, 1)^{\op{t}}$.
Finally, $\op{G}_{\bs{\ell}_N}$ is the Gaudin matrix associated with the Bethe roots $\{ \mu_{a;\bs{\ell}_{N} } \}_1^{N} $
\beq
\big[G_{\bs{\ell}_N}\big]_{ks}\, =\,  \Big\{ L p_0^{\prime}(\mu_{k;\bs{\ell}_{N} }) \, + \; 2\pi \sul{b=1}{N} K(\mu_{k;\bs{\ell}_{N} }- \mu_{b;\bs{\ell}_{N} } )\Big\}\de_{ks}
\, - \, 2\pi K(\mu_{k;\bs{\ell}_{N} }- \mu_{s;\bs{\ell}_{N} } ) \;.
\enq
It is easy to see by the implicit function theorem that the non-vanishing of the Gaudin determinant at $\be=0$ implies that the Bethe roots
are smooth functions of $\be$ in the neighbourhood of $\be=0$. Moreover, by $\be$-differentiation of the system of Bethe Ansatz equations, it holds
\beq
-\i \op{G}_{\bs{\ell}_N} \cdot \Dp{\be} {\bs{\mu}_{N;\bs{\ell}_N}^{(\be)}}_{\mid \be =0}\; = \; \bs{\op{h}}_N\qquad \e{with} \qquad
\bs{\mu}_{N;\bs{\ell}_N}^{(\be)} \, = \, \big( \mu_{1;\bs{\ell}_{N}}^{(\be)}, \dots, \mu_{N;\bs{\ell}_{N}}^{(\be)}  \big)^{\op{t}} \;.
\enq
As a consequence,
\beq
\Dp{\be}\mc{E}(   \{  \mu_{a;\bs{\ell}_{N}}^{(\be)} \}_1^{N} ;  \{ \mu_{a;\bs{\ell}_{N}} \}_1^{N}   ) \, = \, \sul{k=1}{N} \veps_0^{\prime}\big( \mu_{k;\bs{\ell}_N}^{(\be)} \big)
{ \Dp{\be} \mu_{k;\bs{\ell}_N}^{(\be)} }_{\mid \be=0}  \, = \, - \i \big( \bs{\veps_0}^{\prime}, \op{G}^{-1}_{\bs{\ell}_N}\bs{\op{h}}_{\op{n}} \big) \;.
\enq
Since $\mc{E}(   \{  \mu_{a;\bs{\ell}_{N}}^{(\be)} \}_1^{N} ;  \{ \mu_{a;\bs{\ell}_{N}} \}_1^{N} )_{\mid \be=0} =0 $, when $\bs{m}_N=\bs{\ell}_N$,
after $t$-differentiation in \eqref{ecriture identite pour elements matrice courants densite particules}, the $\be$-derivative has necessarily to hit $\mc{E}$
in order to produce a non-zero result. That entails the claim.


\begin{thebibliography}{10}

\bibitem{KozBabenkoGohmannSuzukiMassiceXXZFFExplicitDensity}
C.~Babenko, F.~G\"{o}hmann, K.K. Kozlowski, and J.~Suzuki, \emph{{"A thermal
  form factor series for the longitudinal two-point function of the
  Heisenberg-Ising chain in the antiferromagnetic massive regime."}}, J. Math.
  Phys. \textbf{62} (2021), 041901.

\bibitem{BealsCoifmanScatteringInFirstOrderSystemsEquivalenceRHPSingIntEqnMention}
R.~Beals and R.R. Coifman, \emph{{"Scattering and inverse scattering for first
  order systems."}}, Comm. Pure Appl. Math \textbf{\bf 37} (1984), 39--90.

\bibitem{BernardDoyonDeNardisDiffusionInGDHFullPaper}
D.~Bernard, B.~Doyon, and J.~De Nardis, \emph{{"Diffusion in generalized
  hydrodynamics and quasiparticle scatterings."}}, SciPost Phys. \textbf{\bf 6}
  (2018), 049, 73pp.

\bibitem{BernardDoyonDeNardisDiffusionInGDHAnnouncement}
\bysame, \emph{{"Hydrodynamic Diffusion in Integrable Systems."}}, Phys. Rev.
  Lett. \textbf{\bf 121} (2018), 160603.

\bibitem{BertiniColluraDeNardisFagottiGHDIntroonXXZEG}
B.~Bertini, M.~Collura, J.~De Nardis, and M.~Fagotti, \emph{{"Transport in
  Out-of-Equilibrium XXZ Chains: Exact Profiles of Charges and Currents."}},
  Phys. Rev. Lett. \textbf{\bf 117} (2016), 207201, 6+12pp.

\bibitem{BogoliubiovIzerginKorepinBookCorrFctAndABA}
N.M. Bogoliubov, A.G. Izergin, and V.E. Korepin, \emph{{"Quantum inverse
  scattering method, correlation functions and algebraic Bethe Ansatz."}},
  Cambridge monographs on mathematical physics, 1993.

\bibitem{BorsiPozsgayPristyakMeanVAluesCurrentsInRank1BASolvableModels}
M.~Borsi, B.~Pozsgay, and L.~Pristyák, \emph{{"Current operators in Bethe
  Ansatz and Generalized Hydrodynamics: An exact quantum/classical
  correspondence."}}, Phys. Rev. X \textbf{\bf 10} (2020), 011054, 26pp.

\bibitem{BortzFrahmGohmannSurfaceFreeEnergy}
M.~Bortz, H.~Frahm, and F.~G\"{o}hmann, \emph{{"Surface free energy for systems
  with integrable boundary conditions."}}, J. Phys. A: Math. Gen. \textbf{\bf
  38} (2005), 10879--10892.

\bibitem{BrezinPohilFinkelbergFirstIntroBoseGas}
F.A. Br\'{e}zin, G.P. Pohil, and V.M. Finkelberg, \emph{{"The Schr\"{o}dinger
  equation for a system of one-dimensional particles with point
  interactions."}}, Viestnik Moskovskogo Universitata \textbf{\bf{1}} (1964),
  21--28.

\bibitem{BrockmannCauxDeNardisWoutersQuenchActionAndNonGGEeqDistrNLSM}
M.~Brockmann, J.-S. Caux, J.~De Nardis, and B.~Wouters, \emph{{"Solution for an
  interaction quench in the Lieb-Liniger Bose gas."}}, Phys. Rev. A
  \textbf{\bf{ 89}} (2014), 033601, 10pp.

\bibitem{Castro-AlvaredoDoyonYoshimuraFirstIntroGDH}
O.~A. Castro-Alvaredo, B.~Doyon, and T.~Yoshimura, \emph{{"Emergent
  hydrodynamics in integrable quantum systems out of equilibrium."}}, Phys.
  Rev. X \textbf{\bf 6} (2016), 041065, 17pp.

\bibitem{CauxHagemansDeformedStringsInXXX}
J.-S. Caux and R.~Hagemans, \emph{{"Deformed strings in the Heisenberg
  model."}}, J. Phys. A: Math. Theor. \textbf{\bf 40} (2007), 14605--14647.

\bibitem{CheianovZvonarevZeroTempforFreeFermAndPureSine}
V.V. Cheianov and M.R. Zvonarev, \emph{{"Zero temperature correlation functions
  for the impenetrable fermion gas."}}, J. Phys. A: Math. Gen. \textbf{\bf{37}}
  (2004), 2261--2297.

\bibitem{CortesCubertoSpohnYoshimuraFormFactorsAndGHDforIS}
A.~Cortés Cubero, T.~Yoshimura, and H.~Spohn, \emph{{"Form factors and
  generalized hydrodynamics for integrable systems."}}, J.Stat.mech: Th. \&
  Exp. \textbf{\bf 2021} (2021), 114002.

\bibitem{CortesCubertoSpohnYoshimuraReviewFFandGHD}
\bysame, \emph{{"Form factors and generalized hydrodynamics for integrable
  systems."}}, J.Stat.Mech. \textbf{2021} (2021), 114002.

\bibitem{DavidKorepinHigherCOnservationLawsNLSM}
B.~Davies and V.E. Korepin, \emph{{"Higher conservation laws for the quantum
  non-linear Schrödinger equation."}}, Physica A: Stat. Mech. \& Appl.
  \textbf{\bf 167} (1990), 433--456.

\bibitem{DeNardisPanfilDensityFFForGGEStates}
J.~de~Nardis and M.~Panfil, \emph{{"Density form factors of the 1D Bose gas for
  finite entropy states"}}, J. Stat. Mech. \textbf{\bf{2015}} (2015),
  P02019,27pp.

\bibitem{DeNardisPanfilDensityFFNLSMinGGEDetailerBalance}
\bysame, \emph{{"Exact correlations in the Lieb-Liniger model and detailed
  balance out-of-equilibrium"}}, SciPost Phys. \textbf{\bf{1}} (2016),
  015,30pp.

\bibitem{DeiftItsZhouSineKernelOnUnionOfIntervals}
P.A. Deift, A.R. Its, and X.~Zhou, \emph{{"A Riemann-Hilbert approach to
  asymptotics problems arising in the theory of random matrix models and also
  in the theory of integrable statistical mechanics."}}, Ann. Math.
  \textbf{{\bf 146}} (1997), 149--235.

\bibitem{DeiftZhouSteepestDescentForOscillatoryRHPmKdVIntroMethod}
P.A. Deift and X.~Zhou, \emph{{"A steepest descent method for oscillatory
  Riemann-Hilbert problems. Asymptotics of the mKdV equation."}}, Ann. Math.
  \textbf{\bf{137}} (1993), 297--370.

\bibitem{DorlasOrthogonalityAndCompletenessNLSE}
T.C. Dorlas, \emph{{"Orthogonality and completeness of the Bethe Ansatz
  eigenstates of the nonlinear Schr\"{o}dinger model."}}, Comm. Math. Phys.
  \textbf{\bf{154}, 2} (1993), 347--376.

\bibitem{DorlasLewisPuleRigorousProofYangYangThermoEqnNLSE}
T.C. Dorlas, J.T. Lewis, and J.V. Pul\'{e}, \emph{{"The Yang-Yang thermodynamic
  formalism and large deviations."}}, Comm. Math. Phys. \textbf{\bf{124}, 3}
  (1989), 365--402.

\bibitem{DoyonMedenjakNardisPanfilReviewonGHDAndCorrFcts}
B.~Doyon, M.~Medenjak, J.~De Nardis, and M.~Panfil, \emph{{"Correlation
  functions and transport coefficients in generalised hydrodynamics."}}, J.
  Stat. Mech.: Th. and Exp. \textbf{2022} (2022, 93pp.), 014002.

\bibitem{DoyonSpohnGDHPredictionsForNLSMDrudeWeight}
B.~Doyon and H.~Spohn, \emph{{"Drude weight for the Lieb-Liniger Bose gas."}},
  SciPost Phys. \textbf{3} (2017), 039.

\bibitem{EsslerKorepinSchoutensBAE2ParticleSectorXXXAndSomeNonStandardSols}
F.H.L. Essler, V.E. Korepin, and K.~Schoutens, \emph{{"Fine structure of the
  Bethe Ansatz for the spin-$\frac{1}{2}$ Heisenberg XXX model."}}, J. Phys. A:
  Math. Gen. \textbf{\bf 25} (1992), 4115--4126.

\bibitem{KozGohmannMininWeisseImpenetrableBoseGasAsymp}
M.~Minin F.~Göhmann, K.K.~Kozlowski and A.~Wei{\ss}e, \emph{{"Asymptotic
  analysis of the dynamical field-field correlation function of the
  impenetrable {B}ose gas."}}, cond-mat.stat-mech: 2611.08350, 43 pp.

\bibitem{KozFaulmanGohmannLowTNLIERigourousAnalysisForQTMMasslessRegime}
S.~Faulmann, F.~G\"{o}hmann, and K.K. Kozlowski, \emph{{"Low-temperature
  spectrum of the quantum transfer matrix of the XXZ chain in the massless
  regime."}}, Probability and Mathematical Physics \textbf{6} (2025),
  1507--1658.

\bibitem{GaudinTBAXXZMassiveInfiniteSetNLIE}
M.~Gaudin, \emph{{"Thermodynamics of a Heisenberg-Ising ring for $\De \geq
  1$."}}, Phys. Rev. Lett. \textbf{\bf 26} (1971), 1301--1304.

\bibitem{GaudinMcCoyWuNormXXZ}
M.~Gaudin, B.M. McCoy, and T.T. Wu, \emph{{"Normalization sum for the Bethe's
  hypothesis wave functions of the Heisenberg-Ising chain."}}, Phys. Rev. D
  \textbf{\bf 23} (1981), 417--419.

\bibitem{KozGohmannGoomaneeSuzukiRigorousApproachQTMForFreeEnergy}
F.~G\"{o}hmann, S.~Goomanee, K.K. Kozlowski, and J.~Suzuki,
  \emph{{"Thermodynamics of the spin-1/2 Heisenberg-Ising chain at high
  temperatures: a rigorous approach."}}, Comm. Math. Phys. \textbf{377} (2020),
  623--673.

\bibitem{KozGohmannSpaceLikeAEFor2PtsFctsThermalXXZ}
F.~G\"{o}hmann and K.K. Kozlowski, \emph{{"Space-like asymptotics of the
  thermal two-point functions of the XXZ spin-1/2 chain."}}, J. Stat. Mech:
  Theor and Exp. (2024), 033101, 21 pages.

\bibitem{KozGohmannSuzukiXXTransverseDynamicInfiniteT}
F.~G\"{o}hmann, K.K. Kozlowski, and J.~Suzuki, \emph{{"High-temperature
  analysis of the transverse dynamical two-point correlation function of the XX
  quantum-spin chain."}}, J. Math. Phys. \textbf{61} (2020), 013301.

\bibitem{KozGohmannSuzukiXXTransverseSpaceLikeAsymptotics}
\bysame, \emph{{"Long-time large-distance asymptotics of the transverse
  correlation functions of the XX chain in the space-like regime."}}, Lett.
  Math. Phys. \textbf{110} (2020), 1783–1797.

\bibitem{GranetEsslerGGEFFSumIncMinus1}
E.~Granet and F.H.L. Essler, \emph{{"A systematic $1/c$-expansion of form
  factor sums for dynamical correlations in the Lieb-Liniger model."}}, SciPost
  Phys. \textbf{\bf 9} (2020), 082, 76pp.

\bibitem{KozGohmannMininTimeAndSpaceDpdtThermalGSK}
F.~Göhmann, K.K. Kozlowski, and M.~Minin, \emph{{"Asymptotic analysis of a
  Fredholm determinant occurring in the description of the dynamical
  correlation functions of the Lieb–Liniger Bose gas."}}, math-ph:
  2608.16727, 87 pp.

\bibitem{ItsIzerginKorepinCorrRadiusImpBosons}
A.R. Its, A.G. Izergin, and V.E. Korepin, \emph{{"Correlation radius for the
  one-dimensional impenetrable bosons."}}, Phys. Lett. \textbf{\bf{A141}}
  (1989), 121--125.

\bibitem{ItsIzerginKorepinTemperatureLongDistAsympBoseGas}
\bysame, \emph{{"Long-distance asymptotics of temperature correlators of the
  impenetrable Bose gas."}}, Comm. Math. Phys. \textbf{\bf{130}} (1990),
  471--488.

\bibitem{ItsIzerginKorepinSlavnovDifferentialeqnsforCorrelationfunctions}
A.R. Its, A.G. Izergin, V.E. Korepin, and N.A. Slavnov, \emph{{"Differential
  equations for quantum correlation functions."}}, Int. J. Mod. Physics
  \textbf{\bf{B4}} (1990), 1003--1037.

\bibitem{ItsIzerginKorepinSlavnovTempCorrFctSpinsXY}
\bysame, \emph{{"Temperature correlations of quantum spins."}}, Phys. Rev.
  Lett. \textbf{\bf{70}} (1993), 1704--1706.

\bibitem{ItsIzerginKorepinVarzuginTimeSpaceAsymptImpBoseGaz}
A.R. Its, A.G. Izergin, V.E. Korepin, and G.G. Varzugin, \emph{{"Large time and
  distance asymptotics of field correlation function of impenetrable bosons at
  finite temperature."}}, Physica D \textbf{{ \bf 54}} (1992), 351.

\bibitem{ItsSlavnovNLSTimeAndSpaceCorrDualFields}
A.R. Its and N.A. Slavnov, \emph{{"On the Riemann-Hilbert approach to the
  asymptotic analysis of the correlation functions of the Quantum Nonlinear
  Schr\"{o}dinger equation. Non-free fermion case."}}, Theor. Math. Phys.
  \textbf{\bf{119}:2} (1999), 541--593.

\bibitem{IzerginKorepinLatticeVersionsofQFTModelsABANLSEandSineGordon}
A.G. Izergin and V.E. Korepin, \emph{{"Lattice versions of quantum field theory
  models in two dimensions."}}, Nucl. Phys. B \textbf{\bf{205}} (1982),
  401--413.

\bibitem{IzerginKorepinReshetikhinCommRelationFieldOpAndMonoMatrix}
A.G. Izergin, V.E. Korepin, and N.Yu. Reshetikhin, \emph{{"Correlation
  functions in a one-dimensional Bose gas."}}, J. Phys. A: Math. Gen.
  \textbf{\bf{20}} (1987), 4799--4822.

\bibitem{KozBabilonGohmannMininFFcurrentOperator}
K.~K.~Kozlowski J.~Babilon, F.~G\"ohmann and M.~D. Minin, \emph{{"{F}orm
  factors and correlation functions of the particle current in the
  {L}ieb--{L}iniger {B}ose gas."}}, In preparation, 2026., .

\bibitem{JieLargetABofDynCorrFctsinXX}
X.~Jie, \emph{{"The large time asymptotics of the temperature correlation
  functions of the XX0 Heisenberg ferromagnet: The Riemann-Hilbert
  approach."}}, Ph.D. thesis, Indiana University Purdue University Indianapolis
  (1998), 84pp.

\bibitem{KozKitMailSlaTer6VertexRMatrixMasterEquation}
N.~Kitanine, K.K. Kozlowski, J.-M. Maillet, N.A. Slavnov, and V.~Terras,
  \emph{{"On correlation functions of integrable models associated with the
  six-vertex R-matrix."}}, J. Stat. Mech. (2007), P01022.

\bibitem{KozKitMailSlaTerXXZsgZsgZAsymptotics}
\bysame, \emph{{"Algebraic Bethe Ansatz approach to the asymptotics behavior of
  correlation functions."}}, J. Stat. Mech: Th. and Exp. \textbf{04} (2009),
  P04003.

\bibitem{KozKitMailSlaTerEffectiveFormFactorsForXXZ}
\bysame, \emph{{"On the thermodynamic limit of form factors in the massless XXZ
  Heisenberg chain."}}, J. Math. Phys. \textbf{50} (2009), 095209.

\bibitem{KozKitMailSlaTerRHPapproachtoSuperSineKernel}
\bysame, \emph{{"The Riemann-Hilbert approach to a generalized sine kernel and
  applications."}}, Comm. Math. Phys. \textbf{{\bf 291}} (2009), 691--761.

\bibitem{KozKitMailSlaTerThermoLimPartHoleFormFactorsForXXZ}
\bysame, \emph{{"Thermodynamic limit of particle-hole form factors in the
  massless XXZ Heisenberg chain."}}, J. Stat. Mech. : Th. and Exp.
  \textbf{1105} (2011), P05028.

\bibitem{KitanineMailletTerrasFormfactorsperiodicXXZ}
N.~Kitanine, J.-M. Maillet, and V.~Terras, \emph{{"Form factors of the XXZ
  Heisenberg spin-$1/2$ finite chain."}}, Nucl. Phys. B \textbf{554} (1999),
  647--678.

\bibitem{KlumperNLIEfromQTMDescrThermoXYZOneUnknownFcton}
A.~Kl\"{u}mper, \emph{{"Thermodynamics of the anisotropic spin-$1/2$ Heisenberg
  chain and related quantum chains."}}, Z. Phys. B: Cond. Mat. \textbf{\bf 91}
  (1993), 507--519.

\bibitem{KlumperShiroishiTakahashiEquivalenceTBAandQTM}
A.~Kl\"{u}mper, M.~Shiroishi, and M.~Takahashi, \emph{{"Equivalence of TBA and
  QTM."}}, J. Phys. A: Math. Gen. \textbf{34} (2001), L187.

\bibitem{KojimaKorepinSlavnovNLSEDeterminatFormFactorAndDualFieldTempeAndTime}
T.~Kojima, V.E. Korepin, and N.A. Slavnov, \emph{{"Determinant representation
  for dynamical correlation functions of the quantum nonlinear Schr\"{o}dinger
  equation."}}, Comm. Math. Phys. \textbf{\bf 188} (1997), 657--689.

\bibitem{KorepinAnalysisofBoundStateConditionMassiveThirring}
V.E. Korepin, \emph{{"Direct calculation of the $S$-matrix in the massive
  Thirring model."}}, Theor. Math. Phys. \textbf{\bf 41} (1979), 953--967.

\bibitem{KorepinNormBetheStates6-Vertex}
\bysame, \emph{{"Calculation of norms of Bethe wave-functions."}}, Comm.
  Math.Phys. \textbf{\bf 86} (1982), 391--418.

\bibitem{KorepinFirstIntroDualFields}
\bysame, \emph{{"Dual field formulation of quantum integrable models."}},
  Comm.Math.Phys. \textbf{\bf 113} (1987), 177--190.

\bibitem{KorepinSlavnovTimeDepCorrImpBoseGas}
V.E. Korepin and N.A. Slavnov, \emph{{"The time dependent correlation function
  of an impenetrable Bose gas as a Fredholm minor I."}}, Comm. Math.Phys.
  \textbf{\bf 129} (1990), 103--113.

\bibitem{KorepinSlavnovRHPForNLSMOpertorValuedSettingInto}
\bysame, \emph{{"The Riemann--Hilbert problem associated with the quantum
  Nonlinear Schr\"{o}dinger equation."}}, J.Phys. \textbf{A30} (1997),
  8241--8255.

\bibitem{KorepinSlavnovFormFactorsNLSEasDeterminants}
\bysame, \emph{{"The form factors in the finite volume."}}, Int. J. Mod. Phys.
  B \textbf{\bf 13} (1999), 2933--2942.

\bibitem{KorffHilbertSpaceStructureXXZAtRootOfUnity}
C.~Korff, \emph{{"Auxiliary matrices for the six-vertex model at $q^N = 1$ and
  a geometric interpretation of its symmetries."}}, J. Phys. A: Math. Gene.
  \textbf{36} (2003), 5229, 38pp.

\bibitem{KozProofOfStringSolutionsBetheeqnsXXZ}
K.K. Kozlowski, \emph{{"On string solutions to the Bethe equations for the XXZ
  chain: a rigorous approach."}}, to appear.

\bibitem{KozFFConjFieldNLSELatticeSpacingGoes0}
\bysame, \emph{{"On form factors of the conjugated field in the non-linear
  Schr\"{o}dinger model."}}, J. Math. Phys. \textbf{52} (2011), 083302.

\bibitem{KozTimeDepGSKandNatteSeries}
\bysame, \emph{{"Riemann--Hilbert approach to the time-dependent generalized
  sine kernel."}}, Adv. Theor. Math. Phys. \textbf{15} (2011), 1--89.

\bibitem{KozReducedDensityMatrixAsymptNLSE}
\bysame, \emph{{"Large-distance and long-time asymptotic behavior of the
  reduced density matrix in the non-linear Schr\"{o}dinger model."}}, Ann.
  Henri-Poincar\'{e} \textbf{16} (2015), 437--534.

\bibitem{KozProofOfAsymptoticsofFormFactorsXXZBoundStates}
\bysame, \emph{{"Form factors of bound states in the XXZ chain."}}, J. Phys. A:
  Math. Theor. Special Issue "Emerging talents" \textbf{50} (2017), 184002.

\bibitem{KozMasslessFFSeriesXXZ}
\bysame, \emph{{"On the thermodynamic limit of form factor expansions of
  dynamical correlation functions in the massless regime of the XXZ spin $1/2$
  chain."}}, J. Math. Phys. "Ludwig Faddeev memorial volume" \textbf{59}
  (2018), 091408.

\bibitem{KozTerrasNatteSeriesNLSECurrentCurrent}
K.K. Kozlowski and V.~Terras, \emph{{"Long-time and large-distance asymptotic
  behavior of the current-current correlators in the non-linear Schr\"{o}dinger
  model."}}, J. Stat. Mech.: Th. and Exp. (2011), P09013.

\bibitem{LiebExcitationStructureBoseGas}
E.H. Lieb, \emph{{"Exact analysis of an interacting Bose gas II: the excitation
  spectrum."}}, Phys. Rev. \textbf{\bf{130}} (1963), 1616--1624.

\bibitem{LiebLinigerCBAForDeltaBoseGas}
E.H. Lieb and W.~Liniger, \emph{{"Exact analysis of an interacting Bose gas. I.
  The general solution and the ground state."}}, Phys. Rev. \textbf{\bf{130}}
  (1963), 1605--1616.

\bibitem{Panfil2phDSFGGENLSM}
M.~Panfil, \emph{{"The two particle-hole pairs contribution to the dynamic
  correlation functions of quantum integrable models."}}, J. Stat. Mech.
  \textbf{\bf 2021} (2021), 013108, 31pp.

\bibitem{SakhnovichFirstStepsOfElementsOfIntegrableIntOps}
L.A. Sakhnovich, \emph{{"Operators which are similar to unitary operators with
  absolutely continuous spectrum."}}, Funct. Anal. and App. \textbf{\bf 2}
  (1968), 48--60.

\bibitem{SimonsInfiniteDimensionalDeterminants}
B.~Simons, \emph{{"Notes on infinite determinants of Hilbert space
  operators."}}, Adv. Math. \textbf{\bf{24}:2} (1977), 244--273.

\bibitem{SlavnovScalarProductsXXZ}
N.A. Slavnov, \emph{{"Calculation of scalar products of wave-functions and
  form-factors in the framework of the algebraic Bethe Ansatz."}}, Theor. Math.
  Phys. \textbf{\bf 79} (1989), 502--508.

\bibitem{SlavnovFormFactorsNLSE}
\bysame, \emph{{"Non-equal time current correlation function in a
  one-dimensional Bose gas."}}, Theor. Math. Phys. \textbf{\bf 82} (1990),
  273--282.

\bibitem{SlavnovPDE4MultiPtsFreeNLSM}
\bysame, \emph{{"Differential equations for multipoint correlation functions in
  a one-dimensional impenetrable Bose gas."}}, Theor. Math. Phys. \textbf{\bf
  106} (1996), 131--142.

\bibitem{SlavnovComputationDualFieldVaccumExpLongTimeDistTempeRedDensNLSE}
\bysame, \emph{{"Integral equations for the correlation functions of the
  quantum one-dimensional Bose gas."}}, Theor. Math. Phys. \textbf{\bf 121}
  (1999), 1358--1376.

\bibitem{SpohnGGEBook}
H.~Spohn, \emph{{"Hydrodynamic Scales of Integrable Many-Body Systems."}},
  World Scientific, 2024.

\bibitem{TakahashiTBAforXXZFiniteTinfiniteNbrNLIE}
M.~Takahashi, \emph{{"One-dimensional Heisenberg model at finite
  temperature."}}, Prog. Theor. Phys. \textbf{\bf 42} (1971), 1289.

\bibitem{TakahashiThermodynamics1DSolvModels}
\bysame, \emph{{"Thermodynamics of one dimensional solvable models."}},
  Cambridge university press, 1999.

\bibitem{TakahashiSuzukiFiniteTXXZandStrings}
M.~Takahashi and M.~Suzuki, \emph{{"One-dimensional anisotropic Heisenberg
  model at finite temperatures."}}, Prog. Theor. Phys. \textbf{\bf 48} (1972),
  2187--2209.

\bibitem{YangYangNLSEThermodynamics}
C.N. Yang and C.P. Yang, \emph{{"Thermodynamics of a one-dimensional system of
  bosons with repulsive delta-interactions."}}, J. Math. Phys. \textbf{\bf 10}
  (1969), 1115--1122.

\end{thebibliography}
\end{document}